\documentclass[11
pt]{article}

\usepackage[a4paper,margin=1in]{geometry}
\usepackage{amsmath,amssymb}
\usepackage{microtype}
\usepackage{hyperref}

\title{Renormalization Group and String Loops }
\author{A.\ A.\ Tseytlin\\
\\
{\it  P.N.\ Lebedev Physical Institute, Moscow 117924, USSR}}
\date{\small May 1989}

\begin{document}
\maketitle

\begin{abstract}
The fixed points of the 2d renormalization group flow are known to correspond to the tree level string vacua. We analyze how the ``renormalization group'' (or ``sigma model'') approach can be extended to the string loop level. The central role of the condition of renormalizability of the generating functional for string amplitudes with respect to both ``local'' and ``modular'' infinities is emphasized. Several one- and two-loop examples of renormalization are discussed. It is found that in order to ensure the renormalizability of the generating functional one is to use ``extended'' (e.g.\ Schottky-type) parametrizations of moduli space. An approach to resummation of the string perturbative expansion based on operators of insertion of topological fixtures is suggested.
\end{abstract}

\

\

\noindent\textbf{\ \ \ \ \ Contents} 

1.\ Introduction 

2.\ Tree approximation

3.\ String loop divergences and their renormalization 

4.\ Generating functional (or $\sigma $ model) approach to renormalization of string loops 

5.\ Correspondence with effective action 

6.\ String generating functional in Schottky parametrization

7.\ Renormalization at one Loop: torus and disc topologies 

8.\ Genus-two examples of renormalization 

9.\ Approach based on operators of insertion of topological fixtures 

10.\,Concluding remarks 

  \ \ \  \ \  Appendix \\

\

\

{\it Lectures at    ICTP   Trieste School and Workshop on Superstrings,   
3-14 April 1989

Published  in Int. J. Mod. Phys. A 5 (1990) 589--658  }

\def \ve {\varepsilon}
\def \vepsilon {\varepsilon}  \def \del {\partial} \def \foot {\footnote} 

\newpage
\section{Introduction}

The remarkable role played by the renormalization group (RG) in determining the string equations of motion and the effective action at the string tree level [1-6] (for  reviews   and extensive list of references see [7,8])
 suggests that the RG approach may shed light on (or even be a part of) a more fundamental formulation of string theory. In fact, in order for the RG to operate consistently at string loop level, the relative weights of string diagrams should take some particular values. Also, an understanding of a role of RG and 2d cutoffs is important for an off-shell extension of string theory.

The idea that the equivalence between the string equations of motion and the vanishing of some $\beta$-functions may be generalized to string loop level, first suggested in Refs.~1 and 9, was studied, e.g.\ in Refs.~10--19. If true it would imply that one is able to determine the true ``loop-corrected'' string vacua by classifying some ``nonperturbatively conformal invariant'' 2d theories. Below we are going to discuss the basic issues related to the possibility of extension of the $\sigma$ model approach to string loop level.

To understand if (and how) the RG acts in string loops we should first determine a basic object of the theory which should be RG invariant, i.e.\ renormalizable. The object is the (properly defined) generating functional for string amplitudes $\hat Z$ [2] which can be expressed in terms of the $\sigma$ model partition function $Z$ (integrated over moduli parameters). Introducing a universal 2d short distance cutoff (which regularizes both the $\sigma$ model and integrals over the moduli) we find the ``local'' and ``modular'' divergences in $\hat Z$. 

The renormalizability condition implies that we should be able to absorb these divergences into the $\sigma$ model couplings $\varphi^i$ which can be interpreted as the space-time fields corresponding to the massless string models. We would like to stress that it is only if $\hat Z$ is renormalizable that we can define the ``loop corrected'' $\beta$-functions ($\beta^i=\mathrm{d}\varphi^i/\mathrm{d}\log\vepsilon$) and hence may question their relation to string equations of motion. As we shall see, it is nontrivial to define $\hat Z$ in such a way that it satisfies the renormalizability condition with respect to all  (``local'' and ``modular'')  infinities.\foot{It is important to have the combined ``local" plus ``modular'' (plus ``mixed'') renormalisability in order for the corresponding ``local'' and ``modular'' pieces to appear on equal footing in the beta functions, as they do in the effective action   and hence in  the string equations of motion.    } 

The crucial difference between the string theory object $\hat Z$ and the $\sigma$ model object $Z$ is that for the on-shell values of its arguments $\hat Z$ should reproduce the usual expressions for the on-shell string amplitudes and hence the definition of $\hat Z$ should include a subtraction of the M\"obius infinities at the tree level and an integration over the moduli at higher loop orders. It turns out that the procedures of regularization and fixing  of the M\"obius symmetry do not ``commute''.

 To implement the RG in a consistent way it is necessary to treat all the infinities including the M\"obius ones on an equal footing. This implies in particular that to ensure the renormalizability property of $\hat Z$ one should use an ``extended'' (e.g.\ Schottky) set of moduli parameters with which one has formal on-shell projective invariance, i.e.\ the formal $\Omega^{-1}=(\mathrm{vol}\,SL(2,\mathbb C))^{-1}$ M\"obius factor for all terms in the expansion in genus. Regularizing the corresponding modular integrals, one is then to replace the $\Omega^{-1}\sim (\log \vepsilon$)$^{-1}$ factor [20,21]
 by the RG invariant  operator $\del/\del \log \ve$ [21,22]. 
 
If we  first fix the M\"obius gauge reducing the number of modular parameters to the usual one ($3n-3$, $n\ge 2$) the $\hat Z$ we get will not be {\it   not}  be  renormalizable with respect to both local and modular infinities. The crucial point is that the M\"obius infinities can be interpreted as a subclass of local or model infinities and hence should be regularized on an equal footing with all other local and modular infinities.

\

We shall start (Sec.\ 2) with a brief review of the $\sigma$ model approach at the string tree level. In Sec.\ 2 we shall also discuss the nature of ``local'' infinities present in the string correlations and give an argument which demonstrates the renormalizability of the string generating functional starting from the assumption that the massless sector of the string $S$-matrix can be reproduced by an effective field theory.

In Sec.\ 3 we shall first recall the classification of divergences which appear in string loop amplitudes (see Refs.~23--26), emphasizing that the ``non-dividing'' tachyonic singularities should be avoided by using a kind of analytic continuation prescription  [27, 28] and hence should be ignored in the RG approach. Then we shall analyze the structure of the ``tadpole'' logarithmic infinities, which may be interpreted in terms of propagation of the zero momentum dilaton and (scalar part of) graviton. We shall find that, in principle, they may be cancelled out by renormalizing the slope ($\alpha'$) and the string coupling ($g$) parameters of the string theory in a trivial (flat) vacuum.

In Sec.\ 4 we introduce the generating functional for string amplitudes $\hat Z$ and describe how one may eliminate the tadpole (and external leg or self-energy insertion) infinities by renormalizing the $\sigma$ model couplings (the space-time metric and the dilaton) thus reinterpreting the result of Sec.\ 3 about the renormalization of $\alpha'$ and $g$. The corresponding ``modular'' terms in the $\beta$-functions are field independent (tadpole) and  ``linear in $\varphi$'' (external leg) corrections. Their complicated dependence on the string coupling, i.e.\ on the constant part of the dilaton suggests that the $\beta$-functions should also contain other ``modular'' contributions which are of higher orders in the fields $\varphi^i$. These terms correspond, in fact, to the momentum dependent singularities of loop amplitudes which are due to the massless poles in transferred momentum (which contribute to the Weyl or BRST anomaly). We stress that the inclusion of these higher order terms is necessary for a correspondence between the $\beta$-functions and the effective action. We also discuss the higher order $\log^n\vepsilon$ counterterms whose presence is dictated by the RG.

The correspondence with the effective action (EA) is analyzed in detail in Sec.\ 5. We explain how the assumption of an existence of a EA reproducing (by the tree diagrams) the massless loop-corrected string amplitudes imposes constraints on relative weights of string loop corrections and also implies renormalizability of $\hat Z$. We formulate the higher loop generalization of the tree level relations between the generating functional $\hat Z$, the effective action $S$ and the $\beta$-functions, which already are assumed to contain contributions from all ``dividing'' (momentum dependent as well as momentum independent) singularities corresponding to the massless poles in the amplitudes. As an application, we discuss a simple solution of the $\beta=0$ or $\delta S/\delta\varphi=0$ equation in the leading approximation, accounting only for the one-loop cosmological term in the EA. While the solution with (anti) de Sitter metric [9] and constant   dilaton 
formally exists if $D\neq 26$, there is a problem of its interpretation since the one-loop cosmological constant is complex because of the analytic continuation prescription used to regularize the tachyonic loop infinity [28].

In Secs.\ 6--9 we  try  to check the condition of renormalizability of the generating functional, analyzing systematically the loop contribution to $\hat Z$. As we already mentioned above, to realize RG in string loops in a consistent way it is necessary to use an ``extended'' parametrization of the moduli space, in which the M\"obius volume factor $\Omega^{-1}$ appears in front of all loop corrections to the amplitudes. A distinguished parametrization of this type is the Schottky parametrization [23--26, 29,31].
 In Sec.\ 6 we present the expression for the generating functional $Z$ in the Schottky parametrization.

The renormalization of $\hat Z$ in the one-loop approximation is studied in detail in Sec.\ 7, where we use the Schottky-type parametrization (for the torus and the disc) to isolate the modular infinities from the local ones. We check the renormalizability of $\hat Z$ with respect to the modular tadpole infinities. To obtain the consistent result in the case of the disc topology one is to use the ``corrected'' expression for the modular measure [15].
In particular, we discuss the renormalization of the $\alpha'R\log\vepsilon$ infinity in the one-loop contribution to $\hat Z$ (see also Refs.~19 and 22) and show that one is to use a special definition of $\hat Z$ (a special prescription for subtracting of M\"obius infinities) in order to satisfy the requirement of renormalizability with respect to the sum of local and modular infinities.

In Sec.\ 8 we consider the renormalization of the derivative-independent part of $\hat Z$ (i.e.\ of the ``string partition function'') on the two-loop examples (annulus and genus 2 surface). The logarithmic divergence in the genus 2 string partition function is found explicitly in the Schottky parametrization. It is shown how the condition that this divergence should be cancelled out by the renormalization of the metric in the torus contribution to $\hat Z$ fixes the overall coefficient in the two-loop expression. We also interpret the results in terms of the tree diagrams of the corresponding effective action.

In the final Sec.\ 9 we rederive and generalize the results of the previous sections using the approach based on a representation of $\hat Z$ in terms of the operators of insertion of ``topological fixtures'' like holes, handles, etc. (see also Ref.~32).
For the RG to actually operate in string theory the divergences should ``exponentiate'' to become counterterms in the string ($\sigma$  model) action. 

This suggests a resummation of the standard expansion in genus, which is analogous to a ``renormalization group improved'' perturbation theory or may be interpreted as a ``dilute handle gas'' approximation (valid only in some regions of moduli spaces). For the vacuum string partition function we propose an exact expression in terms of the expectation value on the sphere of the exponent of the sum of ``subtracted'' (or ``induced'') topological fixture operators. We demonstrate how to construct these modified operators for the genus 2 example. This approach seems to be very promising. A possible point of view  is that consistently imposing the requirement that the renormalization group should act  on  string loops may eventually lead us to a ``nonperturbative'' formulation of string theory.

Section 10 contains some concluding remarks. In the Appendix we discuss the computation of the string generating functional within the weak field expansion.

\def \td {\tilde}  

\section{Tree approximation}

To motivate the attempt to generalize the ``renormalization group'' or ``sigma model'' approach to string loop level it is useful first to recall briefly some basic facts known in the tree approximation (sigma model approach at the string tree level was  discussed in the review  [7]  where an extensive list of references can be found).

\subsection*{1.}

\def \pvarphi {\phi}   \def \D {{\cal D}} 

Let $V_i$ be the dimension 2 interaction vertices present in the action of a general renormalizable bosonic sigma model and $\varphi^i$ be the corresponding dimensionless coupling constants. The partition function for the sigma model [33] 
defined on a compact 2-space (of an arbitrary Euler number $\chi$) with some regular 2-metric $g_{ab}$ is
\begin{equation}
Z=\int Dx\,e^{-I}\,,
\qquad
I=I_0+I_{\mathrm{int}}\,,
\tag{2.1}
\end{equation}
where
\begin{equation}
I_0=\frac{1}{4\pi\alpha'}\int d^2z\,\sqrt{g}\,\partial_a x^\mu\partial^a x^\nu G_{\mu\nu}\,,
\qquad
I_{\mathrm{int}}=\varphi^i V_i\,,
\tag{2.2}
\end{equation}
i.e.
\begin{equation}
I=\frac{1}{4\pi\alpha'}\int d^2z\,\sqrt{g}\,\partial_a x^\mu\partial^a x^\nu G_{\mu\nu}(x)
+\frac{1}{4\pi}\int d^2z\,\sqrt{g}\,R^{(2)}\phi(x)\,.
\tag{2.3}
\end{equation}
The measure $Dx$ should be defined so that the theory is covariant under the general coordinate transformations of $x$. Naively,
\begin{equation}
Dx=\prod_{z} dx(z)\sqrt{G(x(z))}\,,
\qquad
G=\det G_{\mu\nu}\,.
\tag{2.4}
\end{equation}
A particular definition of the product in (2.4) depends on a regularization chosen (see Refs.~7 and 22 for details). One may introduce a short distance 2d cutoff through the free propagator, e.g.
\begin{equation}
\mathcal{G}(z,z')=-\frac{1}{4\pi}\log\!\Big(a^{-2}|z-z'|^2+\vepsilon^2 e^{-\rho(z)-\rho(z')}\Big),
\qquad
g_{ab}=e^{2\rho}\delta_{ab},\quad
\vepsilon^2=(a\Lambda)^{-2}\to 0\,,
\tag{2.5}
\end{equation}
or
\begin{equation}
\mathcal{G}(z,z')=\sum_{n\neq 0} e^{- \vepsilon^2\lambda_n}\lambda_n^{-1}\psi_n(z)\psi_n(z')\,,
\qquad
\vepsilon=\Lambda^{-1}\,,
\tag{2.5$'$}
\end{equation}
\begin{equation}
\Delta \mathcal{G}=\delta^{(2)}(z,z')-\frac{1}{V}\,,
\qquad
V=\int d^2z\,\sqrt{g}\,,
\qquad
\Delta=-\nabla^2\,,
\tag{2.6}
\end{equation}
where $\lambda_n$ and $\psi_n$ are the eigenvalues and eigenfunctions of the Laplacian and $a\sim V^{1/2}$ is a ``scale'' of the 2-space. Under a proper choice of the measure all power (quadratic) infinities cancel out and one finds [22]\footnote{See also Refs.~33, 34 and 35 for other discussions of the measure in the $\sigma$-model.}
\begin{equation}
Z=\int d^D y\,\sqrt{G}\,e^{-\chi\phi-W}\,,
\qquad
\chi=\frac{1}{4\pi}\int d^2z\,\sqrt{g}\,R^{(2)}\,.
\tag{2.7}
\end{equation}
Here $G$ and $\phi$ are the bare couplings and $W(G,\phi,g_{ab},\vepsilon)$ is given by the sum of 1-PI diagrams constructed with the propagator in which the constant (zero mode) part $y^\mu$ of $x^\mu$ is ``projected out''. $W$ is a covariant function of $G$ and $\phi$ depending on their derivatives at the point $y^\mu$. In general
\begin{equation}
W=\gamma_1+\gamma_2\log\vepsilon+\gamma_3\log^2\vepsilon+\cdots\,,
\tag{2.8}
\end{equation}
where $\gamma_i$ are functionals of $G$, $\phi$ and $g_{ab}$, e.g.
\begin{equation}
\gamma_1=-\frac{1}{16\pi} \td  \beta^\phi\, \int  R^{(2)}\Delta^{-1}R^{(2)}+\cdots\,.
\tag{2.9}
\end{equation}
$\td \beta^{\pvarphi}$ is the basic Weyl anomaly coefficient related to the $\beta$-functions by
\begin{equation}
\td \beta^{\phi}=\bar\beta^{\varphi}-\frac{1}{4}\,  K^{\mu\nu}  \bar\beta_{\mu\nu}^{G},
\qquad
K^{\mu\nu}=G^{\mu\nu}+\cdots\,,
\tag{2.10}
\end{equation}
\begin{equation}
\bar\beta^{\phi}=\beta^{\phi}+\xi^\mu\partial_\mu\phi,
\qquad
\bar\beta_{\mu\nu}^{G}=\beta_{\mu\nu}^{G}+\D_\mu\xi_\nu+\D_\nu\xi_\mu.
\tag{2.11}
\end{equation}
$\bar\beta^{i}$ are the Weyl anomaly coefficients which appear in the operator expression for the trace of the 2d energy momentum tensor [36].  Power counting and the proper choice of the measure guarantees the renormalizability of $Z$ within the loop $(\alpha')$ expansion
\begin{equation}
Z(\varphi(\vepsilon),\vepsilon)=Z_R(\varphi_R)\,,
\tag{2.12}
\end{equation}
\begin{equation}
\frac{dZ}{d\lambda}=\frac{\partial Z}{\partial\lambda}-\beta^{i}\frac{\partial Z}{\partial\varphi^{i}}=0,
\qquad
\beta^{i}(\varphi)=-\frac{d\varphi^{i}}{d\lambda},
\qquad
\lambda=\log\vepsilon\,.
\tag{2.13}
\end{equation}
Direct computation shows that [22]\footnote{To make contact with string theory we included  the standard ``-26''   contribution of the ghosts [37].}
\begin{equation}
W=W_0-\frac{1}{2}\alpha'\log\vepsilon\,\Big(R+(\nabla\phi)^2\Big)+\cdots\,,
\tag{2.14}
\end{equation}
\begin{equation}
W_0=-\frac{1}{6}(D-26)\bigl(\chi\log\vepsilon+\cdots\bigr)
+\frac{1}{16\pi}\int R^{(2)}\Delta^{-1}R^{(2)}+\text{const.}
\nonumber
\end{equation}
Hence (see (2.7)  and also  the Appendix) 
\begin{equation}
Z=a_{\chi}e^{\frac{1}{6}(D-26)\chi\log\vepsilon}
\int d^D y\,\sqrt{G}\,e^{-\chi\phi}\Big[1+\frac{1}{2}\alpha'\log\vepsilon\Big(R+\chi \D^2\phi \Big)+\cdots\Big].
\tag{2.15}
\end{equation}
The coefficients of the $\log\vepsilon$ terms in (2.14) are consistent with the renormalizability of $Z$ (Eq.\ (2.13)) as one can check using the known expressions for the $\beta$-functions [33, 36, 38]
\begin{equation}
\beta^{G}_{\mu\nu}=\alpha' R_{\mu\nu}+\frac{1}{2}\alpha'^2 R_{\mu\alpha\beta\gamma}R_{\nu}{}^{\alpha\beta\gamma}+O(\alpha'^3)\,,
\tag{2.16}
\end{equation}
\begin{equation}
\beta^{\phi}=\frac{1}{6}(D-26)-\frac{1}{2}\alpha'\D^2\phi+\frac{1}{16}\alpha'^2 R_{\mu\nu\alpha\beta}R^{\mu\nu\alpha\beta}+O(\alpha'^3)\,.
\tag{2.17}
\end{equation}
Substituting the expressions for the bare fields
\begin{equation}
G_{\mu\nu}=G_{R\,\mu\nu}-\log\vepsilon\, \alpha' R_{\mu\nu}+\cdots\,,
\qquad
\phi=\phi_R-\frac{1}{6}(D-26)\log\vepsilon+\cdots
\tag{2.18}
\end{equation}
into $Z$ (2.7) we indeed find the cancellation at the leading $\log\vepsilon$ terms.

\def \no {\nonumber }  \def \vphi {\varphi}

It is possible to prove that the $\bar\beta$-functions can be expressed in terms of the derivatives
of some functional $S$ [39--41,7]
\begin{equation}
\bar\beta^{i}=\kappa^{ij}\,\frac{\delta S}{\delta \varphi^{j}} \, ,
\tag{2.19}
\end{equation}
where the matrix $\kappa$ is nondegenerate (within the $\alpha'$-expansion)
\begin{equation}
\kappa=\kappa_{0}+O(\alpha') \, , \qquad
\kappa^{GG}_{0\, \mu\nu,\alpha\beta}= k\, G^{-1/2} G_{\mu(\alpha}G_{\beta)\nu} \, ,\no 
\end{equation}
\begin{equation}
\kappa^{G\phi}_{0\, \mu\nu}= \kappa^{\phi G}_{0\, \mu\nu}
= \frac{1}{4} k\, G^{-1/2} G_{\mu\nu} \, , \qquad
\kappa^{\phi\phi}_0= \frac{1}{16} k \, G^{-1/2}(D-2) \, .
\tag{2.20}
\end{equation}
Under a particular choice of the couplings (i.e.\ in a particular renormalization scheme) $S$
can be represented in the form [3,40]
\begin{equation}
S = a \int d^{D}y\,\sqrt{G}\,e^{-2\phi}\,\td \beta^{\phi} \, , \qquad
a=4k^{-1} \, ,
\tag{2.21}
\end{equation}
 \begin{equation}
\td\beta^{\phi}=\frac{1}{6}(D-26)
-\frac{1}{4}\alpha'\bigl(R+4\D^{2}\phi -4(\del\phi)^{2}\bigr)
+\frac{1}{4}\alpha'^2 R_{\mu\nu\lambda\rho}^{2}
+\ldots \, .
\tag{2.22}
\end{equation}
The relations (2.19), (2.21) were checked by explicit perturbative computations up to
rather high order in the expansion in $\alpha'$.
The leading $\alpha'$-independent term in $\kappa$ is essentially ``one'':
the nondiagonal terms in $\kappa_{0}$ are due to the ``mixing'' between the graviton
and the dilaton in the kinetic term in $S$ (2.21).
If we redefine the couplings to diagonalize the kinetic term [2]
\begin{equation}
G_{\mu\nu}=G'_{\mu\nu}\exp\!\Big(\frac{\phi'}{\sqrt{D-2}}\Big),
\qquad
\phi=\frac{1}{4}\sqrt{D-2}\,\phi' \, ,
\tag{2.23}
\end{equation}
\begin{equation}
S = a \int d^{D}y\,\sqrt{G'}\,
\Big[
\frac{1}{6}(D-26)\exp\!\Big(\frac{D}{2\sqrt{D-2}}\phi'\Big)
-\frac{1}{4}\alpha'\Bigl(R' - \frac{1}{4}(\partial\phi')^{2}\Bigr)
+\ldots
\Big] ,
\tag{2.24}
\end{equation}
then we find
\begin{equation}
\bar\beta^{i}=\kappa'^{ij}\frac{\delta S}{\delta \varphi'^j},
\qquad
\kappa'=\kappa'_0+O(\alpha') \, ,
\tag{2.25}
\end{equation}
\begin{equation}
\kappa'^{GG}_{0\,\mu\nu,\alpha\beta}
=k  G^{-1/2}
\Big[ G_{\mu(\alpha}G_{\beta)\nu}-(D-2)^{-1}G_{\mu\nu}G_{\alpha\beta} \Big],
\tag{2.26}
\end{equation}
\begin{equation}
\kappa'^{G\phi}_{0\, \mu\nu}=\kappa'^{\phi G}_{0\, \mu\nu}=0,
\qquad
\kappa'^{\phi\phi}_0=\kappa G^{-1/2}.
\tag{2.27}
\end{equation}
Expanding near the constant background $(G_{\mu\nu}=\delta_{\mu\nu}+h_{\mu\nu})$
\begin{equation}
S=\frac{1}{2}\,\vphi^{i}\Delta_{ij}\vphi^{j}+O(\vphi^{3}) \, ,
\tag{2.28}
\end{equation}
\begin{equation}
\Delta_{ij}=P_{ij}\Delta, \qquad \Delta=-\Box \, ,
\tag{2.29}
\end{equation}
we find by comparing the leading order terms in (2.19)
\begin{equation}
\kappa^{ij}_{0}P_{jk}=\frac{1}{2}\alpha'\delta^{i}_{k},
\qquad
\kappa_{0}=\frac{1}{2}\alpha' P^{-1} , 
\tag{2.30}
\end{equation}
where we have used that  $\beta^G_{\mu\nu}= {1\over 2} \Delta h_{\mu\nu} + ..., \  \beta^\phi= {1\over 2} \alpha' \Delta \phi + ... $.
Thus $\kappa_0^{-1}$ is proportional to the kinetic   matrix in $S$. 

\subsection*{2.}

A relation to string theory is established through the observation [2]  that $Z$ (2.1) is the
generating functional for the correlators of the massless vertex operators (we consider the
tree approximation and hence assume the spherical topology, $\chi=2$)
\begin{equation}
Z=\sum_{N=0}^{\infty}\frac{(-1)^{N}}{N!}\,T_{N}\varphi^{N},
\qquad
T_{N}=\langle V_{1}\ldots V_{N}\rangle .
\tag{2.31}
\end{equation}
The on-shell string tree amplitudes $A_{N}$ differ from $T_{N}$ by the infinite factor
of the SL$(2,\mathbb{C})$ M\"obius group volume, $A_{N}\sim\Omega^{-1}T_{N}$.
Introducing a short distance cutoff $\vepsilon$, one finds that (up to power divergent terms)
$\Omega\sim\log\vepsilon$  [20,21].
However, this is not a correct prescription since it does not treat all the short distance
divergences in $T_{N}$ on an equal footing.

$T_{N}$ contains the following types of logarithmic infinities:

(1) singularities corresponding to the region of integration where $N$ or $N-1$ integration
points $z_{i}$ are close to each other;  these are ``momentum  independent''  (i.e. present for arbitrary  on-shell external momenta)   M\"obius   infinities,  which are absent  in $A_N$   computed with 3 Koba-Nielsen (KN) points fixed

(2) ``momentum dependent'' massless pole singularities corresponding to regions where $M=2,\ldots,N-2$
points are close to each other.

As discussed in [21], 
the correct RG invariant relation between the regularized expressions for $A_{N}$ and $T_{N}$ is\foot{The difference between $\del/\del \log \ve$ and $(\log \ve)^{-1}$  prescriptions     becomes  important at  the $\log^2 \ve$  level.  For example, the 4-point  correlator contains  a $\log^2 \ve$   piece  in which one  $\log \ve$  factor is the M\"obius   divergence  and the other  corresponds to a massless  pole.    } 
\begin{equation}
A_{N}\sim \frac{\partial}{\partial\log\vepsilon}\,T_{N} .
\tag{2.32}
\end{equation}
Hence the generating functional for the massless string tree amplitudes
$\hat Z=\sum_{N}A_{N}\varphi^{N}$ can be represented as
\begin{equation}
\hat Z=\frac{\partial}{\partial\log\vepsilon}\,Z .
\tag{2.33}
\end{equation}
Since the partition function $Z$ is renormalizable, i.e.\ satisfies (2.13),
the same is true for $\hat Z$ 
\begin{equation}
\frac{d\hat Z}{d\lambda}=0,
\qquad
\hat Z(\vphi(\vepsilon),\vepsilon)=\hat Z_{R}(\vphi_{R}) , \qquad \lambda =\log \ve \ . 
\tag{2.34}
\end{equation}

We would like to stress that $\hat Z$ is the basic object which defines the
first-quantized string theory. In particular, it is $\hat Z$ and not the
amplitudes which satisfies the renormalizability property. Since a
renormalization of logarithmic divergences in $\hat Z$ corresponds to a
subtraction of the massless poles in the amplitudes [1,42], 
the renormalized value $\hat Z_R$ of $\hat Z$ should coincide [40] 
with the effective action (EA) $S$ which reproduces the massless sector of
the string $S$-matrix. 

Let $\mu$ be a renormalization parameter,
$Z = Z_R(\varphi_R(\mu),\, \mu)$. Then
\begin{equation}
\frac{dZ_{R}}{dt}=0,
\qquad
\hat Z_{R}=\frac{\partial}{\partial t}Z_{R}
=-\beta^{i}(\vphi_{R})\frac{\partial}{\partial\vphi^{i}_{R}}Z_{R}(\vphi_{R}),
\qquad
t=\log\mu .
\tag{2.35}
\end{equation}
Using (2.7) (for $\chi=2$) and redefining the fields to absorb the finite part of $W$, we get
\begin{equation}
Z_{R}\sim \int d^{D}y\,\sqrt{G_{R}}\,e^{-2\vphi_{R}}.\no 
\end{equation}
Substituting this into $\hat Z_R$ (2.35) we finish with [21]
   \[
\hat Z_R(\varphi) = S(\varphi)
= a \int d^D y \, \sqrt{G} \, e^{-2\phi} \, \tilde\beta^\phi \, ,
\qquad
\tilde\beta^\phi = \bar\beta^\phi
- \frac{1}{4} \, \bar\beta^G_{\mu\nu} G^{\mu\nu} .
\tag{2.36}
\]
(we used $\varphi$ instead of $\varphi_R$ to simplify the notation).

\def \v {\nu}

Comparing (2.21) and (2.36) we conclude that EA which reproduces the
massless string amplitudes coincides with the functional appearing in the
relation (2.19) for the Weyl anomaly coefficients. Since $\kappa^{ij}$ is
nondegenerate (within the $\alpha'$-expansion) the string effective low
energy equations of motion $\delta S / \delta \varphi^i = 0$ are thus
equivalent to the conditions of the Weyl invariance of the corresponding
renormalizable sigma model.

In view of (2.35) $S = \hat Z_R$ can be represented in terms of the
scale anomaly [{21}]  ($V_i$  are ``vertex operators''   that  appear in (2.2)) 
   \[
S = \frac{\partial}{\partial t} Z_R
= \Big\langle \int d^2 z \, \sqrt{g} \, \theta \Big\rangle ,
\tag{2.37}
\]
   \[
\langle \theta(z) \rangle
= \frac{2}{\sqrt{g}} \, g^{ab}
\frac{\delta Z_R}{\delta g^{ab}}
\sim \bar\beta^i \, v_i  ,
\qquad V_i = \int d^2 z \, \sqrt{g} \, v_i .
\]
While a straightforward generalization of the scale anomaly action (2.37)
may not reproduce the full string equations with all massive modes included
(see Ref.~43) it certainly gives the correct effective equations of motion
in the massless sector. 

Let us note also that
$\partial \hat Z / \partial \varphi^i
= -(\partial / \partial \lambda) \langle V_i \rangle$
is proportional to the massless tadpole computed in a background, i.e.
   \[
\frac{\partial S}{\partial \varphi_R^i}
= - \Big(
\frac{\partial}{\partial \lambda}
\langle V_i \rangle
\Big)_R
= A_{1i} .
\tag{2.38}
\]
Hence the effective equations of motion are indeed equivalent to the vacuum
stability condition, i.e. to the vanishing of the massless tadpoles in a
background.

Equation (2.37) gives the explicit representation for the effective action
in terms of the partition function of the $\sigma$ model. It is easy to
derive, for example, the Einstein term in the EA. According to (2.7),
(2.15) ($\chi = 2$; we set $D = 26$ for simplicity)
   \[
Z \sim \int d^D y \, \sqrt{G} \, e^{-2\phi}
\Big( 1 + \frac{1}{2} \alpha' R \log \vepsilon + \cdots \Big).
\tag{2.39}
\]
Since the order $R$ term corresponds to the (order $k^2$ term in)
3-graviton correlator on the sphere,
$T_3 = \langle V_1 V_2 V_3 \rangle_0$,
the $\log \vepsilon$ divergence in (2.39) can be interpreted as the Möbius
divergence present in the regularized expression for $T_3$.
Differentiating over $\log \vepsilon$ we thus obtain the ($k^2$ piece of)
usual 3-graviton (Möbius gauge fixed) amplitude which corresponds to the
order $R$ term in the EA\foot{Had we  not set $D=26$   we would obtain  (see (2.15)) the order $D-26$   ``cosmological'' term in (2.40), in agreement with (2.21),(2.36).}   \[
S = \Big( \frac{\partial Z}{\partial \log \vepsilon} \Big)_R
= d_0 \int d^D y \, \sqrt{G} \, e^{-2\phi}\, 
\Big( \alpha' R + \cdots \Big).
\tag{2.40}
   \]To summarize, there is a deep connection between the low energy
(massless sector) string dynamics and the renormalizable $\sigma$ model
at the string tree level which is expressed by the following relations   \[
\frac{\delta S}{\delta \varphi^i} = 0\ \ \ 
\;\rightarrow\; \ \ \ 
\bar{\beta}^i = 0\ \ \ 
\;\rightarrow\;\ \ \ 
A_{1i} = 0 ,
\tag{2.41}
\]
where $S$ is the effective action reconstructed from the massless string
amplitudes, $\bar{\beta}^i$ are the Weyl anomaly coefficients
(``$\beta$-functions'') of the $\sigma$ model and $A_{1i}$ are the massless
1-point amplitudes computed in a massless background.
Moreover, $S$ can be explicitly represented in terms of the Weyl anomaly
coefficients (see (2.36) and (2.37)).\foot{Let us stress that though to compute EA one uses the string $S$-matrix
corresponding to a particular (flat space) string vacuum, being extended
off-shell the EA ``forgets'' about this particular vacuum and, in fact,
contains  information about all ``nearby'' vacua. For example, it
``interpolates'' between the flat space, group manifold vacua (assuming
that the $D - 26$ term is included as in (2.22)) and, in fact, all other
``weak-coupling'' conformal field theories.    }

\

It would be very interesting to generalize the above string theory
$\sigma$ model correspondence to the string loop level. One of the
consequences of the equivalence between the string equations of motion and the 
conditions of conformal invariance of the $\sigma$ model is that it makes
possible to avoid the difficult problem of determining the stationary
points of the EA (reconstructed order by order from string amplitudes),
replacing it by the problem of classification of 2-d conformal field
theories. Similar equivalence at the loop level would provide a way of
studying exact string vacua solutions by solving some ``generalized
conformal invariance'' conditions.

\subsection*{3.}

The basic property of the string generating functional $\hat Z$ which makes it 
possible to establish the connection with $\beta$-functions  is its
\textit{renormalizability} with respect to the local world sheet infinities.
To be able to extend the string-$\sigma$ model relation to string loop level
it is necessary to have the corresponding renormalizability property of the
loop-corrected generating functional. Before turning to the analysis of
string loop corrections it is instructive to add some remarks about
renormalization of $\hat Z$ at the string tree level.

Consider the partition function   \[
Z = \langle e^{-\varphi^i V_i} \rangle_0
\]
expanded in powers of $\varphi$, see (2.31). The regularized correlators
$T_N$ contain divergences proportional (in view of the factorization
property) to $T_M$ with $M < N$. In fact, the divergences come from the
regions where 2 or more integration points $z_i$ coincide.
If   we consider
only the low momentum massless particles as the external ones, the operator
product $V_i V_j$ closes on dimension 2 massless vertex operators\footnote{A
``physical'' explanation of the perturbative renormalizability of $Z$ is that
if the external momenta (or derivatives of $\varphi$) are small compared to
$\alpha'^{-1/2}$, we get only the singularities corresponding to massless
poles (there is no energy enough to produce massive poles).}
and we get the correlators already present in the sum (2.31),   \[
Z \sim 1 + T_1 \varphi + T_2 \varphi^2 + \ldots
\]
Iin more detail,   \[
v_i(z_i) v_j(z_j)
\sim |u|^{\alpha' k_i k_j - 2}
\, c_{ij}{}^{k} \, v_k(z),
\qquad
u = z_i - z_j ,
\]   \[
\int_{\ve}
\frac{d^2 u}{|u|^2}
\, |u|^{\alpha' k_i k_j}
\sim \log \vepsilon
+ \frac{\alpha'}{2} k_i k_j \log^2 \ve
+ \ldots ,
\]
where we expand in $\alpha' k_i k_j$, etc. 
Hence we should be able to
cancel the divergences by redefining $\varphi^i$ order by order,   \[
\varphi^i
=
\varphi_R^i
+
\log \ve
\Big(
B^i{}_{2jk}\,\varphi_R^j \varphi_R^k
+
B^i{}_{3jkl}\,\varphi_R^j \varphi_R^k \varphi_R^l
+
\ldots
\Big)
+ O(\log^2 \ve) .
\tag{2.42}
   \]The usual argument for the perturbative renormalizability of $Z$ is based
essentially on dimensional (or power counting) considerations. The
coefficients which appear in the counterterms (2.42) are,  in fact,  proportional
to the string scattering amplitudes (see [1, 44, 8] and references
therein). 

To see this one should introduce a background $\bar x$ and compute
the counterterms, extracting the local, order $V_i[x] \log \ve$, term in
$W[\bar x] = - \log Z[\bar x]$. Expanding in power of $\varphi^i$ it is
possible to check that the ``overall'' $\log \ve$ singularity in
$\langle V_{i_1}[x + \bar x] \ldots V_{i_N}[x + \bar x] \rangle$
comes from the region where all $N$ points are close to each other.
Factorizing on the pole one finds that the coefficient is the usual
$N+1$-point amplitude $A_{N+1}$ (computed with the Möbius gauge fixed,
e.g. $z_1 = 0$, $z_2 = 1$, $z_{N+1} = \infty$).
Thus the $\beta$-function is   \[
\beta^i \sim \sum_{N>2} B_N \varphi^N,
\qquad
B_N = (A_{N+1})_{\text{subtr}}
\]
(to obtain the $\beta$-function we should subtract all subleading
singularities). This result is obviously consistent with the
$\beta \sim \delta S / \delta \varphi$ relation (2.19) since the EA is   \[
S \sim \sum_N (A_N)_{\text{subtr}} \ \varphi^N .
   \]\def \ss  {{\rm{s}}}

Let us now give another argument for the renormalizability of the string
generating functional $\hat Z$ which is based on the usual assumption that
the combinatorics of string amplitudes is such that the massless sector of
the string $S$-matrix can be reproduced by an effective field theory.
By this one implies that $\hat Z$ should be equal to the generating
functional $\ss$ for the field theory $S$ matrix. 

The idea is to establish the
renormalizability property of $\ss$ introducing a cutoff through the space-time
propagator and assuming the proportionality (2.19) between the $\beta$-functions
and the derivative of the EA $S$. Let   \[
S(\phi) = \frac{1}{2} \phi \Delta \phi + U(\phi) .
\tag{2.43}
\]
Consider the functional   \[
\ss(\varphi) = W(J)\big|_{J \to \Delta \varphi} ,
\tag{2.44}
\]
where $W$ is the generating functional for connected Green functions.
In the tree approximation   \[
W(J) = S(\phi) - \phi J ,
\qquad
\frac{\del S}{\del \phi} = J .
\tag{2.45}
\]
The on-shell $S$-matrix generating functional is   \[
\ss(\varphi_{\text{in}}) = S(\varphi_{\text{cl}})
- \varphi_{\text{in}} \Delta \varphi_{\text{cl}} ,
\tag{2.46}
\]   \[
\varphi_{\text{cl}}
= \varphi \big|_{\varphi_{\text{in}}}
= \varphi_{\text{in}} + \ldots ,
\qquad
\Delta \varphi_{\text{in}} = 0 ,
\qquad
\big( \frac{\partial S}{\partial \varphi} \Big)_{\varphi_{\text{cl}}} = 0 .
\]
It is convenient not to specify $\varphi$ to be equal to $\varphi_{\text{in}}$
(and hence $\phi$ to be equal to $\varphi_{\text{cl}}$) at the intermediate
stages (see also Ref.~45). Then   \[
\ss(\varphi)
=
- \frac{1}{2} \varphi \Delta \varphi
+ U(\phi(\varphi))
+ \frac{1}{2}
\Big( U' \Delta^{-1} U' \Big)_{\phi(\varphi)} ,
\tag{2.47}
\]   \[
\phi
= \varphi - \Delta^{-1} U'(\phi) ,
\qquad
U' = \frac{\partial U}{\partial \varphi} ,
\tag{2.48}
\]   \[
\ss(\varphi_{\text{in}})
=
U(\vphi_{\text{cl}}(\varphi_{\text{in}}))
+ \frac{1}{2}
\Big( U' \Delta^{-1} U' \Big)_{\vphi_{\text{cl}}(\varphi_{\text{in}})}
=
U(\varphi_{\text{in}})
- \frac{1}{2}
\Big( U' \Delta^{-1} U' \Big)_{\varphi_{\text{in}}}
+ \ldots .
\]
Now let us introduce a cutoff by a formal substitution   \[
\Delta^{-1}
\rightarrow
\Delta^{-1}(\vepsilon, \Delta)
\sim
\log \ve
+ O(\Delta \log^2 \vepsilon) .
\tag{2.49}
\]
For example, one may put (we use $\alpha'$ to match the dimensions)   \[
(\alpha' \Delta)^{-1}
\rightarrow
\int_\ve^{1} \frac{dt}{t} \, t^{\alpha' \Delta}
=
(\alpha' \Delta)^{-1}
-
(\alpha' \Delta)^{-1} \ve^{\alpha' \Delta}
=
- \log \ve
- \frac{1}{2} \alpha' \Delta \log^2 \ve
+ \ldots
\tag{2.50}
\]
In general (cf. (2.29))   \[
\frac{\partial \Delta^{-1}}{\partial \log \ve}
=
- \pi(\ve, \Delta) ,
\qquad\qquad 
\pi
=
\alpha' P^{-1}
+
O(\log \ve) .
\tag{2.51}
\]
Note that $\Delta^{-1}$ appears in $\ss$ (2.47) explicitly as well as implicitly
through $\phi$. Differentiating $\ss$ with respect to $\log \ve$ we find   \[
\frac{\partial \ss}{\partial \lambda}
=
U' \dot{\phi}
(1 + \Delta^{-1} U'')
+
\frac{1}{2}
U' \dot{\Delta}^{-1} U' \ ,
\qquad 
(\dot{\ }) = \frac{\partial}{\partial \lambda},
\qquad
\lambda \equiv \log \ve .
\tag{2.52}
\]
Using that   \[
\dot{\phi}
=
-
(1 + \Delta^{-1} U'') \dot{\Delta}^{-1} U'
\]
we finish with   (similar relations appeared in Ref.~46)   \[
\frac{\partial \ss}{\partial \lambda}
=
-
\frac{1}{2}
\big( U' \dot{\Delta}^{-1} U' \big)_\phi
=
\frac{1}{2}
\Big( U' \pi U' \Big)_\phi .
\tag{2.53}
\]
On the other hand, if $\ss$ were  renormalizable, we would have
(here $\varphi^i$ are  bare couplings)   \[
\frac{d\ss}{d\lambda}
=
\frac{\partial \ss}{\partial \lambda}
-
\beta^i
\frac{\partial \ss}{\partial \varphi^i}
=
0 ,
\qquad \ \ \qquad 
\beta^i(\varphi)
=
-
\frac{d \varphi^i}{d\lambda} .
\tag{2.54}
\]
Let us assume that the $\beta$-functions satisfy the relation (2.19)
with $S$ given by (2.43). Then (2.54) implies that\foot{Note that the ``diffeomorphism terms'' on which
$\bar{\beta}^i$ differ from $\beta^i$ (see (2.11))
drop out from the products if $S$ (and $\ss$) are covariant.}
   \[
\frac{\partial \ss}{\partial \lambda}
=
\beta^i
\frac{\partial \ss}{\partial \varphi^i}
=
\frac{\partial S}{\partial \varphi^k}
\, \kappa^{kj}
\frac{\partial \ss}{\partial \varphi^j}
=
\Big[
\Delta_{ik} \varphi^k + U'_i(\varphi)
\Big]
\kappa^{ij}
\Big[
- \Delta_{jl} \varphi^l + U'_j(\phi)
\Big] .
\tag{2.55}
\]
If we go on shell ($\varphi = \varphi_{\text{in}}$) and compare (2.55)
with (2.53) we get   \[
\frac{\partial \ss(\varphi_{\text{in}})}{\partial \lambda}
=
\frac{1}{2}
\Big[
U'_i(\phi)\, \pi^{ij}\, U'_j(\phi)
\Big]_{\phi = \vphi_{\text{cl}}(\varphi_{\text{in}})} ,
\tag{2.53$'$}
\]
   \[
\frac{\partial \ss(\varphi_{\text{in}})}{\partial \lambda}
=
\Big[
U'_i(\varphi_{\text{in}})\, \kappa^{ij}\, U'_j(\phi)
\Big]_{\phi = \vphi_{\text{cl}}(\varphi_{\text{in}})} .
\tag{2.55$'$}
\]
Since $\varphi_{\text{cl}} = \varphi_{\text{in}} + O(\log \vepsilon)$ and
$\kappa = \frac{1}{2}\alpha' P^{-1} + \ldots = \frac{1}{2}\pi + \ldots$
(see (2.30) and (2.51)) we conclude that Eqs.~(2.53$'$) and (2.55$'$)
are in agreement to the leading order. This implies that $\ss$ is indeed
renormalizable at least to the $\log  \vepsilon$ order. In general,
(2.53$'$) would be equivalent to (2.55$'$) if $\kappa^{ij}$ were such that   \[
\kappa^{ij}(\varphi_{\text{in}})
\, U'_j(\varphi_{\text{in}})
=
\frac{1}{2}
\pi^{ij}(\vepsilon, \Delta)
\, U'_j(\vphi_{\text{cl}}(\varphi_{\text{in}})) .
\tag{2.56}
\]
It seems plausible that this relation can be satisfied by properly
adjusting $\kappa$ order by order in $\alpha'$.

The argument for the renormalizability of $\hat Z$ we have just given
should generalize to string loop level. It is natural to assume that
the massless sector of the full loop corrected string $S$-matrix can
be reproduced by the tree $S$-matrix corresponding to some effective
action. This is so if a string field theory exists and should also follow
in general from the conditions of factorization and unitarity. We
shall return to this topic in Sec.~5.

\section*{3.\ String loop divergences and their renormalization}


 To understand how the renormalization group can be defined to act in string loops
we are first to analyze the divergences which may appear in string loop corrections.
We shall limit the discussion to the case of the massless external particles.
Let us recall that the tree string amplitudes do not contain momentum independent
singularities. Namely, they diverge in some regions of momentum space but are convergent
in the others and hence can be defined to be finite (modulo poles) for all momenta
using the analytic continuation.

To make contact with the $\sigma$-model and RG approach it is necessary, however,
not to use the analytic continuation but to introduce a short distance cutoff in the
2d propagator (and also to expand in external momenta if we are interested in the
low-energy RG realized on the massless fields corresponding to the renormalizable couplings
of the $\sigma$ model).

To regularize the string loop amplitudes it is not sufficient to introduce a cutoff in the
2d propagator: we should regularize the divergences which may come from the boundaries of
the integration regions for both the Koba--Nielsen and the moduli parameters. Such a
regularization can be introduced in a more or less systematic way in the parametrizations
in which the moduli are represented as coordinates of points on a complex plane (the examples
are the branch point  [47] and Schottky   [23b,30,31] parametrizations).
In these parametrizations all the divergences can be regularized by imposing the restriction
that any two points on a  complex plane  must be separated by a distance larger than $\vepsilon$.\footnote{Let us note, however, that such a restriction does not introduce a ``built-in'' cutoff in the theory: there still remains a difference between the regularization of ``local'' infinities which can be done by inserting a cutoff in the propagator and a regularization of ``modular'' infinities which is done ``by hand''. For example, nothing prevents us from using different cutoffs for different parameters (say $\vepsilon$ for local and $k \ve $ for modular). This may lead to ambiguities in the coefficients of some divergences.}

\

Let us now recall the classification of possible singularities in closed string loop
amplitudes  [23--26].
 There are two basic types of singularities which
appear a general $n$-loop $N$-point amplitude:

(1) {\it Singularities corresponding to shrinking of a ``nondividing'' (nontrivial) cycle(s).}

 The ``momentum-dependent'' singularities of this type are the usual unitary
singularities (discontinuities) which are due to loops of internal particles. The
``momentum-independent'' ``nondividing'' singularities in the vacuum amplitude (string partition function)
and the massless amplitudes may be attributed to the tachyon loop, i.e.\ interpreted as being due to
the tachyon present in the tree level spectrum of the bosonic  string theory in the standard flat space vacuum.

Similar singularities are found in the scattering amplitudes for massive states in the superstring theory  [{48,28}].
In this case they are due to the massless particles which are the lowest mass particles in the superstring spectrum
on which massive particles can decay. The way to deal with these singularities, which is consistent with unitarity,
is not to regularize (and then to renormalize) them but to trade them for the imaginary parts by defining the amplitudes
using a special analytic continuation prescription  [{28}].
This is very natural from the physical point of view:
for example, the imaginary part of the vacuum amplitude in the bosonic string is a signal of expansion near an unstable (tachyonic) vacuum.\footnote{Still, from geometrical point of view, it is not at all obvious that one should not regularize the ``nondividing'' singularities. In fact, if we introduce a ``universal'' (e.g.\ geodesic distance) cutoff on the 2-surface, it automatically regularizes also the ``nondividing'' singularities (since the lengths of all cycles then are greater than $ \vepsilon$). For a recent attempt to regularize the ``nondividing'' singularity of the one-loop partition function see Ref.\ 50.}

It is possible to give also other arguments why the ``nondividing'' singularities should not be accounted for in the RG approach since they do not produce Weyl (or BRST) anomalies [{18,49}].   In the ``nondividing'' limit the $n$ loop amplitude reduces to the infinity times the $n-1$ loop amplitude with two zero momentum tachyon insertions at two different points. Hence it is not possible to cancel this divergence by adding a local counterterm to the string action. The final argument why we should not bother about the ``nondividing'' singularities follows from the expected correspondence between the loop corrected $\beta$-functions and the effective action. The contributions to the effective action are proportional to the amplitudes with the massless poles subtracted out. Since the ``nondividing'' singularities are not related to the massless poles, they should not produce contributions to the $\beta$-functions.

(2) {\it  Singularities corresponding to shrinking of ``dividing'' (trivial) cycles.}

 These  can in turn be split into the
``momentum dependent'' and ``momentum independent'' singularities. The first appear when the numbers of external particles
on the both sides of the string diagram are greater than two. They are the usual physical pole singularities present only
for special values of external momenta. The pole singularities corresponding to the massless internal propagators should be
subtracted in constructing the EA and should be accounted for in computing the $\beta$-functions.

 The momentum independent
singularities include the ``tadpole'' and ``external leg'' divergences present for arbitrary values of external (on-shell) momenta.
If one ignores the momentum dependent singularities (assuming the analytic continuation in external momenta) and also ignores the tachyonic
tadpole (quadratic) infinities\footnote{Using, e.g., a kind of analytic continuation prescription  [{27}] or renormalizing the tachyon coupling in the string action (cf.\ the discussion of ``massless'' renormalizations below).}
then the only remaining infinities are the logarithmic divergences which are due to the massless scalar dilaton and trace of graviton tadpoles
$\bigl((\alpha' k^{2})^{-1}_{k_\mu\to 0}\to -\log\vepsilon\bigr)$ and the external leg divergences
$\bigl((\alpha' k^{2})^{-1}_{k^2\to 0}\to -\log\vepsilon\bigr)$ (we consider only the massless external particles).

 The external leg infinities correspond
to the factorization
\begin{equation}
A_{N}^{(n)}\to \sum_{l=0}^{n} A_{N}^{(l)}\,\log\vepsilon\, A_{2}^{(n-l)},
\qquad
A_{N}^{(n)}\sim \langle V_{1}\ldots V_{N}\rangle_{n} \, ,
\tag{3.1}
\end{equation}
and are usually assumed to be absorbable into the renormalization of the vertex operators [{51}]
\begin{equation}
V_{iR}=z^{j}{}_{i}(\vepsilon)\,V_{j},
\qquad
z^{j}{}_{i}=\delta^{j}{}_{i}+\sum_{p} M^{j}{}_{pi}\,\log^{p}\vepsilon \, .
\tag{3.2}
\end{equation}
The idea is that properly choosing $M_{p}$ and using (3.1) one can make the total (summed over genera) amplitude free of the external leg divergences.
Since $\langle V_{i}V_{j}\rangle_{n}$ itself contains self-energy subdivergences
\begin{equation}
\langle V_{i}V_{j}\rangle_{n}\sim
\sum_{l,k}\langle V_{i}V_{k}\rangle_{l}\,\log\vepsilon\,\langle V_{k}V_{j}\rangle_{n-l}
+O(\log^{2}\vepsilon),
\tag{3.3}
\end{equation}
it is natural to anticipate the exponentiation of the ``rudimental'' self-energy corrections,
\begin{equation}
z^{i}{}_{nj}=\exp\!\Big(\sum_{r}\mu^{i}{}_{rj}\,\log\vepsilon\Big),
\qquad\qquad 
\mu^{i}{}_{nj}\sim \langle V_{i}V_{j}\rangle_{n}^{\mathrm{subtr}},
\tag{3.4}
\end{equation}
where $\mu_{n}$ is the massless 2-point amplitude on genus $n$ with all self-energy subdivergences subtracted out.
This exponentiation was checked on the example of the one-loop tachyon leg correction in Ref.\ 18. Note that (3.4) can be true only if the relative weights
of string loop corrections take particular values. As we shall see, the exponentiation of divergences is crucial in order for the RG to act in string loops
(see also Ref.\ 14).

\

\def \O {{\cal O}}

Let us now consider the logarithmic tadpole divergences. A careful analysis of factorization of the closed string amplitudes in the ``tadpole'' limit gives the following
result (see Ref.\ 15 and references therein; in the subsequent sections we shall check this result on several particular examples, see also Refs.\ 13, 14 and 16)
\begin{equation}
A_{N}^{(n)}\sim \sum_{l=1}^{n-1}\hat A_{N}^{(n-l)}\,\log\vepsilon\,\hat d_{l},
\tag{3.5}
\end{equation}
\begin{equation}
\hat A_{N}^{(n-l)}\sim \langle V_{1}\ldots V_{N}\,\O_{l}\rangle_{n-l},
\qquad
\hat d_{l}\sim \langle 1\rangle_{l},
\tag{3.6}
\end{equation}
\begin{equation}
\O_{n}=\frac{1}{4\pi\alpha'}\int d^{2}z\,\sqrt{g}\,
\Bigl[2n:\partial_{\alpha}x^{\mu}\partial^{\alpha}x_{\mu}:
+\alpha'(1-n)R^{(2)}\Bigr],
\tag{3.7}
\end{equation}
where normal ordering is with respect to the propagator on the sphere,
\begin{equation}
:\partial_{\alpha}x^{\mu}\partial^{\alpha}x_{\mu}:\ 
=\partial_{\alpha}x^{\mu}\partial^{\alpha}x_{\mu}
+\frac{1}{4}\alpha' D\,  R^{(2)}.
\tag{3.8}
\end{equation}
We assume that a regular metric is introduced on the world surface, so that the Euler number is
\begin{equation}
\chi_{n}=\frac{1}{4\pi}\int d^{2}z\,\sqrt{g}\,R^{(2)}=2(1-n).
\tag{3.9}
\end{equation}
$\hat d_{l}$ is proportional to a massless tadpole or, equivalently, to the vacuum amplitude. 

It is possible to rewrite (3.5) in the following way
\begin{equation}
A_{N}^{(n)}\sim -4\sum_{l=1}^{n-1}
\Big\langle V_{1}\ldots V_{N}\Big(\frac{1}{D}V_{g}\langle V_{g}\rangle_{l}
-\frac{1}{8}\chi_{n}\chi_{n-l}\langle 1\rangle_{l}\Big)\Big\rangle_{n-l}
\log\vepsilon,
\tag{3.10}
\end{equation}
where
\begin{equation}
V_{g}\equiv \frac{1}{4\pi\alpha'}\int d^{2}z\,\sqrt{g}:\partial_{\alpha}x^{\mu}\partial^{\alpha}x_{\mu}:
\tag{3.11}
\end{equation}
is the (trace of) soft graviton vertex operator, and (see Refs.\ 14--16)
\begin{equation}
\langle V_{g}\rangle_{l}=\frac{1}{4}D(\chi_{l}-2)\langle 1\rangle_{l}.
\tag{3.12}
\end{equation}
The $\log\vepsilon$ which appears in (3.5) and (3.10) originates from the graviton and dilaton propagators at zero momentum. In fact, one can represent $\O_{n}$ (3.7) as the combination
of the vertex operators for the trace of the soft graviton and the soft dilaton
\begin{equation}
\O_{n}=2V_{g}-\chi_{n}V_{d}=2V_{g}+2(n-1)V_{d},
\tag{3.13}
\end{equation}
\begin{equation}
V_{d}\equiv \frac{1}{4\pi\alpha'}\int d^{2}z\,\sqrt{g}\Big(:\partial_{\alpha}x^{\mu}\partial^{\alpha}x_{\mu}:-\frac{1}{2}\alpha'R^{(2)}\Big)
=\frac{1}{4\pi\alpha'}\int d^{2}z\,\sqrt{g}\Big(\partial_{\alpha}x^{\mu}\partial^{\alpha}x_{\mu}+\frac{1}{4}\alpha'(D-2)R^{(2)}\Big).
\tag{3.14}
\end{equation}
Note that the dilaton couples to the Euler number (cf.\ (2.3)). We shall also use another equivalent representation for $\O_{n}$
\begin{equation}
\O_{n}=\frac{1}{4\pi\alpha'}\int d^{2}z\,\sqrt{g}\Big[\partial_{\alpha}x^{\mu}\partial^{\alpha}x^{\nu}\delta_{\mu\nu} f_{1n}+\alpha'R^{(2)}f_{2n}\Big],
\tag{3.15}
\end{equation}
\begin{equation}
f_{1n}=2n,
\qquad
f_{2n}=1+\frac{1}{2}n(D-2).
\tag{3.16}
\end{equation}
Observing that (3.15) looks similar to the string action (2.3) in a trivial vacuum
\begin{equation}
G_{\mu\nu}=\delta_{\mu\nu},
\qquad
\phi=\log g=\text{const},
\qquad
g=\text{string coupling},
\tag{3.17}
\end{equation}
one may try to cancel the tadpole divergences (3.5) in the total amplitude $A_{N}=\sum_{n}A_{N}^{(n)}$ by adding the counterterms to the vacuum string action [9].

To illustrate the idea of renormalization consider, e.g., the genus 3 example. Different tadpole factorizations of 3-loop amplitude give the following divergences
\begin{equation}
(A_{N}^{(3)})_{\mathrm{tadpole}}\sim
\langle\ldots(\hat d_{1}\O_{1}\log\vepsilon)^{3}\rangle_{0}
+\langle\ldots(\hat d_{1}\O_{1}\log\vepsilon)(\hat d_{2}\O_{2}\log\vepsilon)\rangle_{0}
+\langle\ldots(\hat d_{3}\O_{4}\log\vepsilon)\rangle_{0}\no 
\end{equation}
\begin{equation}
\hspace*{1.2cm}
+\langle\ldots(\hat d_{1}\O_{1}\log\vepsilon)^{2}\rangle_{1}
+\langle\ldots(\hat d_{2}\O_{2}\log\vepsilon)\rangle_{1}
+\langle\ldots(\hat d_{1}\O_{1}\log\vepsilon)\rangle_{2},
\tag{3.18}
\end{equation}
where $\ldots$ stands for $V_{1},\ldots,V_{N}$. $\hat d_{n}\sim \langle 1\rangle_{n}$ are, in general, divergent, containing tadpole subdivergences,
\begin{equation}
\hat d_{1}=d_{1},
\qquad
\hat d_{2}=e_{1}d_{1}^{2}\log\vepsilon+d_{2},
\tag{3.19}
\end{equation}
\begin{equation}
d_{3}=e_{2}d_{1}^{3}\log^{2}\vepsilon+e_{3}d_{2}d_{1}\log\vepsilon+d_{3},
\qquad
d_{n}=(\hat d_{n})_{\mathrm{finite}},
\qquad
e_{i}=\text{const}.\no
\end{equation}
Suppose now that the correlators are computed with the ``bare'' string action
$I=I_{R}+\delta I$, where the counterterm is a power series in the (renormalized, see below)
string coupling and $\log\vepsilon$. Expanding in the (renormalized) string coupling we then get insertions of $\delta I$
in the correlators and hence may hope to cancel the tadpole divergences between different terms in $A_{N}$. For example,   \[\langle\ldots e^{-\delta I}\rangle_{0}+\langle\ldots\rangle_{1}+\cdots=\cdots+\langle\ldots(-\delta I)\rangle_{0}+\langle\ldots\O_{1}d_{1}\log\vepsilon\rangle_{0}+\cdots = \text{finite} . \]
The resulting expression for the counterterm is
\begin{align}
\delta I=\sum_{n=1}^{\infty}\hat b_{n}(\vepsilon)\O_{n}\log\vepsilon &+O(\log^{2}\vepsilon)
\tag{3.20}
=\sum_{n=1}^{\infty} b_{n}\O_{n}\log\vepsilon+O(\log^{2}\vepsilon),
\\
&\hat b_{n}\sim \hat d_{n},
\qquad
b_{n}\sim d_{n}\sim g^{2n}.
\tag{3.21}
\end{align}
Though it may seem that the factorization suggests that the $\log^{m}\vepsilon$, $m\ge 2$ terms are absent in (3.21), the consistency of the RG implies that they must be present (and have coefficients
 related to $(d_{n})_{fin})$.\footnote{We have recently checked explicitly the presence of the $\log^{2}\vepsilon$ term in (3.21) by factorizing the 2-loop string amplitudes [77]. The coefficient of this term is precisely the one dictated by the RG.}
Using the expression for $\O_{n}$ (3.15) one can rewrite $\delta I$ as
\begin{equation}
\delta I=\frac{1}{4\pi\alpha'}\int d^{2}z\,\sqrt{g}\Big[\partial_{\alpha}x^{\mu}\partial^{\alpha}x_{\mu}\, q_{1}+\alpha'R^{(2)}\, q_{2}\Big],
\tag{3.22}
\end{equation}
\begin{equation}
q_{i}=a_{i}^{(1)}\log\vepsilon-a_{i}^{(2)}\log^{2}\vepsilon+\cdots,
\tag{3.23}
\end{equation}
where $a_{i}^{(k)}$ are power series in the (renormalized) string coupling constant. 

Comparing (3.22) with the vacuum string action (2.3), (3.17) we conclude that the tadpole divergences may be absorbed into a renormalization of the constant vacuum values of the metric and the dilaton, or, equivalently, into a renormalization of the two basic constants: $\alpha'$ and the string coupling $g$
\begin{equation}
(\alpha'(\vepsilon))^{-1}=(\alpha'_{R})^{-1}+q_{1}(g_{R},\vepsilon),
\qquad
\log g(\vepsilon)=\log g_{R}+q_{2}(g_{R},\vepsilon).
\tag{3.24}
\end{equation}

The consistency of this renormalization procedure depends crucially on whether the relative weights of string diagrams are such that the tadpole divergences actually exponentiate. One may consider the exponentiation to be a consequence of the condition of factorization which together with unitarity  [{52}] fixes the weights of string amplitudes. Alternatively, one may impose the requirement of renormalizability (which implies exponentiation) and, 
 as a consequence, fix the relative weights of string diagrams.


\section*{4.\,Generating functional (or $\sigma$ model) approach
\\  \hspace*{12pt} to renormalization of string loops}

The results of the previous section about the elimination of the tadpole and external leg divergences through the renormalization of $\alpha'$ and $ g$ and the vertex operators can be reinterpreted in a more fundamental way as the renormalizations of the couplings of the $\sigma$ model which appears in the generating functional for string amplitudes $\hat Z$. $\hat Z$ is the basic object which defines the theory
\begin{equation}
\hat Z=\Omega^{-1}\sum_{n=0}^{\infty} c_{n}\int d\mu_{n}(m)\int_{\mathcal{M}_{n}} Dx\,\, e^{-I},
\qquad
I=I_{0}+\vphi^{i}V_{i},
\qquad
\hat Z\sim \sum_{N}A_{N}\vphi^{N}.
\tag{4.1}
\end{equation}
In general, $I_{0}$ is the string action in a particular vacuum and $V_{i}$ are the corresponding ``massless'' vertex operators. We assume that ``extended'' sets of moduli $\{m\}$ are used so that the M\"obius group volume factor $\Omega^{-1}$ is present for all genera. $c_{n}$ are normalization factors (weights) which we prefer to indicate explicitly. Expanding $\hat Z$ in powers of $\vphi^{i}$ and putting them on shell ($\Box\vphi^{i}=0$) we get the on-shell amplitudes as the coefficients.

$\hat Z$ is formally defined for arbitrary (``off-shell'') values of the fields $\vphi^{i}$. In (4.1) we do not integrate over the conformal factor of 2-metric, fixing a Weyl gauge. The basic consistency requirement is that $\hat Z$ evaluated for the ``true'' vacuum values of couplings $\vphi^{i}$ (which solve the string equations of motion or generalized Weyl invariance conditions ``$\beta=0$'') should be Weyl gauge independent
 (i.e.\ Weyl invariant).\footnote{There is, of course, an ambiguity in splitting the metric on a conformal factor and moduli for each particular $n$. It would be nice to have a universal prescription, which relates the splittings for different $n$. We anticipate that this ambiguity can be ``absorbed'' into redefinitions of the couplings $\vphi^{i}$.}

\def \vphi {\varphi}
\def \p {\phi}

Suppose now that a short distance cutoff is introduced in (4.1), which regularizes, in particular, the ``local'', ``tadpole'' and ``external leg'' divergences. We can cancel the local divergences separately for each genus by choosing the bare fields to correspond to the ``local'' $\sigma$ model counterterms (see Sec.\ 2). To cancel the ``modular'' divergences we are to combine the contributions of different genera. Since it is the product $\vphi^{i}V_{i}$ that appears in (4.1) we trade the renormalization of the vertex operators (3.2) for the multiplicative renormalization of the couplings. The counterterm (3.20), (3.22) corresponds directly to the renormalization of the metric and the dilaton in the $\sigma$ model action (2.3). The corresponding bare couplings are
\begin{equation}
\vphi^{i}=\vphi^{i}_{R}+\delta_{\mathrm{tad}}\vphi^{i}+\delta_{\mathrm{e.l.}}\vphi^{i},
\tag{4.2}
\end{equation}
\begin{equation}
\delta_{\mathrm{tad}}\vphi^{i}=a_{1}^{i}\log\vepsilon+a_{2}^{i}\log^{2}\vepsilon+\cdots,
\qquad
\delta_{\mathrm{e.l.}}\vphi^{i}=z^{j}{}_{i}(\vepsilon)\vphi^{j}_{R}=\vphi^{i}_{R}+M^{i}{}_{lj}\vphi^{l}_{R}\log\vepsilon+\cdots.
\tag{4.3}
\end{equation}
The coefficients which appear here are power series in the string coupling (see also Ref.\ 14). The crucial point is that if $\hat Z$ is renormalizable
\begin{equation}
\hat Z(\vphi( \vepsilon), \vepsilon)=\hat Z_{R}(\vphi_{R}),
\tag{4.4}
\end{equation}
it is possible to introduce the $\beta$-functions (see (2.34) and (2.35))
\begin{equation}
\frac{\partial \hat Z}{\partial \log\vepsilon}-\beta^{i}(\phi)\frac{\partial \hat Z}{\partial \vphi^{i}}=0,
\qquad
\beta^{i}(\phi)=-\frac{d\vphi^{i}}{d\log\vepsilon}.
\tag{4.5}
\end{equation}
In general, the relation between the bare and renormalized couplings is
\begin{equation}
\vphi^{i}=\vphi^{i}_{R}+T^{i}_{1}(\vphi_{R})\log \vepsilon+T^{i}_{2}(\vphi_{R})\log^{2} \vepsilon+\cdots \, .
\tag{4.6}
\end{equation}
The basic restriction imposed by the RG is that
$\beta^{i}=-d\vphi^{i}/d\log \vepsilon$ should depend only on $\vphi^{i}$ but not explicitly on $\vepsilon$.
This implies that
\begin{equation}
\beta^{i}(\vphi)=-T^{i}_{1}(\vphi),
\qquad
T^{i}_{2}(\vphi)=\frac{1}{2}\,T^{j}_{1}\frac{\partial}{\partial \vphi^{j}}T^{i}_{1}(\phi),
\ \ \text{etc.}
\tag{4.7}
\end{equation}
We get from (4.2)
\begin{equation}
\beta^{i}=-a^{i}_{1}-M^{i}{}_{j}\vphi^{j}.
\tag{4.8}
\end{equation}
Thus the tadpole counterterm corresponds to the ``inhomogeneous'' term in the $\beta$-function
while the external leg one -- to the term linear in the fields. The coefficients in (4.8)
have the following structure (see (3.4) and (3.21))
\begin{equation}
a^{i}_{1}=\sum_{n=1}^{\infty} g^{2n} b^{i}_{n},
\qquad
M^{i}{}_{j}=\sum_{n=1}^{\infty} g^{2n}\mu^{i}{}_{nj},
\tag{4.9}
\end{equation}
where $b^{i}_{n}$ and $\mu^{i}{}_{nj}$ are proportional to the finite parts of the massless tadpoles and
2-point functions on genus $n$ (they are both proportional to the genus $n$ vacuum amplitude).

Recalling the structure of the counterterm (3.22) we can write the metric $G$ and dilaton $\phi$ 
$\beta$-functions in the following symbolic form (we set  $G_{\mu\nu}=\delta_{\mu\nu}+h_{\mu\nu}$)
\begin{equation}
\beta^{G}_{\mu\nu}=A^{G}\delta_{\mu\nu}+B^G_{\phi \mu\nu}\p +B^{G\alpha \beta} _{G\mu\nu}h_{\alpha\beta},
\qquad
\beta^{\phi}=A^{\phi}+B^{\phi}\phi+B_{\phi}^{\phi \mu\nu}h_{\mu\nu},
\qquad
A\sim a_{1},
\ \ B\sim M_{1}.
\tag{4.10}
\end{equation}
If the theory is defined (regularized) in a way consistent with general covariance in $D$ dimensions,
one should be able to rewrite (4.10) as
\begin{equation}
\beta^{G}_{\mu\nu}=A^{G}G_{\mu\nu},
\qquad
\beta^{\phi}=A^{\phi}+B^{\phi}_\p \phi.
\tag{4.11}
\end{equation}
Note that $A$ and $B$ should be constants; the terms in (4.11) are the only covariant terms
to the linear order in the fields $h_{\mu\nu}$ and $\phi$.

In (4.11) $G_{\mu\nu}$ and $\phi$ are the full nonconstant fields (constant vacuum values plus fluctuations).
An important observation is that $\beta^{i}$ are nontrivial functions of the constant part of $\phi$ since
it is directly related (see (2.3) and Refs.\ 53, 2 and 54) to the string coupling constant which appears in (4.9).
Since we renormalize $\phi$ we should also renormalize $g$ and it is, in fact, the renormalized value of $g$
(or $\phi$) which appears in the coefficients in (4.3). If we split $\phi$ on the constant and nonconstant parts
\begin{equation}
\phi=\phi_{c}+\tilde\phi,
\qquad\qquad 
\phi_{c}=\log g=\text{const},
\tag{4.12}
\end{equation}
we get the following expressions for the $\beta$-functions
\begin{equation}
\beta^{G}_{\mu\nu}=\sum_{n=1}^{\infty}\lambda_{n}e^{2n\phi_{c}}\,G_{\mu\nu},
\tag{4.13}
\end{equation}
\begin{equation}
\beta^{\phi}=\sum_{n=1}^{\infty}\nu_{n}e^{2n\phi_{c}}
+\sum_{n=1}^{\infty}\mu_{n}e^{2n\phi_{c}}\,\tilde\phi.
\tag{4.14}
\end{equation}
They look unnatural because of the different dependence on $\phi_{c}$ and $\tilde\phi$ and cry for a generalization.

The complicated dependence on $\phi_{c}$ suggests that the modular counterterms (4.3) and the $\beta$-functions (4.8)
should, in fact, contain all powers of the fields (in particular, of the dilaton).
Then e.g. the  $\tilde\phi$ term  in (4.14)   should  correspond to a linearization  of the exact expression for $\beta^\p$   which is non-linear in $\p$. 
Thus a    consistent dependence on $\p$  implies that $\beta^{i}$ should have the form\footnote{ Note that   to be able to rewrite (4.14) in the form (4.15)  one should  have a relation   between the dilaton  tadpole  and self-energy  coefficients $\nu_n$  and $\mu_n$   in (4.14).
The existence of such a relation follows from  the ``exponential'' dependence of $\hat Z$ on the dilaton.}
\begin{equation}
\beta^{G}_{\mu\nu}=F_{1}(\phi)\,G_{\mu\nu}+\cdots,
\qquad
\beta^{\phi}=F_{2}(\phi)+\cdots,
\tag{4.15}
\end{equation}
\begin{equation}
F_{1}=\sum_{n=1}^{\infty}\lambda_{n}e^{2n\phi},
\qquad
F_{2}=\sum_{n=1}^{\infty}\nu_{n}e^{2n\phi},
\tag{4.16}
\end{equation}
where dots in (4.15) stand for other possible terms depending on derivatives of the fields.
If we further use the explicit expression for the tadpole counterterm (3.20), (3.15) we get
\begin{equation}
F_{1}=-\sum_{n=1}^{\infty}2n\,b_{n}e^{2n\phi},
\tag{4.17}
\end{equation}
\begin{equation}
F_{2}=-\sum_{n=1}^{\infty}\Big[1+\frac{1}{2}n(D-2)\Big]b_{n}e^{2n\phi}.
\tag{4.18}
\end{equation}

To get a deeper  understanding  of the structure of the $\beta$-functions  (4.15),(4.17)  and (4.18)  
let us consider the ``zero-momentum'' part of the generating functional $\hat Z$ in (4.1) or the string partition function 
computed for the constant vacuum values of $G_{\mu\nu}$ and $\phi$.
Assuming that the string path integral is defined in a way preserving $D$-dimensional general covariance
(i.e.  the functional measure  is consistent with a regularization, etc.,  see  [16,22]) 
we find  (cf.\ (2.7))
\begin{align}
\hat Z= &\hat Z_{0}+\hat Z_{1}+\hat Z_{2}+\cdots\no \\
= & d_{0}\!\int d^{D}y\,\sqrt{G}\,e^{-2\phi}(\alpha'R+\cdots)
+\hat d_{1}\!\int d^{D}y\,\sqrt{G}(1+\cdots)\no \\
&+\hat d_{2}\!\int d^{D}y\,\sqrt{G}\,e^{2\phi}(1+\cdots)
+\hat d_{3}\!\int d^{D}y\,\sqrt{G}\,e^{4\phi}(1+\cdots)+\cdots .
\tag{4.19}
\end{align}
This can be written as
\begin{equation}
\hat Z=S_{0}+\int d^{D}y\,\sqrt{G}\,\omega(\phi)+\cdots,
\tag{4.20}
\end{equation}
\begin{equation}
\omega=\sum_{n=1}^{\infty} d_{n}e^{(2n-2)\phi},
\qquad
d_{n}\sim (\alpha')^{-D/2},
\tag{4.21}
\end{equation}
where $\hat d_{n}$ ($d_{n}$) are the (finite parts of the) higher genus contributions to the vacuum amplitude
(see (3.19)).\footnote{Let us note that $d_n$ is   complex because  we assume  that an  analytic continuation  prescription [28]
is used  in order to get rid of the tachyonic ``nondividing'' singularities (see Sec. 3).} 
In (4.19) we have included the tree level contribution (see (2.36), (2.21));
$S_{0}$ denotes the tree level piece of the effective action.

Consider now the divergent parts of $\hat Z$
\begin{equation}
\hat Z(\vphi,\vepsilon)=\hat Z_{R}(\vphi)+\hat Z^{(1)}(\vphi)\log\vepsilon+\hat Z^{(2)}(\vphi)\log^{2}\vepsilon+\cdots.
\tag{4.22}
\end{equation}
The renormalizability implies that (see (4.5))
\begin{equation}
\hat Z^{(1)}=\beta^{i}\frac{\partial}{\partial\vphi^{i}}\hat Z_{R},
\qquad
\hat Z^{(2)}=\frac{1}{2}\beta^{i}\frac{\partial}{\partial\vphi^{i}}\hat Z^{(1)}.
\tag{4.23}
\end{equation}
The renormalized part of $\hat Z_R$ should coincide with the effective
action (since renormalization is simply a subtraction of massless
exchanges). The operator (see (4.15))     \[
\beta^i \frac{\partial}{\partial \varphi^i}
=
F_1 G_{\mu\nu}\frac{\partial}{\partial G_{\mu\nu}}
+
F_2 \frac{\partial}{\partial \phi}
\]
represents the insertion of the tadpole counterterm, so that
(4.23) is in correspondence with the factorization property of the
amplitudes.

Let us now show that the expressions for $F_{1}$, $F_{2}$ which follow from factorization imply that
the $\beta$-functions (4.15) are equivalent to the equations of motion following from the ``cosmological term''
in the loop-corrected effective action (4.20). Namely, 
let us assume that the tree level relation (2.15) between the $\beta$-functions and the effective action holds
also at the loop level
\begin{equation}
\beta^{i}=\kappa^{ij}\frac{\partial S}{\partial\vphi^{j}},
\qquad
\beta^{i}=\sum_{n=0}^{\infty}\beta^{i}_{n},
\qquad
S=\sum_{n=0}^{\infty}S_{n},
\qquad
\kappa^{ij}=\sum_{n=0}^{\infty}\kappa^{ij}_{n}.
\tag{4.24}
\end{equation}
The loop corrections to $\kappa^{ij}$ are not important to the leading
order (recall that $\kappa^{ij}$ (2.20) is necessary in order to account
for the mixing between the metric and the dilaton in the kinetic term of
the tree level EA). Applying (4.24) and (2.20) to the case of $S$ given by
(4.20) one finds  [13,14] (cf.\ (2.16))
\begin{align}
&\beta^i=  \beta^i_0 + \beta^i_{\rm tad} + ...\ , \no \\
&\beta^{G}_{\mu\nu}=\alpha'\big(R_{\mu\nu}+2\D_{\mu}\D_{\nu}\phi\big)+\cdots
+F_{1}(\phi)G_{\mu\nu}+\cdots,
\no \\ 
& \tag{4.25} \beta^{\phi}=\Big[\frac{1}{6}(D-26)-\frac{1}{2}\alpha'\D^{2}\phi+\alpha'(\partial\phi)^{2}+\cdots\Big]
+F_{2}(\phi)+\cdots, \\
&
F_{1}=-\frac{1}{4}e^{2\phi}d_{0}^{-1}(2\omega+\omega'),
\qquad
\omega'=\frac{d\omega}{d\phi},
\tag{4.26}\\
&
F_{2}=-\frac{1}{16}e^{2\phi}d_{0}^{-1}\Big[2D\omega+(D-2)\omega'\Big].
\tag{4.27}
\end{align}
Substituting here the expression (4.21) for the dilaton potential $\omega(\phi)$,
one finds that (4.26), (4.27) coincide with (4.17) and (4.18) if
\begin{equation}
b_{n}=\frac{1}{4}d_{n}d_{0}^{-1}.
\tag{4.28}
\end{equation}
To find the bare couplings corresponding to the $\beta$-functions (4.15) we are to solve the equations
$\beta^{i}(\vphi)=-d\vphi^{i}/d\log\vepsilon$.
If there were no higher order terms in $\beta^{i}$, i.e.\ if (4.8) were  true, we would get
(we take $M^{i}{}_{j}$ diagonal for simplicity)
\begin{equation}
\vphi(\vepsilon)=e^{M_{1}\log\vepsilon}\big(\vphi_{R}+M_{1}^{-1}a_{1}\big)-M_{1}^{-1}a_{1}
=\vphi_{R}+a_{1}\log\vepsilon+M_{1}\vphi_{R}\log\vepsilon+\frac{1}{2}a_{1}M_{1}\log^{2}\vepsilon+O(\log^{3}\vepsilon).
\tag{4.29}
\end{equation}
Thus the RG implies the exponentiation of the divergences (cf.\ (3.4)) and, in particular,
the presence of higher order $\log^{k}\vepsilon$ counterterms with the coefficients fixed in terms of the
coefficients appearing in the $\beta$-function.
The bare fields corresponding to (4.15) are
\begin{align}
&G_{\mu\nu}=G_{R\,\mu\nu}-F_{1}(\phi_{R})G_{R\,\mu\nu}\log\vepsilon
+\frac{1}{2}\bigl(F_{1}^{2}+F_{2}F_{1}'\bigr)G_{R\,\mu\nu}\log^{2}\vepsilon+O(\log^{3}\vepsilon),\no \\
&\phi=\phi_{R}-F_{2}(\phi_{R})\log\vepsilon+\frac{1}{2}F_{2}F_{2}'\log^{2}\vepsilon+O(\log^{3}\vepsilon), 
\tag{4.30}
\end{align}
where the $\log^2  \vepsilon$ terms can be found by applying the general
relations (4.7) and (4.6). It would be a nontrivial check of the consistency
of the RG approach to find that the string loops contain the
$\log^2 \vepsilon$ divergences which can be cancelled by inserting the
counterterms (4.30) (or, equivalently, that the $\log^2 \vepsilon$ part of
$\hat Z$ satisfies (4.23)).\foot{It is interesting to observe that the $\log^2 \vepsilon$ terms in (4.30)
are different from those which follow from a naive expectation that the
full counterterm $\delta I$ can be represented as the sum of the full
(unrenormalized) tadpoles $\hat a_n$ times $\O_n \log \vepsilon$.
Thus there should be $O(\log^2 \vepsilon)$ terms in (3.20).
This, in fact, was checked in Ref.~77 (see also footnote 12 above).       }

Let us now discuss the origin of the higher order $O(\varphi^m)$,
$m \ge 2$, terms in the $\beta$-functions (4.15). Obviously they
correspond to the divergences which appear when an amplitude factorizes
into two parts (with the numbers of external particles $N_1$ and $N_2$)
and all external particles on one part (say, the second) have zero
momentum (the tadpole divergence thus corresponds to the case when
$N_2 = 0$). Clearly, this is a particular case of ``momentum dependent''
singularity due to massless pole in
$(k_1 + \cdots + k_{N_2})^2$ which one may try to eliminate by using an 
analytic continuation in external momenta. However, as we have argued
above, it is necessary to account for (regularize) such divergences in
order to get covariant expressions for the $\beta$-functions.

The continuity in external momenta then suggests that we should
regularize the massless pole momentum dependent singularity also for
nonvanishing external momenta. This is consistent with the fact that the
pole singularities contribute to the Weyl (or BRST) anomaly  [{49}]
and is obviously necessary for the correspondence between the
$\beta$-functions and the effective action.

\section*{5.\ Correspondence with effective action}

We have  already discussed  the role plyaed   by the effective action   in the RG approach  at the tree level (see  Sec. 2).
Most of this  discussion generalizes to loops.
To construct the effective action (EA) for the massless fields we start with the massless string scattering
amplitudes (with loop corrections included) and subtract all massless poles (see also [14]).\foot{Note that we are not assuming an expansion in external momenta since at
the loop level this expansion may lead to additional IR singularities
(due to massless loops). The EA constructed by expanding in powers of
fields will not be manifestly covariant. It may be rewritten in a
covariant form by using appropriate nonlocal structures (like
$R^n \log(\alpha' \D^2 + \alpha' R + \cdots$), etc.) which are
nonanalytic in $\alpha'$ and momenta.}
To subtract the massless exchanges we are to consider all possible ``dividing'' factorizations of a string diagram.

At the same time,  the corrections to the $\beta$-functions which we discussed in Secs.\ 3 and 4 come precisely from
the massless pole singularities which appear in these factorizations.
Hence, qualitatively, we should have
\begin{equation}
S\sim \sum_{N}(A_{N})_{\mathrm{subtr}}\,\varphi^{N},
\qquad
\beta(\varphi)\sim \sum_{N}(A_{N+1})_{\mathrm{subtr}}\,\varphi^{N},
\qquad
\frac{\partial S}{\partial\varphi}\sim \beta.
\tag{5.1}
\end{equation}
In general, we expect to find that:

(1) the generating functional $\hat Z$ (4.1) is renormalizable with
respect to the infinities associated with all ``dividing'' singularities
of massless string amplitudes (``momentum dependent'' as well as
``momentum independent'' ones). As at the tree level, it is necessary
to combine particular amplitudes into $\hat Z$ in order to obtain an
object which is renormalizable with respect to ``momentum--dependent''
infinities.

(2) The EA $S(\varphi)$ is a renormalized (finite) value of $\hat Z$
(since all the divergences in $\hat Z$ are due to the massless exchanges,
a subtraction of the latter should be equivalent to a renormalization
procedure).

(3) The resulting effective equations of motion are equivalent to a
generalized conformal invariance conditions $\beta^i = 0$ and also to
generalized vacuum stability conditions $A_{1i} = 0$ (the vanishing of
tadpoles in a background, see (2.41) and (2.38)).

\

As we have shown in Sec.~2 the renormalizability of $\hat Z$ is closely
connected with the existence of an effective action. The important point
is that the existence of the EA (and hence renormalizability of $\hat Z$)
depends on the values of relative weights of string diagrams. These
weights are usually assumed to be fixed by the condition of factorization
(and unitarity)  [{52,15,55}]
 which in turn should guarantee
the existence of the EA. 

Alternatively, it is interesting to observe that
the assumption of existence of the EA leads to constraints on the
weights. Consider first an  ordinary field theory, e.g.   \[
S_0 = \tfrac{1}{2}\varphi \Delta \varphi   + U(\vphi) \ , \qquad \ \ \   U= \tfrac{1}{6} g \varphi^3 .
\]
It is possible to represent its quantum effective action
\begin{equation}
S(\varphi)=W(J)-\varphi J,
\qquad
\frac{\partial W}{\partial J}=\varphi,
\qquad
e^{-W(J)}=\int D\varphi\,e^{-S_{0}(\varphi)+\varphi J}
\tag{5.2}
\end{equation}
in  the following way (cf.\ (4.1))
\begin{equation}
S(\varphi)=S_{0}(\varphi)+\sum_{\gamma_n}c_{n}\int d\mu_{n}(\tau)\int_{\gamma_n}Dx(\tau)\,e^{-I},
\qquad
I=I_{0}+I_{\mathrm{int}},
\tag{5.3}
\end{equation}
\begin{equation}
I_{0}=\int dt\,e^{-1}\dot x^{2},
\qquad
I_{\mathrm{int}}=\varphi V=g\int dt\,e\,\varphi(x(t)),\no 
\end{equation}
where $\gamma_n$ are 1--PI graphs and $\langle \tau \rangle$ are the proper
time parameters (``moduli'') of a 1-d metric $e(t)$. To arrive at this
expression one should represent $S$ in terms of the background dependent
propagators and to use the path integral representation for the latter.
The ``weights'' $c_n$ or the relative normalizations of the modular
measures are fixed by the functional integral definition of $S$ (5.2).
The generating functional for the quantum $S$ matrix is simply the tree
generating functional for $S$ (see (2.41) and (2.46)). Thus the relative
weights of the terms in the EA automatically determine the weights of
various amplitudes in agreement with factorization property.

The situation in string theory is different since if we do not start with
a string field theory functional integral we do not know \emph{a priori}
the relative weights in the   ``first-quantized''   path integral (4.1).\foot{It is commonly believed that all weights should be ``ones'' if one
properly defines the measure of the Polyakov integral (to be ultralocal,
etc.  [{56}]) and integrates over one copy of the moduli space.}

Suppose now that the massless sector of the string $S$-matrix can be reproduced by an effective field theory,
i.e.\ that $\hat Z(\varphi_{\mathrm{in}})$ can be represented as the generating functional for the tree level
$S$-matrix for some action $S(\varphi)$ (see (2.43), (2.46), (2.48))\foot{A remarkable fact that distinguishes string theory from field theory is
that the path integral representation is true not only for loop correction  terms
but also for the tree level term in $\hat Z$, cf.\ (4.1), (5.3).
}
\begin{equation}
\hat Z(\varphi_{\mathrm{in}})=\ss(\varphi_{\mathrm{in}}),
\qquad
\ss(\varphi_{\mathrm{in}})=S(\varphi_{\mathrm{cl}}(\varphi_{\mathrm{in}}))-\varphi_{\mathrm{in}}\Delta\varphi_{\mathrm{cl}}(\varphi_{\mathrm{in}}),
\tag{5.4}
\end{equation}
\begin{equation}
\ss=U(\varphi_{\mathrm{in}})-\frac{1}{2}\bigl(U'\Delta^{-1}U'\bigr)_{\varphi_{\mathrm{in}}}+\cdots.
\tag{5.5}
\end{equation}
The reason why Eq.~(5.4) imposes constraints on the string weights is
that it fixes the relative coefficients of the ``one particle reducible''
terms in terms of ``one particle irreducible'' ones. However, both the
reducible and irreducible graphs originate from one string diagram.

If we assume that  we know the $n$-loop modular measure (and region of integration)
up to one overall constant $c_n$. $c_n$ will thus appear both as the
coefficient of the irreducible part of $n$-loop string amplitude (which
contributes to $S$) and as the coefficient of the reducible part which
should appear in $\hat Z$ (5.4). If we know already the coefficients of
lower genus contributions to $S$ we thus know (according to (5.5)) the
coefficients of reducible diagrams constructed from them and hence are
able to fix $c_n$ in terms of $c_k$, $k < n$.

Expanding in powers of string coupling we find from (5.5)
\begin{equation}
\ss=\sum_{n=0}^{\infty}\ss_{n},
\qquad
\ss_{n}=U_{n}-\frac{1}{2}\sum_{m=0}^{n} U'_{m}\Delta^{-1}U'_{n-m}+\cdots,
\qquad
U=\sum_{n=0}^{\infty}U_{n}.
\tag{5.6}
\end{equation}
As is clear from (5.6), the first nontrivial relation is
$c_2 \sim c_1^{\,2}$ (we shall discuss it in Sec.~8). It is not possible
to fix $c_1$ using (5.6). To fix $c_1$ we should employ the unitarity
constraint   [{55}]  or compare the coefficient of the one-loop
term in $S$ with the coefficient of the usual
$\tfrac{1}{2}\,\log \det \Box$  term in a field theory EA (properly fixing
the normalization of the background fields, cf.\ Ref.~56).

Since the massless propagators in $\ss$ correspond to the infinities in the regularized $\hat Z$,
Eq.\ (5.4) implies that $S$ coincides with the renormalized (finite) part of $\hat Z$.
The renormalizability of $\hat Z$ and the relation (4.24) between $S$ and $\beta$-functions imply
\begin{equation}
\hat Z(\varphi,\vepsilon)=S(\varphi)+\beta^{i}(\varphi)\frac{\partial S}{\partial\varphi^{i}}\log\vepsilon+O(\log^{2}\vepsilon)
= S+\frac{\partial S}{\partial\varphi^{i}}\kappa^{ij}\frac{\partial S}{\partial\varphi^{j}}\log\vepsilon+O(\log^{2}\vepsilon).
\tag{5.7}
\end{equation}
This gives an ``off-shell'' analog of (5.5). Expanding in genus we have, to the lowest
order (see (4.24))
\begin{equation}
\hat Z=\sum_{n}\hat Z_{n},
\qquad
\hat Z_{n}=S_{n}
+\sum_{m=0}^{\infty}
\frac{\partial S_{m}}{\partial\vphi^{i}}
\kappa^{ij}_{0}
\frac{\partial S_{n-m}}{\partial\vphi^{j}}
\log\vepsilon
+\cdots .
\tag{5.8}
\end{equation}
Hence we can reformulate the above argument about the relation between weights of
different string diagrams in the following way: since the divergent part of $\hat Z_{n}$
should have the same overall coefficient as its finite part $S_{n}$, one can fix this
coefficient in terms of the coefficients of the lower genus terms $S_{m}$, $m<n$.

\

Let us finish this section with a remark on possible solutions of the generalized
conformal invariance equations $\beta^{0}+\beta^{i}=0$ or the effective equations of motion
$\partial S_{0}/\partial\vphi^{i}+\partial S_{1}/\partial\vphi^{i}+\cdots=0$.
Let us consider only the leading ``cosmological'' (dilaton potential) term in the EA
(see (4.20), (4.21)). The corresponding equations of motion are given by (4.25).
Let us try to solve them for $\phi=\text{const}$ keeping only the leading term in the curvature
(i.e.\ assuming $|\alpha' R |\ll 1$). We find
\begin{equation}
\alpha' R_{\mu\nu}=-F_{1}(\phi_{c})G_{\mu\nu},
\qquad\qquad 
\frac{1}{6}(D-26)=-F_{2}(\phi_{c}),
\qquad\qquad 
\phi_{c}=\text{const}.
\tag{5.9}
\end{equation}
where $F_{i}$ are given by (4.26), (4.27) or (4.17), (4.18).
Note that no solution exists unless $D\neq 26$.

\def \Im {{\rm Im\,}}

These equations determine the vacuum values of $G_{\mu\nu}$ and $\phi$
(or $G_{\mu\nu}$ and $D$)  [{1,9}]. The problem with this
solution (and, in fact, with all other solutions of loop corrected
equations of motion in bosonic  string theory) is that the coefficients of
the loop contributions to the EA are, in general, complex because of the
analytic continuation used to eliminate the tachyonic
divergence  [{28}] (see the discussion in Sec.~3). In
particular, the coefficients which appear in $F_i$ (4.17), (4.18) have
imaginary parts. For example, in the one loop case one has
\begin{equation}
F_{1}(\phi)=-\frac{1}{2}\,d_{1}d_{0}^{-1}e^{2\phi},
\qquad
F_{2}(\phi)=-\frac{1}{8}\,d_{1}d_{0}^{-1}D\,e^{2\phi},
\tag{5.10}
\end{equation}
where $d_{1}$ is the coefficient of the one-loop (torus) vacuum amplitude
\begin{equation}
d_{1}=\frac{1}{2}(4\pi^{2}\alpha')^{-D/2}
\int_{\mathcal{F}}\frac{d^{2}\tau}{(\Im\tau)^{14}}\,
|\eta(\tau)|^{4-2D}.
\tag{5.11}
\end{equation}
The prescription of Ref.\ 28 for evaluating this divergent integral by analytic continuation
in the (mass)$^{2}$ of the tachyon gives
\begin{equation}
d_{1}\approx \frac{\pi}{2(D/2)!}\,(\pi\alpha')^{-D/2}\,(0.3+\cdots+i).
\tag{5.12}
\end{equation}
$\mathrm{Im}\, d_i$ can be interpreted as a decay rate of the unstable
(tachyonic) classical vacuum. The small (complex) shifts of the massless
fields\foot{It is interesting to note that we may keep $F_i$ real (and hence get
real $G_{\mu\nu}$ and $D$) at the expense of making complex the vacuum
value of the dilaton or the string coupling constant (see (5.9)).   }
 which we find by solving the effective equations
cannot eliminate the tachyon and hence do not, in fact, determine a
consistent vacuum. This difficulty is absent in superstring theory.

\section*{6.\ String generating functional in Schottky parametrization}

In discussing renormalization of string loop corrections it is important
to use a parametrization of moduli space which is based on the
``extended'' set of moduli parameters such that the SL$(2,\mathbb{C})$
M\"obius group volume $\Omega$ appears as the universal inverse factor
of all (tree and loop) contributions to on-shell string amplitudes. In
this case one has a clear geometrical picture of factorization of the
amplitudes with the moduli and the Koba--Nielsen (KN) parameters playing
different roles in the degeneration limits. The separation of the
universal $\Omega^{-1}$ factor for all genera makes it possible also to
study the ``exponentiation'' of particular types of divergences.

One of such parametrizations is the Schottky parametrization
(SP)   [{23,29,30,31}] which is also distinguished by its
direct relation to the operator formalism   [{23,31}] and
 the resulting explicitness of the corresponding formulae
 (e.g.\ for the 2-d propagator and the string measure). 
 
 Using SP one
represents the genus $n$ Riemann surface (sphere with $n$ handles) in
terms of an  extended complex plane with $n$ pairs of holes cut out and
chooses the $3n$ moduli to be (roughly) the coordinates of the centers
of holes, their radii and the ``twist'' angles which appear in the
identification of boundary circles necessary to represent the handles.
Since both the KN points and the (relevant subset) of moduli appear as
coordinates on $\mathbb{C}$ it is possible to regularize both the
``local'' and ``modular'' divergences (corresponding to shrinking of the
dividing cycles on $\mathbb{C}$) in a universal way, introducing a short
distance cutoff on $\mathbb{C}$. 

Since the mechanism by which the
renormalization group is implemented in string loops involves
cancellations of divergences between contributions of surfaces of
different genera,  it is also important that in SP all surfaces are
represented universally in terms of $\mathbb{C}$ with holes (with more
topological structure, i.e.\ more holes and hence more moduli parameters
for higher genus).\foot{This should be important for a possibility to resum string loop
corrections (see Sec.~9).}  These important advantages of the
SP in extending the RG approach to string loop level were already
appreciated in Refs.~9 and 18.\foot{It should also be noted that it is in SP that the first expressions for
the multiloop corrections to string amplitudes were obtained.  
Moreover these old explicit results prompted the general expression for
the string loop measure for parametrizations based on complex moduli
parameters.} 

\def \C {\mathbb{C}} \def \ov {\over}

\

\def \F  {{\cal F}}

Let us  now
briefly review the representation of the closed bosonic  string amplitudes in SP
(for more information see [23b,30,31]). 
Let $\{T_{a},\,a=1,\ldots,n\}$ be the elements of SL$(2,\mathbb{C})$ acting on
the complex plane $\C$. The infinite discontinuous Schottky group (SG) is generated
by their products:
SG$=\{T_{\alpha}\},\ T_{\alpha}=T_{1}^{n_{1}}... T_{n}^{n_{k}}...$, $n_k$=rational, 
under the restriction that the fixed points of all the elements form a discrete set   and their  multipliers should satisfy  $k_\alpha| < 1$. 

By definition, if $z'=Tz=(Az+B)/(Cz+D)$  where  $AD-BC=1$, $T$  is represented as 
\begin{align}
&{Tz-\eta \ov  Tx - \xi}= k\,\frac{z-\eta}{z-\xi},
\tag{6.1}
\\
A=\Delta(\eta-k\xi),\quad
B=\Delta\xi\eta(k-1),\quad &
C=\Delta(1-k),\quad
D=\Delta(k\eta-\xi),
\quad
\Delta^{-1}=\sqrt{k}(\xi-\eta).\no 
\end{align}
Here $k$ is called the multiplier of $T$ and $\xi$ and  $\eta$ are the (repulsive and
attractive) fixed points.
We also assume that the isometric circles $I_{a}$ and $I'_{a}$ of $T_{a}$ and $T_{a}^{-1}$ 
($|C_{a}z+D_{a}|=1$ and $|C_{a}z-A_{a}|=1$) with the 
 radii and coordinates of the  centers  given by 
\begin{align}
&R=R'=\frac{1}{|C|}=\frac{\sqrt{|k|}}{|1-k|}\,|\xi-\eta|,
\qquad\ \ \ 
J=-\frac{D}{C} = \frac{\xi-k\eta}{1-k},\no 
\\
&J'=\frac{A}{C}=\frac{\eta-k\xi}{1-k}, \qquad \  \ \ J-J'= {1+k \ov 1-k} (\xi-\eta) 
\tag{6.2}
\end{align}
are exterior to each other.

The fundamental domain $\mathcal{F}$ of SG is
the region outside $2n$ isometric circles of $T_a$ and $T_a^{-1}$
(i.e.\ any point of $\mathbb{C}$ can be obtained from a point of
$\mathcal{F}$ by action of an element of SG and no two points of
$\mathcal{F}$ are related by an element of SG). It can be proved that any
genus $n$ Riemann surface can be represented by the fundamental region
of the SG. Thus $z$ and $T_a z$ represent the same point of Riemann
surface. The elements of SG are defined up to a similarity transformation
$T \rightarrow S T S^{-1}$ from SL$(2,\mathbb{C})$. The two elements
$T_a$ and $T_a'$ belong to the same conjugacy class if they can be
obtained from each other by a cyclic permutation of factors. $T_a$ is
called a primitive element if it is not equal to a power of some other
element of SG.

\def \v {{\nu}}

The moduli are thus represented by the parameters of $T_a$, i.e.\ by the
set of $3n$ complex numbers $\{\xi_a,\ \eta_a,\ k_a\}$,
$a = 1,\ldots,n$. The freedom of making the ``overall'' SL$(2,\mathbb{C})$
transformation reduces the number of independent moduli to the usual one
$3n - 3$. Note that while $\xi_a$ and $\eta_a$ transform in the standard
way under the SL$(2,\mathbb{C})$, $k_a$ remain invariant. This implies
that the torus ($n = 1$) is the special case: we cannot fix SL$(2,\mathbb{C})$
completely by fixing the positions of only two points ($\xi$ and $\eta$).
The remaining symmetry (related to the M\"obius U(1) symmetry of the
torus in the usual ``parallelogram'' representation) can be fixed only by
fixing the position of one extra (KN) point. We shall consider the case
of the torus in detail in the next section.

$\xi_a$ and $\eta_a$ lie inside the corresponding isometric circles.
As is clear from (6.2) there are two singular limits in which the surface
looses a handle and which correspond to shrinking the $a_a$ and $b_a$
cycles on the Riemann surface ($a_a$ go around the isometric circles
$I_a$ and $I_a'$, while $b_a$ connects a point on $I_a$ with its image
under $T_a$ on $I_a'$). The first is ``nondividing'' limit, in which
$k_a \rightarrow 0$ and hence the radii of the holes go to zero while
their centers remain at a finite distance $|\xi - \eta|$. The second is ``dividing''
limit in which $k_a \rightarrow 1$. As is clear from (6.1) this implies
that at the same time $\xi_a \rightarrow \eta_a$ (and vice versa). In
this limit the transformation $T_a$ becomes parabolic (see e.g., the
discussion in Ref.~23b)
   \[
\xi = \eta +  \vepsilon , \qquad
k = 1 - c \vepsilon , \qquad
(Tz - \xi)^{-1} = (z - \xi)^{-1} c .
\tag{6.3}
\]
As a result, the invariant circles $I_a$ and $I_a'$ touch each other with
their radii and the distance between their centers remaining finite.
Let us note that making a global SL$(2,\mathbb{C})$ transformation on
points of $\mathbb{C}$ and hence on $(\xi_a,\eta_a)$ one, in general,
changes the picture of the fundamental domain $\mathcal{F}$ and hence
may represent the ``dividing'' limit in some different way.

Let us now give the expressions for the basic quantities in SP.
The $n$ Abelian holomorphic differentials are
   \[
\omega_a
=
{\sum_{\alpha}}'
\Big(
\frac{1}{z - T_\alpha \eta_a}
-
\frac{1}{z - T_\alpha \xi_a}
\Big) dz
\equiv
\omega_a(z)\, dz ,
\tag{6.4}
   \]where the sum goes over the elements of SG which do not have
$T_a^{\,m}$ as their rightmost factor (recall that $\xi$ and $\eta$ are
the fixed points). The period matrix is given by
   \[
2\pi i\, \tau_{ab}
=
\oint_{b_b} \omega_a
=
\int_{z_0}^{T_b z_0} \omega_a(z)\, dz
\]
   \[
=
\delta_{ab}\,  \log k_a
-
{}^{(a)}{\sum_{\alpha}{}^{'(b)}} 
 \log
\frac{(\eta_b - T_\alpha \xi_a)(\xi_b - T_\alpha \eta_a)}
     {(\eta_b - T_\alpha \eta_a)(\xi_b - T_\alpha \xi_a)}
 , \qquad \ \  t_{ab} \equiv \mathrm{Im}\, \tau_{ab} ,
\tag{6.5}
\]
  where $T_\alpha \neq T_a^{\,k} \dots T_b^{\,m}$ and for $a=b$ the sum
does not include the unit element. The prime form is $E(z,w)(dz\,dw)^{-1/2}$
where
   \[
E(z,w)
=
(z - w)
{\prod_{\alpha}}^{\prime}\, 
\frac{(z - T_\alpha w)(w - T_\alpha z)}
     {(z - T_\alpha z)(w - T_\alpha w)}
\equiv
(z - w)\, \tilde  E(z,w) ,
\tag{6.6}
   \]where the product is over all elements of SG except $1$, and $T_\alpha$
and $T_\alpha^{-1}$ are counted only once. 

The above expressions
determine the Green function for the scalar Laplacian (with the constant
zero mode projected out)
   \[
\mathcal{G}(z,w)
=
- \frac{1}{4\pi}  \log |E(z,w)|^2
+
\frac{1}{8\pi^2}\,
\v_a\, t^{-1}_{ab}\, \v_b , \qquad \ \ \ 
\langle x^\mu(z) x^\nu(w) \rangle
=
2\pi \alpha'\, \mathcal{G}(z,w)\, \delta^{\mu\nu} ,
\]
   \[
\v_a
\equiv
\mathrm{Re}\! \int_{z}^{w} \omega_a
=
\sum_{\alpha}
 \log
\frac{(w - T_\alpha \eta_a)(z - T_\alpha \xi_a)}
     {(w - T_\alpha \xi_a)(z - T_\alpha \eta_a)} .
\tag{6.7}
   \]
\[
\Delta \mathcal{G}
=
-4\,\partial_z \partial_{\bar z}\,\mathcal{G}
=
\delta^{(2)}(z - w)
-
\frac{1}{4\pi^2}\,
\omega_a(z)\, t^{-1}_{ab}\, \bar{\omega}_b(\bar z) ,
\tag{6.8}
\]
where $t^{-1}_{ab}$ is the inverse of the imaginary part of the period
matrix. Since the zero mode is assumed to be projected out, there is an
ambiguity of adding $f(z) + f(w)$ term to (6.7). We can also rewrite
(6.7) as
\[
\mathcal{G}
=
\mathcal{G}_0 + \tilde{\mathcal{G}},
\qquad
\mathcal{G}_0
=
- \frac{1}{4\pi}\,  \log |z - w|^2 ,
\tag{6.9}
\qquad \ \ 
\tilde{\mathcal{G}}
=
- \frac{1}{4\pi}\,  \log |\tilde E|^2
+
\frac{1}{8\pi^2}\,
\v_a\, t^{-1}_{ab}\, \v_b .
\]
Note that $\tilde E$, $\v_a$ and $\tilde{\mathcal{G}}$ depend on the
cross-ratios and hence are ``automorphic functions''
($F(z) = F(T_\alpha z)$). Moreover, they are invariant under an
arbitrary SL$(2,\mathbb{C})$ transformation $S$ acting on
$\mathbb{C}$, i.e.\ on $z$, $w$, $\xi_a$ and $\eta_a$. Hence the only
piece of $\mathcal{G}$ which transforms under the M\"obius group is
$\mathcal{G}_0$,
\[
\mathcal{G}_0 \rightarrow \mathcal{G}_0
-
\frac{1}{4\pi}\,  \log |cz + d|^2
-
\frac{1}{4\pi}\,  \log |cw + d|^2 ,
\qquad
Sz = \frac{az + b}{cz + d} .
\]

We are ready now to quote the expression for the $n$-loop closed bosonic
string scattering amplitude derived in Refs.~23 and 30 and recently
improved in Refs.~31 and 26
\[
A^{(n)}_{N}
=
\Omega^{-1} c_n
\int d\mu_n \,
\langle V_1 \ldots V_N \rangle_n ,
\tag{6.10}
\]

\[
\langle \cdots \rangle_n
=
\int [dx]\,
\exp \!\Big(
-\frac{1}{4\pi \alpha'}
\int x\, \mathcal{G}^{-1}\, x
\Big),
\qquad\qquad 
\langle 1 \rangle_n = 1 ,
\tag{6.11}
\]
where the propagator is given by (6.7) and the modular measure is
\[
d\mu_n
=
\prod_{a=1}^{n}
\frac{d^2 \xi_a\, d^2 \eta_a\, d^2 k_a}
     {|\xi_a - \eta_a|^4\, |k_a|^4}
\, |1 - k_a|^4
\, (\det t_{ab})^{-D/2}\, 
\] \[ \quad \ \ \ \  \times
{\prod_{\alpha}}^{\prime}
\prod_{m=1}^{\infty}
|1 - k_\alpha^{\,m}|^{-2(D-2)}
{\prod_{\alpha'}}^{\prime}
|1 - k_{\alpha'}|^{-4} ,
\tag{6.12}
\]
where the product $\prod_{\alpha}^{\prime}$ goes over all primitive
elements of SG.

 The factor in power $D$ is the contribution of the
integral over $x^\mu$, while the contribution of the 2d
reparametrization ghosts is
$
{\prod_{\alpha}}^{\prime}
\prod_{m=2}^{\infty}
\Big| 1 - k_\alpha^{\,m} \Big|^{-4} .
$
The combination
\[
\frac{d^2 \xi\, d^2 \eta\, d^2 k}{|\xi - \eta|^4}
\times
\Big| \frac{1 - k}{k} \Big|^4
\]
has a natural interpretation of the measure for the SL$(2,\mathbb{C})$
group of transformations
\[
T =
\begin{pmatrix}
A & B \\
C & D
\end{pmatrix},
\qquad
\frac{d^2 A\, d^2 B\, d^2 C}{|A|^2},
\]
with $A,B,C,D$ expressed in terms of $\xi,\eta,k$, see Eq.~(6.1)
(note that the transformation $\xi,\eta,k \rightarrow A,B,C,D$ is not,
of course, one-to-one).

The integration over the moduli $(k,\xi,\eta)$ should be restricted
``by hand'' to go over the fundamental region of the modular group
(which is not explicitly known in the SP). One may formally not restrict
the region of integration assuming that the overall normalization
constants can be finally fixed, e.g., from unitarity considerations.

The measure (6.12) is clearly projective-invariant. Recalling the remark
that the SL$(2,\mathbb{C})$ transformation property of $\mathcal{G}(z,w)$
is the same as of the propagator on $\mathbb{C}$ we conclude that the
M\"obius volume factor $\Omega$ in (6.10) cancels out if the external
momenta satisfy the tree level on-shell conditions. One can fix the
M\"obius symmetry by fixing the positions of any of the three parameters
among $\xi_a,\eta_a$ and the KN points $z_i$. The usual expressions for
the loop amplitudes are obtained by fixing $\xi_1,\eta_1$ and $z_1$ for
$n=1$ and, e.g., $\xi_1,\xi_2,\eta_1$ for $n \ge 2$ (note again that
$n=1$ is a special case).\foot{Two Riemann surfaces are conformally equivalent if their SG's are
related by a similarity transformation, i.e.\ by an overall projective
transformation. The remaining $3n-3$ ($n \ge 2$) fixed points and
multipliers parametrize conformally inequivalent Riemann surfaces. The
projective invariance of the multiloop integrand thus implies that one
is integrating over the conformally inequivalent surfaces. }

It is, in principle, straightforward to study the factorization of
string amplitudes in SP. One draws a cycle $\gamma$ on $\mathbb{C}$ which
one wants to pinch and considers the limit in which the points
$z_i,\xi_a,\eta_b$ (which lie inside $\gamma$) uniformly come close to
each other (e.g.\ $\xi_1=\eta_1+u,\ z_1=\eta_1+uw_1,\ u\to0$, etc.).
A discussion of factorization in SP has already appeared in Ref.~18.
We would like to point out, however, that the general analysis of
Ref.~15 implies that to get consistent results, e.g., for the
tadpole factorization one should properly account for the additional
ghost or 2-d curvature insertions. 

The example of the factorization on
the disc (see Refs.~11,13,14,15,19 and Sec.~7) suggests that when the
string amplitudes diverge one may get different results by using
different representations for the world sheet (which are apparently
conformally equivalent on-shell). A consistent approach is to use a
compact curved 2-space representation for the world sheet and to account
for the curvature dependence in the modular measure, originating from
the frame dependence of the coordinates and radii of the holes on the
surface. In general, if we start with $\mathbb{C}$ with a topological
fixture of size $a \rightarrow 0$ at point $w$ we should go to a curved
space representation and define the center of the fixture by  [{15}]
\[
w' =
\frac{\displaystyle
\int_{|z|\le a} d^2 z\, \sqrt{g(z+w)}\, (z+w)}
{\displaystyle
\int_{|z|\le a} d^2 z\, \sqrt{g(z+w)}}
\simeq
w + O(a^2) .
\tag{6.13}
\]
Then
\[
\frac{da}{a^3}\, d^2 w
\;\rightarrow\;
\frac{da'\, d^2 w'\, \sqrt{g(w')}}{a'^3}
\Big[
1 + \frac{1}{4} a'^2 R^{(2)}(w') + O(a^4)
\Big] .
\tag{6.14}
\]
We shall return to the discussion of this point on the example of the
disc in Sec.~7 below.

Equation (6.10) suggests the following expression
for the generating functional for string amplitudes in SP (cf.~(4.1))
\[
\hat Z = \sum_{n=0}^{\infty} \hat Z_n ,
\qquad
\hat Z_n = \Omega^{-1} \bar Z_n ,
\tag{6.15}
\]

\[
\bar Z_n = c_n \int d\mu_n\, Z_n ,
\qquad
Z_n = \int_{M_n} [dx]\, e^{-I} ,
\tag{6.16}
\]

\[
I = I_0 + I_{\text{int}} ,
\qquad
I_{\text{int}} = \varphi^i V_i ,
\tag{6.17}
\]
where $[dx]$ is normalized so that
$\int [dx] \exp(-I_0) = 1$ (i.e.\ the determinant factors are included
into $d\mu_n$). $\bar Z_n$ is the partition function of the $\sigma$-model,
integrated over the moduli. 

We can introduce a short distance
regularization into $\bar Z_n$ by demanding that any of the points
$(\xi_a,\eta_b,z_i)$ cannot come closer than at the distance
$ \vepsilon \to 0$. While it is possible to regularize the local infinities
in a systematic way by using a ``built in'' cutoff in the propagator
$\mathcal{G}_0$ (see (2.5)) we are still to insert the cutoff explicitly
in the integrals over the moduli. Note that we do not introduce a
regularization for the integrals over the multipliers $k_a$ since, as we
have already discussed in Sec.~3, the ``nondividing'' limits
(corresponding to $k_a \to 0$, etc.) do not produce the Weyl or BRST
anomaly  [{18,49}] \foot{Acting by the conformal (or BRST) generator on the integrand of the
$n$-loop amplitude one finds nonzero contributions coming only from the
contour integrals going around $z_i,\ \xi_a,\ \eta_b$ (i.e.\ from the
corresponding boundaries of the moduli space). We assume that small
discs are cut around these points to regularize the corresponding
divergences.}
 (the integrals over $k_a$ should be
defined using an analytic continuation to eliminate the tachyonic
divergences).\foot{
There remains a technical problem of finding a special choice of moduli
$(\xi,\eta,k)$ with which the limits $\xi_a \to \eta_b$ are independent
from $k_a$ (cf.~(6.3)). This problem was ignored in Ref.~18.  }

We are assuming that the regularization we use is 2d covariant,
i.e.\ that $ \vepsilon$ is coupled to the conformal factor of the metric so
that divergences are directly related to Weyl anomalies. Let us note also
that in the ``generalized'' SP in which we use a compact curved 2-space
with $2n$ holes to represent the world ``surface'' the regularized
expression for $\bar Z_n$ depends on the Weyl factor of the metric
chosen. What is expected is that the Weyl anomaly cancels in the total
expression for $\hat Z$ computed at the true vacuum point.

The partition function $Z_n$ in (6.16) has the following useful
representation
\[
Z_n =
\exp \!\Big[
\frac{1}{2}\,(2\pi \alpha')\, \mathcal{G}\,
\frac{\delta^2}{\delta x^2}
\Big]
\exp\!\Big(-I_{\text{int}}[x]\Big)
\Big|_{x=0} , 
\tag{6.18}
\]
\[
\mathcal{G}\cdot \frac{\delta^2}{\delta x^2}
=
\int d^2 z_1\, d^2 z_2\,
\mathcal{G}(z_1,z_2)\,
\frac{\delta^2}{\delta x(z_1)\,\delta x(z_2)} .
\tag{6.19}
\]
Note that $Z_n$ depends on moduli explicitly through $\mathcal{G}$ in
(6.7) (and also in general, through 2-metric dependence in $I_{\text{int}}$)
and also implicitly through the region of integration over $z$
(the $\sigma$ model is defined on a sphere with holes whose positions
and radii depend on moduli).

\

Our aim will be to analyze the renormalizability property of $\hat Z$.
Since to find  ``nonclassical'' string vacua we are to move out of the
tree level conformal point,  we are thus to consider $\hat Z$ for arbitrary
(off-shell) values of the fields $\varphi^i$. In this case the formal
projective invariance of the integrand in $\hat Z$ (present before
regularization and for the classical on-shell values the fields) is
absent. Then we are to specify how the $\Omega^{-1}$ factor in (6.15)
is to be understood. 

We suggest to use the same prescription as at the
tree level, namely, to replace $\Omega^{-1}$ by the derivative over the
cutoff $\partial/\partial \log \vepsilon$ (see Sec.~2 and Ref.~22). As we
shall see in Sec.~7 on the examples of the torus and the disc this
prescription leads to the expression for the generating functional which
is renormalizable with respect to both ``local'' and ``modular''
infinities. Thus finally we can rewrite (6.15), (6.18) as

\[
\hat Z
=
\frac{\partial}{\partial \log \vepsilon}\, \bar Z ,
\qquad\qquad 
\bar Z = \sum_{n=0}^{\infty} \bar Z_n ,
\tag{6.20}
\]
\[
\bar Z_n
=
c_n \int d\mu_n\ \exp \!\Big[
\frac{1}{2}\,(2\pi \alpha')\, \mathcal{G}\,
\frac{\delta^2}{\delta x^2}
\Big]
\exp\!\Big(-I_{\text{int}}[x]\Big)
\Big|_{x=0} .
\]

\section*{7. Renormalization at one loop: torus and disc topologies}

In this section we shall consider the renormalization of the generating
functional $\hat Z$ in the one-loop approximation. We shall analyze
explicitly the divergences present in the torus and disc contributions
to $\hat Z$ and discuss how $\hat Z$ is to be defined in order to be
renormalizable.

\subsection*{1.}

Let us start with recalling the expression for the string amplitudes on
the torus in the usual ``parallelogram'' representation (see e.g.\
Refs.~58, 59 and 56)
\[
A^{(1)}_{N}
=
c_1 \int_{\mathcal{F}} [d\tau]\,
\langle V_1 \ldots V_N \rangle_{1} ,
\tag{7.1}
\]
where the Koba--Nielsen points $u_i$ belong to the parallelogram
$(\tau,1)$ on $\mathbb{C}$ and
\[
[d\tau]
=
\frac{d^2 \tau}{(\mathrm{Im}\,\tau)^{D/2}}
\, |\eta(\tau)|^{-2(D-2)} ,
\tag{7.2}
\quad 
\eta(\tau) = k^{1/24}\prod_{m=1}^{\infty}(1-k^{m}),
\qquad
k = e^{2\pi i \tau},
\qquad
D=26, \no 
\]
\begin{equation}
\langle f(x)\rangle_1
= 
\exp\!\Big[\frac{1}{2}(2\pi\alpha')\mathcal{G}\cdot
\frac{\delta^{2}}{\delta x^{2}}\Big]
f(x)
\Big|_{x=0},
\qquad
\langle 1\rangle_1 = 1, 
\tag{7.3}
\end{equation}
\begin{equation}
\mathcal{G}(u_1,u_2)
= -\frac{1}{4\pi}
\Big[
\log\!\big| u^{1/2}-u^{-1/2}\big|^{2}
+ \sum_{m=1}^{\infty}
\log\Big|
\frac{(1-vk^{m})(1-v^{-1}k^{m})}{(1-k^{m})^{2}}
\Big|^{2}
- \frac{2\pi}{\Im\tau}(\Im u_{12})^{2}
\Big],
\tag{7.4}
\end{equation}
where $v=\exp(2\pi i u_{12})$   and $u_{12}=u_1-u_2$.

We assume that the corresponding M\"obius group (U(1)) gauge is fixed
by fixing the position of one of the KN points. The tadpole factorization
corresponds to the limit in which all the KN points $u_i$ come close to
each other. Hence in the parallelogram parametrization there is no clear
distinction between ``local'' and ``modular'' infinities.

To make the analysis of renormalization more transparent it is useful to
go to a Schottky-type parametrization based on the ``extended'' set of
moduli. To this end let us consider the $N \ge 3$ amplitude on the torus
and make the two conformal transformations: first map the parallelogram
into the annulus (with coordinates $w = e^{2\pi i u}$ and the radii of
the boundary circles $1$ and $|k| = e^{-2\pi\,\mathrm{Im}\,\tau}$) and
then map the annulus into the complex plane with the two holes cut out,
by the transformation $w \rightarrow z$  [{18}]
\[
\frac{z - z_1}{z - z_3}\,
\frac{z_2 - z_3}{z_2 - z_1}
=
\frac{w - w_1}{w - w_3}\,
\frac{w_2 - w_3}{w_2 - w_1} .
\tag{7.5}
\]
Introducing the two parameters (the corresponding points lie inside the
holes)
\[
\xi = z(w=0), \qquad \eta = z(w=\infty)
\tag{7.6}
\]
which can be expressed in terms of $w_1,w_2,w_3,z_1,z_2,z_3$, we thus
finish with the map
$\{u_2,\ldots,u_N\} \rightarrow \{\xi,\eta,z_4,\ldots,z_N\}$
(we may assume, e.g., that $u_1$ is fixed as an U(1) M\"obius gauge in
the parallelogram representation and $z_1,z_2,z_3$ are fixed as an
SL$(2,\mathbb{C})$ M\"obius gauge). Since the on-shell amplitude is
formally invariant under the conformal transformation we can rewrite
(7.1) as

\[
A^{(1)}_{N}
=
c_1\, \Omega^{-1}
\int d\mu_1 \,
\langle V_1 \ldots V_N \rangle_{1} ,
\tag{7.7}
\]
where   now $V_i$ are defined on the 2-holed   plane   and instead of (7.2) and (7.40  we have 
\begin{equation}
d\mu_1
= \frac{d^{2}\xi\, d^{2}\eta\, d^{2}k}
{(2\pi)^{2}|\xi-\eta|^{4}|k|^{4}}\,
t^{-D/2}
\prod_{m=1}^{\infty}|1-k^{m}|^{-2(D-2)}
= [d\tau]\,
\frac{d^{2}\xi\, d^{2}\eta}{|\xi-\eta|^{4}}, 
\tag{7.8}
\end{equation}
\begin{equation}
\mathcal{G}(z_1,z_2)
= -\frac{1}{4\pi}\Big[ \log|z_1-z_2|^{2}
+
\sum_{m=1}^{\infty}
\log\Big|
\frac{(1-\lambda k^{m})(1-\lambda^{-1}k^{m})}
{(1-k^{m})^{2}}
\Big|^{2}
+ \frac{(\log|\lambda|)^{2}}{\log|k|}\Big]
\equiv \mathcal{G}_0+\tilde{\mathcal{G}},
\tag{7.9}
\end{equation}
\begin{equation}
\lambda = \frac{(z_1-\xi)(z_2-\eta)}{(z_1-\eta)(z_2-\xi)},
\qquad
k=e^{2\pi i\tau},
\qquad
\Im\tau = -\frac{\log|k|}{2\pi}.\no 
\end{equation}
In (7.7) it is assumed that the SL$(2,\mathbb{C})$ M\"obius gauge is
fixed on the KN points $z_i$. As in the general case of the Schottky
parametrization (see Sec.~6) the integrand in (7.7) is formally
SL$(2,\mathbb{C})$ invariant for the on-shell values of momenta (for
which the noninvariance of $\mathcal{G}_0$ in (7.9) does not matter;
note that $\tilde{\mathcal{G}}$ depends on the invariant cross-ratio).
This explains why the same expressions (7.7)--(7.9) can be formally
obtained by using different conformal maps from the annulus to the
2-holed plane. 

For example, instead of (7.5) one could use
$(z-\eta)(z_1-\xi)/(z-\xi)(z_1-\eta)=w$ or (see Ref.~32)
$z=\kappa(w-\xi)/(w-\eta)$.\foot{In the latter case $|k| < |\kappa| < 1$ and the radii of the holes are
\[
r_1
=
|\kappa|\,
\frac{|\xi-\eta|}{|1-|\kappa|^{2}|},
\qquad
r_2
=
\frac{|\kappa||k|\,|\xi-\eta|}
     {|\kappa|^{2}-k^2},
\]
i.e.\ $r_1 \neq r_2$ and $\xi \to \eta$ implies $r_1 \to 0$, $r_2 \to 0$
for arbitrary value of $k$ (so that the boundaries of the holes are not
isometric circles).
   } 
It is necessary, however, to keep in mind that different conformal maps
lead to different integration regions for the parameters in (7.7).

Leaving aside the question of integration regions one can rewrite (7.7)
in several formally equivalent ways by choosing different
SL$(2,\mathbb{C})$ M\"obius gauges. For example, fixing $\xi$, $\eta$
and $z$, we return to the expression (7.1) in the parallelogram
representation (note that for $\xi=0$, $\eta=\infty$ (7.9) reduces to
(7.4)).

To compare (7.8) and (7.9) with the general expressions (6.12) and (6.7)
in the Schottky parametrization let us note that for the torus the
Schottky group has only one generator $T=(k,\xi,\eta)$ and hence
(see (6.6) and (6.8))
\[
\tilde{E}
=
\prod_{m=1}^{\infty}
\frac{(z_1 - T^{m}z_2)(z_2 - T^{m}z_1)}
     {(z_1 - T^{m}z_1)(z_2 - T^{m}z_2)},
\qquad
\omega(z)
=
\frac{1}{z-\eta}
-
\frac{1}{z-\xi},
\tag{7.10}
\]
\[
\frac{T^{m}z - \eta}{T^{m}z - \xi}
=
k^{m}\,
\frac{z - \eta}{z - \xi}.
\]
It is then straightforward to check (employing e.g.\ the projective
invariance of $\tilde{\mathcal{G}}$) that (7.9) is in agreement with
(6.7) and (6.8). In  particular, $\nu=\log|\lambda|$,  and 
\[
\tilde{E}
=
\prod_{m=1}^{\infty}
(1-\lambda k^{m})(1-\lambda^{-1}k^{m})(1-k^{m})^{-2}.
\]
Using that
\[
\log(1-x)=-\sum_{m=1}^{\infty}\frac{1}{m}x^{m},
\qquad
\frac{1}{1-x}=\sum_{m=0}^{\infty}x^{m},
\]
(and dropping $f(z_1)+f(z_2)$ terms) we can also rewrite the Green
function (7.9) in the following way
\[
\mathcal{G}(z_1,z_2)
=
\mathcal{G}_0
+\frac{1}{4\pi}\sum_{m=1}^{\infty}
\Big[
\frac{1}{m}\frac{k^{m}}{1-k^{m}}
\Big(\lambda^{m}+\lambda^{-m}-2\Big)
+\text{c.c.}
\Big]
-\frac{(\log|\lambda|)^{2}}{4\pi\log|k|}.
\tag{7.11}
\]
It is easy to see that only the last nonholomorphic term in
$\mathcal{G}$ (7.11) gives the leading nontrivial (nonholomorphic)
contribution in the $\xi\to\eta$ (or $\lambda\to1$) limit (we again drop
the terms depending only on $z_1$ or $z_2$)
\[
\mathcal{G}\big|_{\xi\to\eta}
\simeq
\mathcal{G}_0
+
\frac{|\zeta|^{2}}{8\pi\log|k|}
\Big[
\frac{1}{(\xi-z_1)(\bar{\xi}-\bar{z}_2)}
+\text{c.c.}
\Big]
+O(\zeta^{2},\bar{\zeta}^{2})
\tag{7.12}
\]
\[
\qquad\qquad \ \ \ =
\mathcal{G}_0
+
\frac{2\pi|\zeta|^{2}}{\log|k|}
\Big[
\partial_{\xi}\mathcal{G}_0(\xi,z_1)\,
\bar{\partial}_{\xi}\mathcal{G}_0(\xi,z_2)
+\text{c.c.}
\Big]
+O(\zeta^{2},\bar{\zeta}^{2}),  \qquad \zeta \equiv \xi-\eta . 
\tag{7.13}
\]
Consider now the expression for the generating functional for the
amplitudes (7.7) (see (6.16)--(6.19))
\[
\hat{Z}_1
=
c_1\,\Omega^{-1}
\int d\mu_1
\exp\!\Big[
\frac{1}{2}(2\pi\alpha')\,
\mathcal{G}\cdot\frac{\delta^{2}}{\delta x^{2}}
\Big]
e^{-I_{\text{int}}[x]}
\Big|_{x=0}.
\tag{7.14}
\]
The tadpole factorization corresponds to the limit $\xi\to\eta$ in
which the handle degenerates. Note that in contrast to the general
case of the Schottky parametrization the parameters $\xi$ and $\eta$
defined by (7.6) are formally independent of the modulus $k$ (their
integration region, in fact, depends on $k$) and hence the limit
$\xi\to\eta$ can be taken for arbitrary $k$. Employing the expansion
(7.13) we find in this limit
\[
\hat{Z}_1
=
c_1\,\Omega^{-1}
\Big\langle
\int_{\varepsilon}
\frac{d^{2}\zeta}{|\zeta|^{2}}
\Big[
t_1 V_t
-\frac{1}{4}\lambda_1 {\mathcal{O}}_1 |\zeta|^{2}
+O(\zeta^{2},\bar{\zeta}^{2})
\Big]
\Big\rangle_{0},
\tag{7.15}
\]
\[
\langle F[x]\rangle_{0}
=
\exp\!\Big[
\frac{1}{2}(2\pi\alpha')\,
\mathcal{G}_0\cdot\frac{\delta^{2}}{\delta x^{2}}
\Big]
F[x]
\Big|_{x=0}, 
\tag{7.16}
\qquad 
t_1=\int [d\tau]\,, \qquad
V_t=\int d^2 z \cdot 1 \,,
\]
\[
\lambda_1=\int (d\tau)
=\int \frac{d^2 k}{(2\pi)^2 |k|^4}
(\Im \tau)^{-D/2-1}
\prod_{m=1}^{\infty}
\Big|1-k^m\Big|^{-2(D-2)} ,
\tag{7.17}
\qquad 
(d\tau)=\frac{[d\tau]}{\Im \tau}\, ,
\]
\[
{\mathcal O}_1
=
\frac{1}{2\pi\alpha'}
\int d^2 z :
\partial_a x^\mu \partial^a x_\mu :\ 
=
\frac{2}{\pi\alpha'}
\int d^2 z :
\partial x^\mu \bar{\partial} x_\mu :
\; .
\tag{7.18}
\]
We have used that the insertion of the operator ${\mathcal O}_1$
into a correlator on the sphere can be represented as
$\sim \int \partial \mathcal G_0 \bar{\partial} \mathcal G_0
\cdot (\delta^2/\delta x^2)$.
The operator ${\mathcal O}_1$ is the same as in (3.7) for $n=1$.
$t_1$ and $V_t$ are the zero momentum tachyon tadpole and the vertex
operator. $\lambda_1$ is proportional to the massless tadpole (or vacuum
amplitude) on the torus (in contrast to $t_1$, $\lambda_1$ has modular
invariant integrand for $D=26$). One can check (using e.g., the
parallelogram representation) that
\[
\langle {\mathcal O}_1 \rangle_1
=
- D \lambda_1 
\tag{7.19}
\]
(cf. (3.12); note that ${\mathcal O}_1$ is \emph{twice} the soft
graviton operator (3.11)). Equation (7.15) thus implies that the
logarithmically divergent (tadpole) part of the generating functional
on the torus is given by (we set 
$\int_\varepsilon d^2\zeta/|\zeta|^2 = -2\pi \log \varepsilon + \cdots$)
\[
\hat Z_{1\infty}
=
c_1 \Omega^{-1}
\frac{1}{2}\,\pi \lambda_1 \log \varepsilon \,
\langle {\mathcal O}_1 e^{-I_{\rm int}} \rangle_0 .
\tag{7.20}
\]
Thus to cancel this divergence in the total expression
$\hat Z=\hat Z_0+\hat Z_1+\cdots$ we are to add the following counterterm
to the string action in $\hat Z_0$
\[
\hat Z_0
=
c_0 \Omega^{-1}
\langle e^{-\delta I - I_{\rm int}} \rangle_0 ,
\tag{7.21}
\]
\[
\delta I
=
\frac{1}{2}\, c_1 c_0^{-1} \pi \lambda_1 \log \varepsilon \,
{\mathcal O}_1 .
\tag{7.22}
\]
This result is consistent with the general expressions (3.21) and (4.28)
if
\[
b_1
=
\frac{1}{2}\, c_1 c_0^{-1} \pi \lambda_1
=
\frac{1}{4}\, d_1 d_0^{-1} ,
\tag{7.23}
\]
where $d_n$ are the constants which appear in $\hat Z$ (4.19). If we fix
$d_1$ by comparing $\hat Z_1$ with the usual field theory result
$\tfrac12 \log \det (\alpha' \Box) + \cdots$ then $d_1$ is given by
(5.10) (see Ref.~56), i.e.
\begin{equation}
d_1 = \frac{1}{2}(4\pi^2\alpha')^{-D/2}\lambda_1.
\tag{7.24}
\end{equation}
Here $(2\pi\alpha')^{-D/2}$ comes from the constant in the string action and
$(2\pi)^{-D/2}$ from the Gaussian integral normalization; the total factor
$(4\pi^2\alpha')^{-D/2}\sqrt{G}$ in (4.19) is the contribution of the zero
mode of $x^\mu$. The choice of $d_0 \sim (4\pi^2\alpha')^{-D/2}$ is a
matter of convention, being related to the choice of the string coupling.

\subsection*{2.}

 Let us now analyze how the combined ``local'' plus ``modular''
renormalization can be carried out in $\hat Z_1$. Let us first consider
the generating functional for the massless amplitudes in the
``parallelogram'' representation of the torus \foot{Note that it is not necessary to fix explicitly the compact $U(1)$ Möbius group. }
\[
\hat Z_1 = c_1 \int (d\tau)\, Z_1 ,
\qquad
Z_1 = \int [dx]\, e^{-I} ,
\qquad
I = I_0 + I_{\text{int}} ,
\tag{7.25}
\]
where $(d\tau) = [d\tau]/\Im \tau$. $\hat Z_1$ is thus simply the integral
over $\tau$ of the partition function of the $\sigma$ model (2.3) defined
on a compact torus represented by the parallelogram on $\mathbb{C}$. As we
have already discussed in Sec.~2, the partition function of the $\sigma$
model on a compact space of arbitrary topology is given by (see (2.7),
(2.15) and Ref.~22;  see also the Appendix)

\[
Z_n = a_n \int d^D y \sqrt{G}\,
e^{2(n-1)\phi}
\Big( 1 + \tfrac{1}{2}\alpha' \log \vepsilon\, R + \cdots \Big) .
\tag{7.26}
\]
We put $\chi = 2 - 2n$ and $D = 26$ in (2.15). For all $n$, $Z_n$ is
renormalizable with respect to ``local'' infinities, i.e.\ satisfies
(2.12). 

This leads to the following apparent paradox: if we renormalize
all the ``local'' infinities in $Z_n$, $Z_1$ in (7.25) will be finite and
hence $\hat Z_1$ will also be finite (since, as we have already argued,
the integral over $\tau$ does not give new divergences of the type we are
interested in). At the same time, $\hat Z_1$ should definitely contain
the ``modular'' divergences discussed above. 

To resolve this paradox one
should observe that what appears as a ``local'' infinity in (7.25) is, in
fact, a ``modular'' infinity from the point of view of the string
amplitudes defined on the parallelogram. Consider, for example, the
3-graviton amplitude on the torus in the parallelogram representation.
The $(\text{momentum})^2$-term in the amplitude corresponds to the
$h_{\mu\nu}^3$  piece of the $R$-term in $\hat Z_1$ or in $Z_1$ (7.26).
This amplitude contains the infinities corresponding to the limits in
which two or all three KN points come close to each other.\foot{By using the expression for the propagator on the torus one can check
directly that the result is in agreement with (7.26).}
These infinities are ``local'' if considered from the point of view of the
$\sigma$ model defined on the parallelogram but are ``modular'' external
leg and tadpole infinities from the standard string theory point of view.\foot{Let us recall that the computation of the $\sigma$-model $\beta$-function
on the torus [{34,60}] gives the same result as on the sphere.
To find the ``modular'' corrections to $\beta$ one is to give up the
parallelogram representation, introducing additional modular parameters
(cf.\ Ref.~9) and defining the ``stringy'' $\sigma$-model as the ordinary
one integrated over the extended set of moduli.}


It is instructive to compare this with what happens on the sphere.
The correlator of the three graviton vertex operators on $\C$ is again
divergent in precisely the same limits
$(z_1 \to z_j$ and $ z_1 \to z_3,\; z_2 \to z_3)$, reproducing the $R\log \vepsilon$
term in (7.26) (for $n=0$).
From the string theory point of view one interprets these divergences as
M\"obius divergences (see Sec.~2 and Refs.~20 and 21) and subtracts them
(e.g.\ by fixing the positions of the vertex operators or applying the
$\del/\del\log  \ve $ prescription (2.33)) obtaining thus the finite expression for the
3-graviton amplitude or the $R$ term in the generating functional
$\hat Z_0$ (see (2.39) and (2.40)).

Let us now summarize and generalize the above discussion.
Consider the standard ``M\"obius gauge fixed'' approach in which one uses
the ``restricted'' sets of moduli $\{\tau\}$ and hence
\begin{equation}
\hat Z_n = c_n \int d\mu_n(\tau)\, Z_n,
\qquad
Z_n = \int_{M_n} [dx]\, e^{-I},
\qquad
n \ge 1.
\tag{7.27}
\end{equation}
Since $Z_n$ is renormalizable with respect to the local infinities,
we can rewrite the $O(\log \vepsilon)$ term in it as
$\beta^i_0 \, \partial Z_n / \partial \vphi^i \, \log \vepsilon$
(see (2.13)).
Hence the part of the $O(\log \vepsilon)$ term in $\hat Z_n$
corresponding to factorizations into the genus $0$ and genus $n$
parts can be represented as (cf.\ (4.23))
\begin{equation}
\hat Z_n = \hat Z_{nR}
+ \beta^i_0 \frac{\partial \hat Z_{nR}}{\partial \vphi^i}\,\log \vepsilon
+ \ldots \, .
\tag{7.28}
\end{equation}
Recalling that
$\beta^i_0 = \kappa^{ij}_0\, \partial \hat Z_{0R}/\partial \vphi^j$
(see (2.19)),
we can absorb the divergence (7.28) into the renormalization of the
tree term in \[\hat Z = \sum_{n=0}^\infty  \hat Z_n,\qquad\qquad  \delta_n  \vphi^i=
\kappa^{ij}_0 {\partial \hat Z_{0R}\over \partial \vphi^j } \,\log \vepsilon
+ \ldots,\]
obtaining the leading order modular correction to the $\beta$-function
\begin{equation}
\beta^i_n
= \kappa^{ij}_0 \frac{\partial \hat Z_{nR}}{\partial \vphi^j},
\qquad\qquad 
\hat Z_{nR}
= d_n \int d^D y\, \sqrt{G}\, e^{(2n-1)\phi}
+ \ldots \, .
\tag{7.29}
\end{equation}
Using the expression for $\kappa^{ij}_0$ (2.20) one finds that (7.29)
is, in fact, in agreement with the previous result for the
``momentum independent'' contribution to the modular
$\beta$-functions (see (4.25)).

Equation (7.28) is not, however, what one would expect to find if
$\hat Z$ were renormalizable with respect to the sum of the local and
modular infinities.
One would expect instead
\begin{equation}
\frac{\partial \hat Z}{\partial \log \vepsilon}
= \beta^i \frac{\partial \hat Z}{\partial \vphi^i},
\qquad
\frac{\partial \hat Z_n}{\partial \log \vepsilon}
= \sum_{m=0}^{\infty}
\beta^i_m \frac{\partial \hat Z_{n-m}}{\partial \vphi^i},
\tag{7.30}
\end{equation}
\begin{equation}
\hat Z_n
= \hat Z_{nR}
+ \beta^i_0 \frac{\partial \hat Z_{nR}}{\partial \vphi^i}\,\log \vepsilon
+ \beta^i_n \frac{\partial \hat Z_{0}}{\partial \vphi^i}\,\log \vepsilon
+ \ldots \, .
\tag{7.31}
\end{equation}
In particular,
\begin{equation}
\hat Z_1
= \hat Z_{1R}
+ 2 \beta^i_0 \frac{\partial \hat Z_{1R}}{\partial \vphi^i}\,\log \vepsilon
+ O(\log^2 \vepsilon) \, ,
\tag{7.32}
\end{equation}
where we have used that to leading order
\begin{equation}
\beta^i_0 \frac{\partial \hat Z_{n}}{\partial \vphi^i}= 
= \beta^i_n \frac{\partial \hat Z_{0}}{\partial \vphi^i}
= \frac{\partial \hat Z_{0}}{\partial \vphi^j}
\kappa^{ji}_0
\frac{\partial \hat Z_{1R}}{\partial \vphi^i}.
\tag{7.33}
\end{equation}
The interpretation of (7.31) and (7.32) is clear: ``one half'' of the
$\log \vepsilon$ divergence in $\hat Z_1$ should be ``local'' and
``one-half'', ``modular''.
The ``local'' divergence is to be cancelled by inserting the local
$(\beta_0)$ counterterm directly into $\hat Z_1$ while the
``modular'' should be cancelled by inserting the modular
$(\beta_1)$ counterterm into $\hat Z_0$.

We conclude that the generating functional defined according to the
standard M\"obius gauge fixed prescription does not satisfy the RG
equation (7.30) with the {\it full}  $\beta$-function (which is proportional
to the full effective equations of motion).
This suggests that we need to modify the definition of $\hat Z$ in order
to ensure its renormalizability.

The qualitative reason for the absence of ``doubling'' of the
$\log \vepsilon$ divergence in $\hat Z_1$ is that we have used the
parametrization in which the M\"obius infinities (which, in fact, are a
subclass of local infinities, see Sec.~2) are already dropped out.
However, the tree-level experience suggests that one should first
regularize both the M\"obius and modular infinities and then use a
special prescription of how to ``divide'' by the M\"obius volume in
order to preserve correspondence with the usual results for on-shell
amplitudes.

Consider once again the 3-graviton correlator on the sphere and the
torus, using the Schottky-type (sphere with two holes) representation
for the latter.
Introducing a short distance cutoff we find  (here $\langle ...\rangle_1$  includes integration  over the
moduli)
\begin{equation}
\langle V_1 V_2 V_3\rangle_0 \sim \log \vepsilon,
\qquad
\langle V_1 V_2 V_3\rangle_1 \sim \log^2 \vepsilon
+ O(\log \vepsilon) \, .
\tag{7.34}
\end{equation}
The $\log \vepsilon$ term in $\langle V_1 V_2 V_3\rangle_0$ and one
of the $\log \vepsilon$ factors in
$\langle V_1 V_2 V_3\rangle_1$ can be interpreted as M\"obius (or
local) infinities originating from the limits in which the vertex
operators collide.
The second $\log \vepsilon$ factor in
$\langle V_1 V_2 V_3\rangle_1$ is the modular infinity corresponding
to the limit $(\xi\to \eta)$  in which the holes collide (the handle disappears).

If we use the standard prescription $(\Omega^{-1} \sim \log \vepsilon)$
for subtracting the M\"obius infinity we get
\begin{equation}
A^{(0)}_3 \sim \Omega^{-1}\langle V_1 V_2 V_3\rangle_0
= \text{finite},
\qquad
A^{(1)}_3 \sim \Omega^{-1}\langle V_1 V_2 V_3\rangle_1
\sim \log \vepsilon + \text{finite},\no 
\end{equation}
where the $\log \vepsilon$ term in $A^{(1)}_3$ corresponds to the modular
divergence in $\hat Z_1$ discussed above.

If instead of $\Omega^{-1}\sim \log \vepsilon$ we use the
$\Omega^{-1}\to \partial/\partial\log \vepsilon$ prescription [21,22], 
we get precisely the desired doubling of the coefficient of the
$\log \vepsilon$ term in $A^{(1)}_3$.
What happens is that the derivative $\partial/\partial\log \vepsilon$
counts ``local'' and ``modular'' $\log \vepsilon$ factors in the
$\log^2 \vepsilon$ term in an independent way, in agreement with
(7.31) and (7.32).

Thus we suggest to use the definition of $\hat Z$ already given in
(6.20) and expect to find
\begin{align}
\hat Z
= \frac{\partial}{\partial \log \vepsilon}
&\big( Z_0 + \bar Z_1 + \cdots \big)
= \frac{\partial}{\partial \log \vepsilon}
\Big[ 
2 d_0 \int d^D y \sqrt{G} e^{-2\phi}
\Big(1 + \frac{1}{2}\alpha' \log \vepsilon\, R + \cdots \Big)
\no\\  & + d_1 \log \vepsilon \int d^D y \sqrt{G}
\Big(1 + \frac{1}{2}\alpha' \log \vepsilon\, R + \cdots \Big)
+ \cdots
\Big]
\tag{7.35}\\
&
= d_0 \int d^D y \sqrt{G} e^{-2\phi}
(\alpha' R + \cdots)
+ d_1 \int d^D y \sqrt{G}
(1 + \alpha' \log \vepsilon\, R + \cdots) \, .
\tag{7.36}
\end{align}
Here the overall $\log \vepsilon$ factor in $\bar Z_1$ corresponds to
the ``restored'' M\"obius group volume factor which should have
(to the considered order) the equivalent interpretation of the modular
divergence originating from the integral over the ``extra'' moduli $\xi$ and $\eta$. 

Assuming that $\hat Z_1 = \bar Z_1 + \log \vepsilon\, \partial \bar Z_1/\partial\log \vepsilon$
and using the relations 
$\beta^i_0 = \kappa^{ij}_0 \partial \hat Z_0/\partial \vphi^j$,
$\beta^i_1 = \kappa^{ij}_0 \partial \bar Z_1/\partial \vphi^j$ (see (7.29),(7.33))
it is easy to check the renormalizability of the total $\hat Z$
 (7.30)--(7.32).\foot{Such counting is also necessary  in order to have the sum of the local  and modular pieces  in the total $\beta$-function.}

Let us now explain how the $\log^2 \vepsilon\, R$ term actually appears in
the integrated over the moduli $\sigma$-model partition function
\[
\bar Z_1
=
c_1 \int [d\tau]\,
\frac{d^2\xi\, d^2\eta}{|\xi - \eta|^4}
\Big\langle e^{-I_{\text{int}}} \Big\rangle_{1} .
\tag{7.37}
\]
Recalling the previous result (7.15) and (7.20) for the logarithmic
singularity in the $\xi \to \eta$ limit we get
\[
\bar Z_{1\infty}
=
\frac{1}{2}\, c_1 \pi \lambda_1 \log \vepsilon\,
\Big\langle \O_{1} e^{-I_{\text{int}}} \Big\rangle_{0} .
\tag{7.38}
\]
Next we note that the insertion of $\O_{1}$ (7.18) into the partition
function on the sphere can be represented in the following way\foot{Note that up to normal ordering $\O_{1}$ is twice the free string
action. The normal ordering implies that
$\partial/\partial \alpha'$ does not act on the overall zero mode factor
$(\alpha')^{-D/2}$ in $Z_{0}$ (see Ref.~14). In fact,
$\int \partial x \partial x \sim \alpha' \del/\del\alpha'$ but the insertion
of the $D \int R^{(2)}$ term in (3.8) cancels against the derivative of
the $(\alpha')^{-D/2}$ factor.  }
\[
\Big\langle \O_{1} e^{-I_{\text{int}}} \Big\rangle_{0}
=
2\alpha' \frac{\partial}{\partial \alpha'}
\Big\langle e^{-I_{\text{int}}} \Big\rangle_{0}
\sim
\alpha' \frac{\partial}{\partial \alpha'} Z_{0}
\sim
\int d^{D}y \sqrt{G}
\Big( \tfrac{1}{2}\alpha' R \log \vepsilon + \cdots \Big) ,
\tag{7.39}
\]
where in the last line we have used (7.26) (the dilaton factor cancels
out since this contribution originates from the torus). Substituting
(7.39) into (7.38) we indeed find the $\log^2 \vepsilon\, R$ term. 

The
problem which still remains is an apparent absence of the
$\log \vepsilon \int d^{D}y \sqrt{G}$
term in $\bar Z_{1}$. This term should be present for a consistency of
the prescription (7.35), (7.36).\foot{Such term is to be expected since the soft graviton amplitudes on the
torus are finite only in the standard (M\"obius gauge fixed)
parallelogram representation.
}

 This suggests that
we are still to improve the above qualitative picture. One subtle point
is related to the replacement of $\mathbb{C}$ by a compact 2-sphere for
which we can use the expression (7.26) for the partition function. It
would be better to start directly with a curved compact 2-sphere with
two holes (with identified boundaries). In this case, however, the
expression for the modular measure may change.


\subsection*{3. }

 To get a deeper understanding of related issues let us now analyze
the case of the disc correction to the generating functional for closed
string amplitudes in the theory of open and closed strings. As in the
case of the torus, let us first consider the ``M\"obius gauge fixed''
representation of the disc in terms of the interior of the unit circle
on the complex plane. It is not necessary to fix the $SL(2,R)$ M\"obius
symmetry on the disc explicitly since the corresponding divergences are
power-like (not logarithmic), i.e.\ the $SL(2,R)$ M\"obius group volume
is finite if we drop power infinities  [{61,20}].

Then the
momentum independent logarithmic modular divergences in the closed
string correlators on the disc come from the limits in which all $N$ or
$N-1$ vertex operators collide [11]  (see also Refs.~14
and 19). They are simply the local divergences when considered from the
point of view of the $\sigma$ model defined on the unit disc. The
corresponding generating functional is thus proportional to the
partition function (7.26) (we formally use $n=\tfrac{1}{2}$ to denote
the contribution of the disc since this is in agreement with
$\chi = 1$)
\begin{equation}
\hat Z_{1/2}
= d_{1/2} \int d^D y \sqrt{G} e^{-\phi}
\Big(1 + \frac{1}{2}\alpha' \log \vepsilon\, R + \cdots\Big) ,
\tag{7.40}
\end{equation}
and so  is renormalizable with respect to the local divergences.

Hence
just as in the case of the torus in the parallelogram representation
(see (7.27)--(7.30)) we find that these local divergences

 (i) can be
interpreted as ``modular'' from string theory point of view; 

(ii) can be
absorbed into a renormalization of the couplings in $\hat Z_{0}$ with
the proper ``modular'' counterterm (corresponding to (4.15), (4.17) and
(4.18) with $n=\tfrac{1}{2}$ or (4.26), (4.27) with
$\omega \sim e^{-\phi}$); 

(iii) are only ``one half'' of the divergences
that should be present in $\hat Z_{1/2}$ in order for
$\hat Z = \hat Z_{0} + \hat Z_{1/2} + \cdots$ to be renormalizable with
respect to both local and modular infinities  [19].

 In
order to resolve the latter problem let us go to the Schottky-type
``plane with a hole'' representation of the disc  [{13}].
Conformally transforming (inverting) the unit disc into the plane with a
hole one finds the following expression for the generating functional
for the closed string amplitudes on the disc [{13}]
\begin{equation}
\hat Z_{1/2}
= c_{1/2}\Omega^{-1} \bar Z_{1/2},
\qquad
\bar Z_{1/2}
= \int d\mu_{1/2}\,
\exp\!\Big[\frac{1}{2}(2\pi\alpha')\mathcal G\cdot
\frac{\delta^2}{\delta x^2}\Big]
e^{-I_{\rm int}}
\Big|_{x=0} ,
\tag{7.41}
\end{equation}
\begin{equation}
d\mu_{1/2} = \frac{da}{a^3}\, d^2 w \, ,
\tag{7.42}
\end{equation}
\begin{equation}
\mathcal G(z_1,z_2)
= -\frac{1}{4\pi}\log|z_1-z_2|^2
-\frac{1}{4\pi}
\log\Big|
1-\frac{a^2}{(z_1-w)(\bar z_2-\bar w)}
\Big|^2
= \mathcal G_0 + \tilde{\mathcal G},
\tag{7.43}
\end{equation}
where $a$ and $w$ are the radius of the hole and the position of its
center on $\mathbb C$.
$\Omega$ is the volume of the SL$(2,\mathbb C)$ M\"obius group on
$\mathbb C$.

In the derivation of (7.41) one assumes that the M\"obius gauge is
fixed by fixing the positions of the KN points (e.g.\ $z_1=0$,
$z_2=1$, $z_3=\infty$).
According to (7.41) we are first to compute the partition function on
the plane with a hole, then average over the ``moduli'' $a$ and $w$
and finally subtract the M\"obius infinities.
It is possible to return to the unit disc picture by fixing the formal
on-shell SL$(2,\mathbb C)$ symmetry by the condition $a=1$, $w=0$,
$z_1=0$ and then making the inversion $z\to -1/z$.
The result will be the unit disc representation for the amplitudes in
the SL$(2,\mathbb C)$ M\"obius gauge $u_1=0$
(the Neumann function (7.43) then reduces to the standard expression
on the unit disc
$\mathcal G=-\frac{1}{4\pi}\log|u_1-u_2|^2|1-u_1\bar u_2|^2$).

As in the case of the Schottky-type parametrization of the torus, the
representation (7.41) makes it possible to clearly isolate the
``modular'' divergences from the ``local'' ones.
The tadpole modular infinity corresponds to the small hole limit
$a\to 0$.
In this limit (see (7.43), cf.\ (7.13))
\begin{equation}
\mathcal G
= \mathcal G_0
+ 4\pi a^2
\bigl[
\partial_w \mathcal G_0(w,z_1)\,
 \partial_{\bar w} \mathcal G_0(w,z_2)
+ \text{c.c.}
\bigr]
+ O(a^4).
\tag{7.44}
\end{equation}
and hence (cf.\ (7.15), (7.16))
\begin{equation}
\hat Z_{1/2}
= c_{1/2}\Omega^{-1}
\Big\langle
\int_ \vepsilon^a \frac{da}{a^3}
\Big(
V_{t} + a^2 \O'_{1/2} + O(a^4)
\Big)
e^{-I_{\rm int}}
\Big\rangle_0,
\tag{7.45}
\end{equation}
\begin{equation}
\hat Z_{1/2\,\infty}
= -c_{1/2}\Omega^{-1}\log \vepsilon\,
\langle \O'_{1/2}\, e^{-I_{\rm int}} \rangle , 
\tag{7.46}
\end{equation}
\begin{equation}
\O_{1/2} \equiv V_g
= \frac{1}{4\pi\alpha'}
\int d^2 w :\partial_\alpha x^\mu \partial^\alpha x_\mu: \, .
\tag{7.47}
\end{equation}
The operator $\O'_{1/2}$ is different, however, from the correct
counterterm operator $\O_{1/2}$ (see (3.7)) needed in order to
renormalize the ``modular'' divergence in the amplitudes on the disc
in a way consistent with the effective action.

This problem was pointed out by Fischler, K\"lebanov and Susskind
(FKS)  [13]. 
As we have mentioned above (see also Ref.~19) this problem is absent
in the unit disc parametrization if one carefully accounts for all
local infinities present in the corresponding partition function, or
equivalently, in the correlators of the vertex operators.
This is necessary in order to reproduce the expression (7.40), which
corresponds to the correct ``modular'' renormalization on the disc.

Let us now discuss how the FKS paradox is resolved in the plane with a
hole parametrization.
The basic point is that to define off-shell or divergent quantities
like $\hat Z$ one should use a {\it compact  curved} 2-space representation
for the world surface.
This is necessary in order to account for the topology of the world
surface in a systematic way and, in particular, in order to have a
correspondence with the sigma-model approach in which the topology is
reflected in the dilaton factor $e^{-\chi\phi}$.


\def \td {\tilde}

The problem is then to find the measure $d\mu_{1/2}$ (7.41) in the
case of a curved compact sphere with a hole.
The approach suggested by Polchinski [15]  is to consider a general curved
surface with a hole with metric
\begin{equation}
g_{ab} = e^{2\rho}\delta_{ab},\no 
\end{equation}
and to compute the measure by expanding in $a\to 0$.
He pointed out that the 2-curvature dependent terms may appear in the
measure due to its frame dependence.
In particular, the original parameters $a$ and $w$ in (7.49) are frame
dependent objects.
In the case of a nontrivial metric one should use instead the invariant
radius and center $\tilde a$ and $\tilde w$, defined as follows
(assuming that the coordinate radius $a$ is small)
\begin{equation}
\tilde a = e^{\rho(\tilde w)} a,
\qquad
\tilde w =
\frac{\int_{|u|\le a} d^2u\, \sqrt{g(w+u)}\,(w+u)}
     {\int_{|u|\le a} d^2u\, \sqrt{g(w+u)}}.
\tag{7.48}
\end{equation}
Then the length of the boundary circle in the coordinate plane is equal
to the length of its image and $\tilde w$ is an average position of the
center (which lies inside the disc which is cut out of the coordinate
plane).
Expanding in $a\to 0$ one finds
\begin{equation}\no 
\tilde w = w + \frac{1}{\pi a^2}
\int_{|u|\le a} d^2u\, 2|u|^2\, \bar \partial\rho(w)
+ \ldots
= w + a^2 \bar\partial\rho + \ldots,
\end{equation}
\begin{equation}
w = \tilde w - \tilde a^2 e^{2\rho}\bar\partial\rho(\tilde w) + \ldots .
\tag{7.49}
\end{equation}
Hence
\begin{equation}\no 
d^2 w = d^2 \tilde w - 2 \td a^2 \partial\bar\partial\rho\, d^2\tilde w
+ \ldots
= d^2\tilde w \Big(1 + \frac{1}{4}R^{(2)} \td a^2 + \ldots \Big),
\end{equation}
and thus finally [15]\footnote{A more complicated definition of the invariant radius
$\tilde a = \frac{1}{2\pi}\int d\theta\int^a_0 du 
\exp\bigl(\rho(w+u e^{i\theta})\bigr)$
suggested in Ref.~63 leads to the same result (7.50) since the
$a^3$ correction in $\tilde a$ does not matter to the leading order:
if $a=\tilde a + c \tilde a^3 + \ldots$ then
$\frac{da}{a^3} = \frac{d\tilde a}{\tilde a^3}(1+O(\tilde a^4))$.}
\begin{equation}
d\mu_{1/2}
= \frac{d\tilde a}{\tilde a^3}\, d^2\tilde w \sqrt{g(\tilde w)}
\Big(1 + \frac{1}{4}\tilde a^2 R^{(2)}(\tilde w) + \ldots \Big).
\tag{7.50}
\end{equation}
To put the measure into the covariant form one is to use that
$\rho(w)=\rho(\tilde w) - a^2 |\partial\rho|^2 + \ldots$ [62]. 

Repeating the analysis of the tadpole $a\to 0$ divergence in
$\hat Z_{1/2}$ with the corrected measure (7.50) one obtains (7.46)
with the operator $\O'_{1/2}$ replaced by 
$\O_{1/2}$, 
\begin{equation}
\hat Z_{1/2,\infty}
= - c_{1/2}\,\Omega^{-1}\,\log \vepsilon\,
\langle \O_{1/2}\, e^{-I_{\rm int}}\rangle_0 \ , 
\tag{7.51}
\end{equation}
\begin{equation}
\O_{1/2}
= \frac{1}{4\pi\alpha'}
\int d^2 w \sqrt{g}\,
\Bigl(:\partial_\alpha x^\mu \partial^\alpha x_\mu:
+ \tfrac{1}{2}\alpha' R^{(2)}\Bigr),
\tag{7.52}
\end{equation}
which is the correct ``modular'' counterterm for the disc ($\chi=1$)
(cf.\ (3.7)). Being inserted into the correlators on the sphere this
operator is equivalent to $\alpha'\,\partial/\partial\alpha'+1 $ [13].
Thus the  account of  the additional curvature dependent $a^2$ term in the modular measure
 resolves the FKS paradox [15]. 

We would like to stress that the analysis of the renormalization of the
leading order ``modular'' infinity in $\hat Z_{1/2}$ is independent of
how one subtracts the $SL(2,\mathbb{C})$ M\"obius infinities (since the
$\Omega^{-1}$ appears as a common factor in the tree and loop
contributions to $\hat Z$). However, as was already noted above, one is
to use a particular prescription for subtraction of the M\"obius
infinities in order to ensure the renormalizability of $\hat Z$ with
respect to all infinities. 

We suggest again to employ the
$\partial/\partial \log \vepsilon$ prescription (6.20). Using the corrected
measure (7.50) (and omitting the waves on $a$ and $w$) we get
\begin{align}
\hat Z_{1/2}
= &
c_{1/2}\,\frac{\partial}{\partial \log \vepsilon}
\Big[
\int_{\vepsilon} \frac{da}{a^{3}}
\Big(
\int d^{2}w \sqrt{g}
+
\frac{1}{4} a^{2}
\int d^{2}w \sqrt{g}\, R^{(2)} + \cdots
\Big) \no\\
&\qquad  \qquad \qquad  \times
\int d^{D}y \sqrt{G}\, e^{-\phi}
\Big(
1 + \tfrac{1}{2}\alpha' R \log \vepsilon + \cdots
\Big)
\Big]
\no\\
= &-2\pi c_{1/2}\,
\frac{\partial}{\partial \log \vepsilon}
\Big[
\log \vepsilon
\int d^{D}y \sqrt{G}\, e^{-\phi}
\Big(
1 + \tfrac{1}{2}\alpha' R \log \vepsilon + \cdots
\Big)
\Big]
+ \cdots
\no\\
= &-2\pi c_{1/2}
\int d^{D}y \sqrt{G}\, e^{-\phi}
\Big(
1 + \alpha' R \log \vepsilon + \cdots
\Big)
+ \cdots ,
\tag{7.53}
\end{align}
where we have dropped the quadratic divergence and used that the Euler
number for the sphere is $2$.\foot{We used that the leading order terms in $Z_{1/2}$ do not depend on
moduli and that in the limit $a \to 0$ $w$ runs over the whole sphere.  }

 Thus the coefficient of
the $\log \vepsilon\, R$ term ``doubles'' in the same way as already
suggested in the case of the torus (cf.\ (7.35) and (7.36)) and hence is
consistent with the renormalizability of $\hat Z$ with respect to the
sum of the local and modular infinities. We see that in the case of the
disc the overall $\log \vepsilon$ factor which multiplies the partition
function $Z_{1/2}$ originates from the modular integral and hence may be
interpreted as a modular divergence.

To conclude,  we have found that the presence of the additional ``topological''
term in the modular measure gives the overall $\log \vepsilon$ factor in
the derivative-independent part of the integrated partition function
$\bar Z_{1/2}$ and hence leads to the correct final result for the
generating functional or vacuum amplitude,
$\hat Z_{1/2} = d_{1/2} \int d^{D}y \sqrt{G}\, e^{-\phi} + \cdots$.
In this way we reobtain the usual finite expressions for the $N$ point
soft graviton and dilaton
amplitudes (corresponding to $\sqrt{G}\, e^{-\phi}$) as well as the
divergent parts of the $N = 2$ (off shell) and $N = 3$ graviton
amplitudes (corresponding to $R \log \vepsilon$) on the disc.

\def \G  {{\cal G}}

\subsection*{4.}

 One could expect that a similar curvature dependent
$O(|\xi - \eta|^{2})$ correction term may be present in the modular
measure (7.8) for the torus. It will then produce an overall
$\log \vepsilon$ factor in the small handle limit and hence will generate
(after taking $\partial/\partial \log \vepsilon$) the finite derivative
independent term $\int d^{D}y \sqrt{G}$ in $\hat Z_{1}$. However, it is
possible to use a flat metric on the torus without worrying about
boundary terms and being in conflict with topology. Moreover, if present,
such additional term would modify the operator $\hat O_{1}$ appearing in
tadpole factorization (7.15)--(7.18). However, this operator is already
known to be consistent with the general picture of the renormalization
of the tadpole divergence (as we have seen, there is no FKS paradox on
the torus).

A reason why we have found a problem with reproducing the zero momentum
part of $\hat Z_{1}$ is that the Schottky-type representation for the
$N$-point amplitudes on the torus (7.7), (7.8) was formally derived from
the parallelogram representation only for $N \ge 3$.\foot{This problem is absent if we use the ``hyperelliptic'' or ``branch
point'' parametrization  [{47}] for the torus, integrating
first over the coordinates of the four branch points and then using
$\partial/\partial \log \vepsilon$ to subtract the M\"obius volume.
A drawback of this parametrization is that it cannot be directly applied
for genera higher than $2$.
}
Hence it probably needs a modification in order to be applicable to the
case of the vacuum and tadpole amplitudes. Let us consider, for example,
the computation of the expectation value for the soft graviton vertex
operator.

Let us consider, for example, the computation of the expectation value
for the soft graviton vertex operator
\begin{equation}
V_g^{\mu\nu}
= \frac{1}{4\pi\alpha'}
\int d^2 z \sqrt{g}\,
\partial_\alpha x^\mu \partial^\alpha x^\nu
= \frac{1}{\pi\alpha'}
\int d^2 z\, \partial x^\mu \bar\partial x^\nu ,
\tag{7.54}
\end{equation}
(symmetrization over $\mu,\nu$ is always implied).
Let us first compute $\langle V_g\rangle_n$ in general in any
parametrization in which the Green function has the form (6.7)--(6.9).
Let $::$ denote the normal ordering with respect to the propagator
$\mathcal G_0$ on $\mathbb C$.
Then (we assume here that $\langle 1\rangle_n=1$)
\[
\frac{1}{\pi \alpha'}\,
\langle : \partial x^{\mu}\, \bar{\partial} x^{\nu} : \rangle_{n}
=
\frac{1}{\pi \alpha'} \lim_{w \to z}
\langle : \partial_{z} x^{\mu}(z)\, \bar{\partial}_{w} x^{\nu}(w) : \rangle_{n}
\]
\[
= 2\,\delta^{\mu\nu} \lim_{w \to z}
\partial_{z}\, \bar{\partial}_{w}\, \widetilde{\mathcal{G}}(z,w)
= -\,\frac{1}{8\pi^{2}}\, \delta^{\mu\nu}
\lim_{w \to z}
\omega^{a}(z)\, t^{-1}_{ab}\, \bar{\omega}^{b}(\bar w)
= -\,\frac{1}{8\pi^{2}}\, \delta^{\mu\nu}\,
\omega^{a}(z)\, t^{-1}_{ab}\, \bar{\omega}^{b}(\bar z)
\tag{7.55}
\]
where we have used that
$\partial_{z}\bar{\partial}_{w}\bigl(\log \tilde E(z,w) + \log \tilde E(\bar z,\bar w)\bigr)=0$.
It can be proved in general that
\[
\int d^{2}z\, \omega^{a}(z)\, \bar{\omega}^{b}(\bar z)
=
\frac{1}{2}\sum_{c=1}^{n}
\Big(
\oint_{a_c}\!\omega^{a}\!\oint_{b_c}\!\bar{\omega}^{b}
-
\oint_{b_c}\!\omega^{a}\!\oint_{a_c}\!\bar{\omega}^{b}
\Big)
= 4\pi^{2}t_{ab} .
\tag{7.56}
\]
Hence\foot{We are assuming a regularization prescription in which
$\int \delta^{(2)}(z,z)=0$ (see the Appendix).
}
\[
\Big\langle
\frac{1}{\pi\alpha'} \int d^{2}z : \partial x^{\mu}\bar{\partial}x^{\nu} :
\Big\rangle_{n}
=
-\,\frac{1}{2}\,\delta^{\mu\nu}\, n ,
\tag{7.57}
\]
where the genus $n$ appeared from $t^{-1}_{ab}t_{ab}=n$.
Now we are to use that for a general curved surface there is a curvature
term present in $:\partial x \partial x:$ [65]
(see (3.8)), i.e.
\[
\frac{1}{4\pi\alpha'}
\int d^{2}z \sqrt{g}\, : \partial_{a}x^{\mu}\partial^{a}x^{\nu} :
=
V^{\mu\nu}_{g}
+
\frac{1}{4}\,\delta^{\mu\nu}\,\chi ,
\qquad
\chi = 2 - 2n .
\tag{7.58}
\]
Hence finally (cf.\ (3.12))
\[
\langle V^{\mu\nu}_{g} \rangle_{n}
=
-\,\frac{1}{2}\,\delta^{\mu\nu},
\qquad\qquad 
\frac{1}{4\pi\alpha'}
\Big\langle
\int d^{2}z \sqrt{g}\, \partial_{a}x^{\mu}\partial^{a}x_{\mu}
\Big\rangle_{n}
=
-\,\frac{1}{2}\, D .
\tag{7.59}
\]
As was discussed in Ref.~16 (see also Refs.~15, 20, 14) this universal
result is consistent with the covariant expression for the ``constant''
part of the string partition function\foot{Note that Eq.~(3.8) is true in dimensional regularization in which one
can ignore the contribution of the measure $\delta^{(2)}(0)=0$. In other
regularizations one is to account for the contribution of the measure in
order to reproduce the covariant result (7.60) [{16,22}].
In general, to derive (7.59) and (7.60) one is only to use that
$\Delta{\cal G}=\delta^{(2)}(z,z')-1/V$ and to set $\delta^{(2)}(z,z)=0$.
  } 
\[
Z_{n} \sim \langle e^{-I_{\text{int}}} \rangle_{n}
=
\int d^{D}y \big( 1 - \langle V^{\mu\nu}_{g} \rangle_{n} h_{\mu\nu}
+ \cdots \big)
\]
\[
=
\int d^{D}y \Big( 1 + \tfrac{1}{2} h^{\mu}{}_{\mu} + \cdots \Big)
=
\int d^{D}y \sqrt{G}\,(1+\cdots),
\qquad
G_{\mu\nu} = \delta_{\mu\nu} + h_{\mu\nu}.
\tag{7.60}
\]
In the case of torus $\chi = 0$ and hence the subtlety related to the
use of (3.8), (7.58) is irrelevant (we, of course, drop the quadratic
divergence $\sim \partial\bar{\partial}\G_{0}$). 

In the parallelogram
representation only the last term in (7.4) contributes to (7.55) and we
find again (cf.\ (7.19))
\begin{align} 
\frac{1}{\pi\alpha'}
\Big\langle
\int d^{2}u : \partial x^{\mu}\bar{\partial}x^{\nu} :
\Big\rangle_{1}
=&
-\,\frac{1}{4}\,\delta^{\mu\nu}
\lim_{u_{1}\to u_{2}}
\int d^{2}u\, \partial_{1}\bar \partial_{2}(u_{12}-\bar u_{12})^{2}\no \\
=&
-\,\frac{1}{2}\,\delta^{\mu\nu}
\int d^{2}u\, (\Im \tau)^{-1}
=
-\,\frac{1}{2}\,\delta^{\mu\nu}
\tag{7.61}
\end{align}
(we have used that the area of the parallelogram is $\Im\tau$).
In the case of the Schottky-type parametrization ${\mathcal G}$
(7.9) belongs to the general class (6.7) and hence we get (7.55) with
$
\omega = (\eta-\xi)(z-\eta)^{-1}(z-\xi)^{-1}$
{(see (7.10))}
\begin{equation}
\langle V_g^{\mu\nu}\rangle_1
= \frac{\delta^{\mu\nu}}{8\pi^2\,\Im\tau}
\int d^2 z \Big|
\frac{\eta-\xi}{(z-\eta)(z-\xi)}
\Big|^2 .
\tag{7.62}
\end{equation}
If we formally put $\xi=0$, $\eta=\infty$ and map back to the
parallelogram ($z=e^{2\pi i u}$) we get
\begin{equation}
\int d^2 z\, |z|^{-2}
= 4\pi^2 \int d^2 u
= 4\pi^2 \Im\tau ,\no 
\end{equation}
in agreement with (7.61).
In computing the integrals in (7.62) one should always be careful
about the region of integration which should correspond to the one
used in the parallelogram representation.

If instead we substitute (7.62) into the modular integral (7.7),
(7.8), we get
\begin{equation}
A^{(1)}_1
= c_1 \Omega^{-1}
\int [d\tau]
\int \frac{d^2\xi\, d^2\eta}{|\xi-\eta|^4}
\langle V_g^{\mu\nu}\rangle_1
=- \frac{\delta^{\mu\nu}}{8\pi^2}
c_1 \lambda_1 \Omega^{-1}
\int
\frac{d^2\xi\, d^2\eta\, d^2 z}
{|\xi-\eta|^2\, |\xi-z|^2\, |z-\eta|^2} ,
\tag{7.63}
\end{equation}
where $\lambda_1$ was defined in (7.17).
Thus we find agreement with the result in the parallelogram
representation if
\begin{equation}
\Omega
= \frac{1}{4\pi^2}
\int
\frac{d^2\xi\, d^2\eta\, d^2 z}
{|\xi-\eta|^2\, |\xi-z|^2\, |z-\eta|^2} .
\tag{7.64}
\end{equation}
Equation (7.64) looks like the standard expression for the M\"obius
group volume.
However, in contrast to the tree level case, here the integration
region is not the full $\mathbb C$ \foot{For example, $z$ belongs to the exterior of the two circles and hence
the limits $z \to \xi$, $z \to \eta$ are excluded from the integration
region (unless $\xi \to \eta$).
  } and, in fact;  depends on $k$
(hence the $\Omega^{-1}$ factor should be better placed under the
integral over $k$).

\def \k {{\kappa}}

This suggests that a possible modification of the ansatz for $\hat Z$
which should apply to the case of the vacuum amplitude may look like
\begin{equation}
\hat Z
= c_1 \Omega^{-1}
\int [d\tau]\,
\frac{d^2 \k\, d^2\xi\, d^2\eta}
{|\k-\xi|^2\, |\k-\eta|^2\, |\xi-\eta|^2}\,
\langle e^{-I_{\rm int}}\rangle_1 \, ,
\tag{7.65}
\end{equation}
where we have introduced one extra moduli parameter $\k$ which appears
only in the measure (which is formally projective invariant) but not
in the Green function.
Now the measure gives the $\log \vepsilon$ factor already for the vacuum
amplitude and hence (7.65) gives the finite expression for the
``constant'' part of $\hat Z$ after we differentiate over $\log \vepsilon$.


\section*{8.\ Genus-two examples of renormalization}

Below we are going to explain how the renormalization of the string
generating functional can be carried in the two-loop approximation.
We shall demonstrate the condition of renormalizability of $\hat Z$
(or, equivalently, the condition that the massless sector of the
string S-matrix can be reproduced from an effective action) fixes the
overall coefficients (or relative weights) of string loop corrections
to $\hat Z$.

Let us first consider the disc and annulus corrections to $\hat Z$ in
the open-closed bosonic  string theory.
Since the annulus factorizes on two discs we expect to find the
logarithmic divergence in its contribution to $\hat Z$ (which we shall
denote as $\hat Z'_1$)
\begin{equation}
\hat Z
= \hat Z_{1/2} + \hat Z'_1 + \cdots
= d_{1/2} \int d^D y \sqrt{G} e^{-\phi}
+ d_1 \int d^D y \sqrt{G} + \cdots ,
\tag{8.1}
\end{equation}
\begin{equation}
\hat d_1 = d_1 + d_1^{(1)} \log \vepsilon .
\tag{8.2}
\end{equation}
Here $\phi$ and $G$ are the bare values of the fields which we know
already from the study of the tadpole renormalization on the disc.
We shall consider only the renormalization of the zero momentum part
of $\hat Z$ and hence will ignore the ``local'' counterterms.
According to the general expressions (4.15)--(4.18), (4.26), (4.27)
\begin{equation}
G_{\mu\nu}
= G_{R\mu\nu}
+ \frac{1}{4} d_0^{-1} e^{2\phi_R}
(2\omega + \omega') \log \vepsilon\,
G_{R\mu\nu}
+ \cdots ,
\tag{8.3}
\end{equation}
\begin{equation}
\phi
= \phi_R
+ \frac{1}{16} d_0^{-1} e^{2\phi_R}
(2D\omega + (D-2)\omega') \log \vepsilon
+ \cdots ,\no 
\end{equation}
where for the disc $\omega = d_{1/2} e^{-\phi_R}$.
Substituting (8.3) into the disc contribution to (8.1) we find that
the divergence in the annulus contribution cancels out if
\begin{equation}
d'^{(1)}_1 = -\frac{1}{16}\, d_{1/2}^2\, d_0^{-1}\, (D-2).
\tag{8.4}
\end{equation}
If we start directly with the well-known expression for the string
partition function on the annulus we get
\begin{equation}
\hat d'_1
= c'_1 \int_ \vepsilon^1 \frac{dq}{q^3}
\Big[\prod_{m=1}^{\infty}(1-q^{2m})\Big]^{-(D-2)}
= c'_1 \Big[ \frac{1}{2 \vepsilon^2} - (D-2)\log \vepsilon
+ \text{finite} \Big].
\tag{8.5}
\end{equation}
Dropping the quadratic divergence and comparing with (8.2) and (8.4)
we find agreement if the overall constant $c'_1$ takes the following
value
\begin{equation}
c'_1 = \frac{1}{16}\, d_{1/2}^2\, d_0^{-1}.
\tag{8.6}
\end{equation}
(note that $d_0$, $d_{1/2}$, $d'_1$ and $c'_1$ are proportional to
$\alpha'^{-D/2}$). Equivalently, one may say that it is the
factorization condition that fixes $c'_1$.

Let us now repeat the same analysis in the case of the torus and genus
2 contributions to $\hat Z$.
Since the genus 2 surface may factorize on the two tori we should find
\begin{equation}
\hat Z
= \hat Z_1 + \hat Z_2 + \cdots
= d_1 \int d^D y \sqrt{G}
+ \hat d_2 \int d^D y \sqrt{G} e^{2\phi}
+ \cdots ,
\tag{8.7}
\end{equation}
\begin{equation}
\hat d_2 = d_2 + d_2^{(1)} \log \vepsilon .
\tag{8.8}
\end{equation}
The renormalizability of $\hat Z$ implies that it should be possible to
cancel the divergence in $\hat Z_{2}$ by substituting the expression for
the bare metric (8.3) in the torus term in (8.7). Using that for the
torus $\omega = d_{1}$, we find the following condition on
$d^{(1)}_{2}$
\begin{equation}
d_2^{(1)} = -\frac{1}{4}\, d_1^2\, d_0^{-1}\, D.
\tag{8.9}
\end{equation}
Let us now check this prediction by directly computing the divergent
part of the genus $2$ partition function using the Schottky
parametrization. According to the general expression for the modular
measure in SP (6.12) we have
\begin{equation}
\hat d_2
= c_2 \Omega^{-1}
\int d\mu_2
= c_2 \Omega^{-1}
\int
\frac{d^2\xi_1 d^2\xi_2 d^2\eta_1 d^2\eta_2}
{|\xi_1-\eta_1|^4 |\xi_2-\eta_2|^4}
\frac{d^2 k_1 d^2 k_2}{|k_1 k_2|^2}
\big|
{(1-k_1)(1-k_2)}
\big|^2
(\det t_{ab})^{-D/2}\no 
\end{equation}
\begin{equation}
\times
{\prod_{\alpha}}'\prod_{m=1}^{\infty}
|1-k_\alpha^m|^{-2D}
{\prod_{\alpha}}'\prod_{m=2}^{\infty}
|1-k_\alpha^m|^4.
\tag{8.10}
\end{equation}
This expression for the genus $2$ string partition function was studied
in Ref.~66 where the equivalence between the loop measure in SP and in
the parametrization in terms of the period matrix [{67}]
(in which the modular invariance is explicit) was checked by expanding
the SP measure in powers of the modular parameters.\foot{Let us note that while the formal agreement of the SP measure with the
general Belavin--Knizhnik expression{25}
$d\mu_{n} = |d^{3n-3}y\, f(y)|^{2}\, ({\rm det}\, t)^{-D/2}$
(where $f$ is the holomorphic $(3n-3,0)$ form without zeroes or poles
in the interior of the moduli space but with double poles at the parts
of the boundary corresponding to surface degenerations) is obvious, the
modular invariance of it remains to be proved.}
If we formally fix the M\"obius group by choosing
$\xi_{1}=\infty$, $\eta_{1}=1$, $\eta_{2}=0$, $\xi_{2}\equiv \xi$,
(8.10) reduces to\foot{It  is possible to fix the M\"obius group by a gauge
here instead of regularizing M\"obius infinities since we are going to
discuss only the true ``modular'' divergences.}
\begin{equation}
\hat d_2
= c_2
\int
\frac{d^2 k_1 d^2 k_2 d^2\xi }
{|k_1|^4 |k_2|^4 |\xi|^4}\,
f(k_1,k_2,\xi)^2\,
(\det t)^{-D/2}.
\tag{8.11}
\end{equation}
The ``dividing'' factorization limit corresponds to $\xi \to 0$.
To find the expansions for $f$ and $t=\Im\tau$ in powers of $\xi$
we apply the results of Ref.~66.
The generators of the Schottky group are $T_1$ and $T_2$:
\begin{equation}
\frac{T_1 z - 1}{T_1 z - \infty}
= k_1 \frac{z-1}{z-\infty},
\qquad
T_1 z = k_1 z + 1 - k_1 ,
\tag{8.12}
\end{equation}
\begin{equation}
\frac{T_2 z - 0}{T_2 z - \xi}
= k_2 \frac{z-0}{z-\xi},
\qquad
T_2 z = \frac{z k_2 \xi}{z(1-k_2)-\xi}.
\tag{8.13}
\end{equation}
Analyzing the $\xi$-dependence of the multipliers of the elements of SG
which are products of $N$ factors
$T_{\alpha_N} = T^{\,n_1}_{a_1}\cdots T^{\,n_N}_{a_N}$,
$T_a = \{T_1, T_2\}$ one finds that
$k_{\alpha_N} \sim \xi^{N}$, $N \ge 2$.
Since we are interested in the logarithmic divergence in (8.11) we are
to expand to the order $|\xi|^{2}$. We have
$T_{\alpha_1} = \{T_1, T_2\}$,
$T_{\alpha_2} = \{T_1^{\,n}T_2^{\,m},\ n,m = 1,2,\ldots\}$,
$k_{\alpha_1} = \{k_1, k_2\}$,
$k_{\alpha_2} = 2\xi^{2}k_1^{\,n}k_2^{\,m}/(1-k^n_1)^{2}(1-k^m_2)^{2}$,
etc. Therefore (see (6.5), (6.12); $t_{ab} = \Im \tau_{ab}$)
\begin{align}
&f = \prod_{m=1}^{\infty}
(1-k_1^m)(1-k_2^m)^{-2(D-2)}
\Big[1 + O(\xi^2,\bar\xi^2)\Big] \ , 
\tag{8.14}
\\
&
2\pi i \tau_{11} = \log k_1 + O(\xi^2),
\qquad
2\pi i \tau_{22} = \log k_2 + O(\xi^2),
\tag{8.15}
\qquad 
2\pi i \tau_{12} = \xi + O(\xi^2).
\end{align}
The nontrivial $O(|\xi|^2)$ correction factor thus comes only from the
expansion of $(\det t)^{-D/2}$ in (8.10),
\begin{align}
(\det t)^{-D/2}
&= \Big[ t_{11}t_{22} -\frac{1}{(2\pi)^2} \big({\rm Re}\, \xi\big)^2 \Big]^{-D/2}
\no\\ &
= (t_{11}t_{22})^{-D/2}
\Big[ 1 + \frac{1}{4} D (2\pi t_{11})^{-1}(2\pi t_{22})^{-1}
|\xi|^2 + O(\xi^2,\bar\xi^2) \Big], 
\tag{8.16}
\\
\hat d_2 &= c_2 \int
\frac{d^2 k_1 d^2 k_2 d^2\xi}
{|k_1|^4 |k_2|^4 |\xi|^4}
\prod_{m=1}^{\infty}
|(1-k_1^m)(1-k_2^m)|^{-2(D-2)}
\no \\
&\times
\Big(
\frac{\log|k_1|}{2\pi}\,
\frac{\log|k_2|}{2\pi}
\Big)^{-D/2}
\Big[
1 + \frac{1}{4} D
\big({\log|k_1|}\,
      {\log|k_2|}\big)^{-1}
|\xi|^2
+ O(\xi^2,\bar\xi^2)
\Big].
\tag{8.17}
\end{align}
Thus finally (cf.\ (8.5))
\begin{equation}
\hat d_2
= c_2 \int_ \vepsilon \frac{d^2\xi}{|\xi|^4}
\Big[
(2\pi)^2 t_1^2
+ \frac{1}{4} D (2\pi)^2 \lambda_1^3 |\xi|^2
+ O(\xi^2,\bar\xi^2)
\Big]
\no
\end{equation}
\[
= c_{2}\Big[
\frac{\pi}{2\vepsilon^{2}}(2\pi)^{4} t_{1}^{2}
-
\frac{1}{4} D (2\pi)^{3} \lambda_{1}^{2}\log \vepsilon
+ \text{finite}
\Big],
\tag{8.18}
\]
where the zero momentum tachyon and massless scalar tadpoles on the
torus $t_{1}$ and $\lambda_{1}$ were defined in (7.17)\foot{We are assuming, of course, that the integrals in (8.11) are restricted
to the fundamental region and hence the integrals over $k_{1}$ and
$k_{2}$ in the factorization limit appear restricted to the fundamental
region of the modular group of the torus.   } 
and hence (see (8.8))
\[
d^{(1)}_{2}
=
-\frac{1}{4}\, c_{2}\,(2\pi)^{3}\lambda_{1}^{2} D .
\tag{8.19}
\]
Comparing this with (8.9) we get
\[
(2\pi)^{3}\lambda_{1}^{2} c_{2}
=
d_{1}^{2} d_{0}^{-1} .
\tag{8.20}
\]
This finally determines $c_{2}$ if we use the relation between the
coefficients $d_{1}$ (the vacuum amplitude) and $\lambda_{1}$ (the
tadpole) for the torus (7.24) (in general $d_{1}=c_{1}\lambda_{1}$).
Let us emphasize that having fixed the overall coefficient in
$\hat Z_{2}$ we automatically determine the normalization of the
2-loop amplitudes generated by $\hat Z_{2}$.

The result that the logarithmically divergent part of $\hat d_{2}$ is
proportional to $D$ is consistent with the general factorization formula
(3.10) (note that $\chi_{1}=0$ and $\langle V_{g}\rangle_{1}\sim D$).
To explain why the massless propagator contributes effectively
$D^{-1}\log \vepsilon$ let us rederive the restriction (8.9) starting from
the assumption that $\hat Z$ can be reproduced by an effective field
theory (for similar analysis in the case of the annulus see Refs.~14
and 16).

According to the discussion in Sec.~5 (see 5.6) we should have
\begin{equation}
\hat Z_2 = s_2 =
\Big(
U_2 - \frac{1}{2}
\frac{\partial U_1}{\partial \vphi^i}
\Delta^{-1}_{ij}
\frac{\partial U_1}{\partial \vphi^j}
\Big)_{\vphi_{\rm in}}
+ \cdots ,
\tag{8.21}
\qquad \qquad 
S = \frac{1}{2}\vphi^i \Delta_{ij} \vphi^j
+ \sum_{n=0}^{\infty} U_n(\vphi).
\end{equation}
If we use the $\sigma$-model parametrization of the fields
$G_{\mu\nu} = \delta_{\mu\nu} + h_{\mu\nu}$ and $\phi$, i.e.\ use the
tree action (2.21), (2.22) then $\Delta^{-1}_{ij}$ is nondiagonal and is
expressed in terms of $\k_{0}^{ij}$ (see (2.29), (2.30) and (2.20)).
We need only the ``$GG$'' element of this matrix since in the
$\sigma$-model parametrization
$
U_{1} = d_{1}\int d^{D}y\,\sqrt{G} + \ldots .$
Hence we find
\[
(\partial U_{1}/\partial\varphi^{i})\,\Delta^{-1}_{ij}\,
(\partial U_{1}/\partial\varphi^{j})
\sim d_{1}^{2}\,\delta^{\mu\nu}(\delta_{\mu\alpha}\delta_{\nu\beta})
\,\delta^{\alpha\beta}\,\Delta^{-1}(0)
\sim d_{1}^{2} D \log\vepsilon,
\]
where we have used that $(\alpha'\Delta)^{-1}(0)\sim -\log\vepsilon$,
see (2.50). Equivalently, we can use the ``$S$ matrix
parametrization'' of the fields (2.23)--(2.27) in which the propagator
diagonalizes. The sphere plus torus contributions to the EA are then
given by (for $D=26$; we use the harmonic gauge for the graviton)
\begin{equation}
S=d_{0}\int d^{D}y\,\sqrt{G'}\Big[\alpha'\Big(R'-\frac{1}{4}(\partial\phi')^{2}\Big)+\cdots\Big]
+d_{1}\int d^{D}y\,\sqrt{G'}\,\exp\!\Big(\frac{D}{2\sqrt{D-2}}\,\phi'\Big)+\cdots
\tag{8.22}
\end{equation}
\begin{equation}
=\alpha' d_{0}\int d^{D}y\Big(-\frac{1}{4}\,h'\Delta_{h}h'
-\frac{1}{4}\,\phi'\Delta_{0}\phi' +\cdots\Big)
+d_{1}\int d^{D}y\Big(1+\frac{1}{2}\,h'^{\mu}{}_{\mu}
+\frac{1}{2}\,\frac{D}{\sqrt{D-2}}\,\phi' +\cdots\Big),\no 
\end{equation}
\begin{equation}
\Delta_{h\,\alpha\beta}^{-1\,\mu\nu}
=\Delta_{0}^{-1}\Big(\delta^{\mu}{}_{(\alpha}\delta^{\nu}{}_{\beta)}
-\frac{1}{D-2}\,\delta^{\mu\nu}\delta_{\alpha\beta}\Big),
\qquad
\Delta_{0}=-\Box \ . 
\tag{8.23}
\end{equation}
 Using (8.21) (or directly integrating over $h'$
and $\phi'$) we get for the $O(d_{1}^{2}\log \vepsilon)$-term in $s_{2}$\footnote{We again replace $(1/k^2)_{k\to 0}$ by $-\log \vepsilon$.
This normalization is consistent with the 2d short distance
regularization.}
\[
s_{2} = \hat{s}_{2}\int d^{2}y + \ldots ,
\qquad \ \ \ 
\hat{s}_{2}
= - d_{1}^{2} d_{0}^{-1} \log \vepsilon
\Big(
-\frac{1}{2}\,\frac{D}{D-2}
+ \frac{D^{2}}{4(D-2)}
\Big)
= -\frac{1}{4}\, d_{1}^{2} d_{0}^{-1} \log \vepsilon \, D .
\tag{8.24}
\]
This result is in agreement with (8.9). The two terms in the brackets are the contributions
of the graviton and the dilaton exchanges.

\section*{9.\ Approach based on operators of insertion of topological fixtures}

In this section we shall rederive some of the results of the previous
sections by using the representation of the loop part of the string
generating functional in terms of operators of insertion of holes,
handles, etc.
Similar approach was originally suggested in Ref.~18 and recently
discussed also in Ref.~32 (our formulation is closer to that of
Ref.~32).

\subsection*{1.}

Let
$
I_0 = \frac{1}{2} \int d^2 z \sqrt{g}\,
x \Delta x  \ ( \Delta = -\nabla^2)
$  be the free string action on an arbitrary curved 2-surface (for
notational simplicity we shall often set $2\pi\alpha'=1$ and not
indicate the index of $x^\mu$).
Consider the expectation value
\begin{equation}
\langle F[x]\rangle
= \int [dx] e^{-I_0} F[x],
\qquad
\langle 1\rangle = 1.
\tag{9.1}
\end{equation}
Using the functional Fourier transformation one can prove the validity of the following
Hori-type representation (see, e.g.\ Refs.~68 and~8)
\begin{equation}
\langle F[x]\rangle
=\Big\{
\exp\!\Big(\frac{1}{2}\,\mathcal G\cdot \frac{\delta^{2}}{\delta x^{2}}\Big)
F[x]
\Big\}_{x=0},
\qquad
\mathcal G=\Delta^{-1},
\tag{9.2}
\end{equation}
which we have already used above (see, e.g., (6.18) and (6.19)).
Suppose now that we split the Green function $\mathcal G$ into two parts
\begin{equation}
\mathcal G=\mathcal G_{0}+\tilde{\mathcal G}.
\tag{9.3}
\end{equation}
At this stage the split (9.3) may be arbitrary but actually we will be interested in the case
when the Riemann surface is represented in terms of $\mathbb C$ with topological fixtures (pairs of
holes in the Schottky parametrization). Then $\mathcal G_{0}$ will be the free propagator on the plane
and $\tilde{\mathcal G}$ will contain all dependence on moduli (see (6.9)).
The corresponding split in the Laplace operator is
\begin{equation}
\Delta=\Delta_{0}+\tilde{\Delta},
\qquad
\Delta_{0}=\mathcal G_{0}^{-1},
\tag{9.4}
\end{equation}
\begin{equation}
\tilde{\Delta}
=-(1+\Delta_{0}\cdot \tilde{\mathcal G})^{-1}\cdot \Delta_{0}\cdot \tilde{\mathcal G}\cdot \Delta_{0}
=-\Delta_{0}\cdot \tilde{\mathcal G}\cdot (1+\Delta_{0}\cdot \tilde{\mathcal G})^{-1}\cdot \Delta_{0}.\no 
\end{equation}
Substituting (9.4) into (9.1) and using (9.2) we get
\begin{equation}
\langle F[x]\rangle=\langle F[x]\,Q[x]\rangle_{0},
\tag{9.5}
\end{equation}
\begin{equation}
\langle \cdots \rangle_{0}
=\int [dx]_{0}\exp\!\Big(-\frac{1}{2}\int x\Delta_{0}x\Big)\cdots,
\qquad
\langle 1\rangle_{0}=1,
\tag{9.6}
\end{equation}
\begin{equation}
Q[x]=N\exp\!\Big(-\frac{1}{2}\int x\tilde{\Delta}x\Big),
\tag{9.7}
\end{equation}
\begin{equation}
N=\Big(\frac{\det\Delta}{\det\Delta_{0}}\Big)^{-D/2}
=\bigl[\det(1+\Delta_{0}\cdot \tilde{\mathcal G})\bigr]^{-D/2},
\tag{9.8}
\end{equation}
where $D$ is the number of fields $x^{\mu}$.

 The functional $Q[x]$ has rather complicated form because
of the nonlocal structure of $\tilde{\Delta}$ (9.4). It is possible to get a simpler representation for $Q$
using normal ordering with respect to $\mathcal G_{0}$. By definition of the normal ordering,
\begin{equation}
: F[x] :
=\exp\!\Big(-\frac{1}{2}\,\mathcal G_{0}\cdot \frac{\delta^{2}}{\delta x^{2}}\Big)F[x].
\tag{9.9}
\end{equation}
Equation (9.9) implies that $\langle :F:\rangle =\bigl(F[x]\bigr)_{x=0}$ and that there are no pairings of $x$'s inside
$F[x]$ when $F$ is inserted into a $\Delta_{0}$-correlator. Equation (9.9) and similar relations, e.g.,
\begin{equation}
: F_{1}[x]F_{2}[x] :\ 
= \ :F_{1}[x]:
\exp\!\Big(-\frac{{ \overleftarrow \delta}}{\delta x}\cdot \mathcal G_{0}\cdot \frac{{\overrightarrow\delta}}{\delta x}\Big)
:F_{2}[x]:
\tag{9.10}
\end{equation}
can be proved by using the functional Fourier transformation
$F[x]=\int[dp]\,e^{ip\cdot x}\,F[p]$,
$p\cdot x=\int d^{2}z\,p(z)x(z)$, and thus reducing the problem to that of the normal ordering of the exponents
\begin{equation}
: e^{ip\cdot x} :
= e^{\frac{1}{2}p\cdot \mathcal G_{0}\cdot p}\, e^{ip\cdot x},
\qquad
p \rightarrow -i\,\frac{\delta}{\delta x(z)},
\tag{9.11}
\end{equation}
\begin{equation}\no 
: e^{ip_{1}\cdot x}::e^{ip_{2}\cdot x}:
= e^{p_{1}\cdot \mathcal G_{0}\cdot p_{2}}\, :e^{ip_{1}\cdot x }: \ : e^{ip_{2}\cdot x}:
\end{equation}
Employing (9.9) and (9.11) it is straightforward to prove that if $A$ is some operator and $::$
is the normal ordering with respect to $\mathcal G_{0}$, then
\begin{equation}
e^{-\frac{1}{2}xAx}=C:\,e^{-\frac{1}{2}xBx}: \ , 
\tag{9.12}
\end{equation}
\begin{equation}
B^{-1}=A^{-1}+\Delta_{0}^{-1},
\qquad
B=(1+A\cdot \mathcal G_{0})^{-1}A,
\qquad
\Delta_{0}\cdot \mathcal G_{0}=1,
\no \qquad 
C=\bigl[\det(1-\mathcal G_{0}\cdot B)\bigr]^{D/2}.
\tag{9.13}
\end{equation}
Applying (9.12) and (9.13) to the case of the functional $Q$ (9.7) we find
\begin{equation}
Q[x]=:\exp h[x]: \ ,
\tag{9.14}
\end{equation}
\begin{equation}
h[x]=-\frac{1}{2}\int x\td{\Delta}x,
\qquad
\td{\Delta}=-\Delta_{0}\cdot \tilde{\mathcal G}\cdot \Delta_{0}.
\tag{9.15}
\end{equation}
Here $A=\tilde{\Delta}$ (see (9.4)) so that $B=-\Delta_{0}\cdot \tilde{\mathcal G}\cdot \Delta_{0}=\tilde{\Delta}$
and the factor $N$ in (9.7) cancels against $C$ in (9.12).
The explicit form of the bilinear functional $h$ (9.15) is
\begin{equation}
h[x]=\frac{1}{2}\int d^{2}z_{1}d^{2}z_{2}\,
(\Delta_{0}x)(z_{1})\,\tilde{\mathcal G}(z_{1},z_{2})\,(\Delta_{0}x)(z_{2}),
\tag{9.16}
\end{equation}
where we have already specified $\mathcal G_{0}$ in (9.3) to be the propagator on the plane.

The basic relation (9.5) implies that a Gaussian correlator with respect to a complicated Green function can be
represented as a  correlator with respect to the trivial Green function with the additional insertion of the
functional $Q=:e^{h}:$ (9.14) with $h$ given by (9.16) where $\tilde{\mathcal G}$ is the nontrivial part of the
original Green function.

If we apply (9.5) to the case of the string generating functional (4.1) we get ($F=e^{-I_{\rm int}}$)
\begin{align}
&\hat Z=\Omega^{-1}\sum_{n=0}^{\infty}\bar Z_{n},
\qquad
\bar Z_{n}=c_{n}\int d\mu_{n}\,Z_{n},
\tag{9.17}
\\  &
Z_n = \langle e^{-I_{\rm int}}\rangle_n
= \langle :e^{h_n}: e^{-I_{\rm int}} \rangle_0, \qquad \ \    \langle 1 \rangle_n=1 \ , \ \ \  n=0,1,2,....\ , 
\tag{9.18}\\
&
h_n = \frac{1}{2}\int \Delta_0 x \cdot \tilde{\mathcal G}_n \cdot \Delta_0 x, \qquad \qquad \td  \G_0=0 \ .
\tag{9.19}
\end{align}
The subscript $n$ indicates that $\td \G_n= \G_n - \G_0$   corresponds to a  genus $n$ surface.

The representation (9.17)--(9.19) may be a bit misleading: it may look
as if it is possible to integrate first over the moduli and then to
compute the expectation value on the plane. However, this is not possible
to do in general since the $\sigma$-model action $I_0 + I_{\text{int}}$
and the expectation values depend implicitly on moduli through the
integration region for the 2-d coordinates (for $n \ge 1$ the
$\sigma$-model is, in fact, defined not on a sphere but on a higher
genus surface). Still (9.17)--(9.19) may be useful in the case when we
study particular regions of the moduli spaces corresponding to surface
degenerations (or when we consider the vacuum partition function,
putting $I_{\text{int}} = 0$).

Let us now demonstrate how the representation (9.18) works on the
examples of the disc and the torus. Expanding the nontrivial part
$\tilde{\mathcal G}$ of the Neumann function on the disc in the
``plane with a hole'' representation (7.43) in powers of radius of the
hole, using
$\bar\partial (z-w)^{-1} = \pi \delta^{(2)}(z-w)$ in order to reduce
$\bar\partial (z-w)^{-n}$ to $\partial^{\,n-1}\delta^{(2)}(z-w)$ and
integrating by parts to get rid of the $\delta$-functions, one finds
the following expression for the functional $h$ (9.19) (see also
Refs.~18 and 32;  here  we set $2\pi\alpha' = 1$)
\[
h_{1/2}[x]
=
4\pi \sum_{m=1}^{\infty}
\frac{a^{2m}}{m!(m-1)!}
\Big(\partial^{m} x\, \bar\partial^{\,m} x\Big)(w) .
\tag{9.20}
\]
If we now consider the limit $a \to 0$
($:e^{h_{1/2}}: \ = 1 + 4\pi a^{2}\, :\partial x \bar\partial x: + \cdots$)
and substitute (9.20) together with the correct modular measure given
by (7.50) into (9.17), (9.18) we easily reproduce the consistent result
(7.51) and (7.52) for the logarithmically divergent part of the disc
contribution to the generating functional $\hat Z_{1/2}$. Note that the
normal ordering in (9.18) leads to the normal ordering in the operator
$\O_{1/2}$ in (7.52).

Using the Schottky-type (plane with 2 holes) representation for the
torus, substituting the nontrivial part $\tilde{\mathcal G}$ of the
corresponding Green function (7.11) into (9.16) and going through the
same steps as in the derivation of (9.20) we get the following expression
for the corresponding functional $h_1$ (9.19)  [{18}]\foot{It is useful to note that
$
\lambda
=
\frac{(z_1 - \xi)(z_2 - \eta)}
     {(z_1 - \eta)(z_2 - \xi)}
=
(1 - \frac{\xi - \eta}{z_1 - \eta})
(1 + \frac{\xi - \eta}{z_2 - \xi}).
$
   }
\[
h_1 \equiv h_1(\xi,\eta,k,[x])
=
4\pi \Big\{
\frac{1}{4 \log |k|}
\,\big[x(\xi) - x(\eta)\big]^{2}
-
\sum_{m=1}^{\infty}
\frac{1}{m!(m-1)!}\,
\frac{k^{m}}{1 - k^{m}}  \]
\[
\qquad \qquad \times \Big[
\partial_{\xi}^{\,m-1}\!\big((\xi-\eta)^{m}\,\partial_{\xi}x\big)
\Big](\xi)\,
\Big[
\partial_{\eta}^{\,m-1}\!\big((\xi-\eta)^{m}\,\partial_{\eta}x\big)
\Big](\eta)
+ \text{c.c.}
\Big\}\ . 
\tag{9.21}
\]
The first term here originates from the last nonholomorphic term in
(7.11). Like the second term in (9.21) it can be represented (by using
Taylor expansion) as an infinite series in powers of $\xi-\eta$ and
derivatives of $x$ (cf.\ (9.13))
\begin{align} 
&h_1
=
4\pi \Big[
\frac{1}{2\log|k|}\,|\xi-\eta|^{2}\,
\partial x\,\bar\partial x(\xi)
+ O\!\Big((\xi-\eta)^{2},(\bar\xi-\bar\eta)^{2}\Big)
\Big],\no 
\\
&:\!e^{h_1}\!:
=
1 + \frac{2\pi}{\log|k|}\,|\xi-\eta|^{2}
:\!\partial x\,\bar\partial x\!:(\xi)
+ \cdots .
\tag{9.22}
\end{align}
Considering the tadpole factorization limit $\xi \to \eta$ and
integrating over $\xi$ and $\eta$ (see (7.8), (7.15)) we reproduce the
result (7.20) for the divergent part of the generating functional
$\hat Z_1$.

It is possible also to derive the general expression for $h_n$ and hence
for the operator of insertion of a topological fixture $Q_n$ (9.14) for  an 
arbitrary genus $n$ representing $\tilde E$ and $\omega$ in the Green
function (6.7)--(6.9) in terms of the Laurent series in $z$ and $w$ [{18}].
As we discussed in the previous sections, to have the correct
factorization  [{15}]  and hence renormalization properties of
$\hat Z$ it is necessary also to go from the flat plane with holes to a
curved 2-space. Then the curvature dependent correction terms appear in
the normal ordering relations and in the modular measure.

\subsection*{2.}

 In order for the renormalization group to operate at string loop
level the tadpole divergences found above should, in fact,
\emph{exponentiate} to become counterterms for the string action
(see Secs.~3 and 4, Eq.~(3.20))
\[
\hat Z_n
=
c_n \Omega^{-1} \int d\mu_n
\Big\langle e^{-I - \delta I} \Big\rangle_n ,
\qquad\qquad 
\delta I
=
\sum_{n=1}^{\infty} b_n \O_n \log \vepsilon + \cdots .
\]
Higher powers of $\log \vepsilon$ appearing from the expansion of
$e^{-\delta I}$ in the string coupling should cancel higher order
tadpole infinities in higher genus contributions to $\hat Z$
(we are, of course, to combine [{77}] the powers of the
$\log \vepsilon$ term in $\delta I$ with (powers of) other
$\log^{m}\vepsilon$ terms in $\delta I$ which must be present for
consistency of the RG, see Eqs.~(3.20)--(3.23) and footnote $12$).

For this to happen the relative coefficients of the loop contributions
to $\hat Z$ should take particular values. Let us assume that the
exponentiation does indeed take place; for example, the tadpole
divergence on the torus exponentiates. Consider the regions of the
moduli space (for each genus $n$) corresponding to surfaces with small
and far separated handles. It is then natural to expect that the
exponentiation of the 1-loop tadpole divergence implies that the
contribution to $\hat Z$ coming from the specified regions of moduli
spaces can be effectively represented as
\[
\hat Z
=
c_0 g^{-2}\, \Omega^{-1}
\Big\langle e^{-I_{\text{int}} + g^{2} H_{1}} \Big\rangle_{0}
+ \cdots ,
\tag{9.23}
\]
where (see also Refs.~18 and 32)
\[
H_1[x]
=
p_1 \int d\mu_{1} : e^{h_1} :
=
p_1 \int [d\tau]\,
\frac{d^{2}\xi\, d^{2}\eta}{|\xi-\eta|^{4}}
: e^{h_1(\xi,\eta,k,[x])} : 
\tag{9.24}
\]
\[
\sim
\frac{1}{2}\, p_1 \xi_1 \pi \log \vepsilon\, \O_{1}[x]
+ \cdots ,
\qquad
p_1 = c_1 c_0^{-1} .
\tag{9.25}
\]
(cf.\ (7.15), (7.20) and (7.38)). Equation (9.23) corresponds to the
``renormalization group improved'' perturbation theory or gives a
``dilute handle gas'' approximation for $\hat Z$.\foot{Let us emphasize that the possibility to ``resum'' $\hat Z$ in the form
(9.23) depends crucially on the use of the extended parametrization of
moduli with the $\Omega^{-1}$ appearing as a universal factor for all
contributions to $\hat Z$.
   }
The representation (9.23) may be true only for small handles since it
is only in this case that we can ignore the dependence of $I_{\text{int}}$
on the moduli through the integration region (for small handles we may
approximately consider the $\sigma$-model as being defined on a sphere).

In Eq.~(9.23) we have explicitly indicated the dependence on the
(bare) string coupling constant $g$. Since $g$ is related to the constant
part of the dilaton it should renormalize together with $\phi$
(see Secs.~3 and 4). This observation suggests that there should be
higher order terms in (9.23) which should generate the renormalization
of $g = e^{\phi_c}$ in the exponent. A natural guess is
\[
\hat Z
=
\Omega^{-1}
\sum_{n=0}^{\infty}
g^{2(n-1)} \tilde c_n
\int d\tilde\mu_n
\Big\langle e^{-I_{\text{int}} + g^{2} H_1} \Big\rangle_{n}
+ \cdots ,
\tag{9.26}
\]
where the wave in $d\tilde\mu_n$ indicates that the corresponding
integration regions over moduli do not contain configurations in which
{\it all}   handles are small and far separated (it is assumed that we have
already summed the contributions of such configurations). The
consistency of the RG demands that we should be able to resum (9.26)
further, generating higher order terms in the exponent (which produce
higher order tadpole renormalizations)
\begin{align}
&e^{-I_{\text{int}}}
\;\longrightarrow\;
e^{-I_{\text{int}} + {\mathcal F}},
\qquad
{\mathcal F}
=
g^{2}\hat H_{1}
+ g^{4}\hat H_{2}
+ g^{6}\hat H_{3}
+ \cdots ,
\qquad
\hat H_{1} = H_{1} ,
\tag{9.27}
\\  &
\hat H_{n}
=
p_n \int d\hat \mu_{n} : e^{h_n} : \ ,
\qquad
p_n = c_n c_0^{-1} .
\tag{9.28}
\end{align}
The hat in the modular measure indicates that the measure and the
integration region are chosen in a way that avoids overcounting
equivalent configurations after we expand (9.27) in powers of $g$ and
compare with the usual expansion in genus. $\hat H_n$ contain the lowest
order $O(\log \vepsilon)$ singularities corresponding to the factorization
on a finite part of a genus $n$ tadpole and a correlator on the sphere.
In this way we may reproduce the expression (3.21) for the tadpole
counterterm.

As we already noted, a representation based on (9.23), (9.27) can be
valid only approximately because of the dependence of $I_{\text{int}}$
on moduli through the integration region. However, $I_{\text{int}}$ is
absent in the case of the vacuum partition function. In this case we may
expect that $\hat Z$ can be  represented  exactly as a correlator on the
sphere
\[
\hat Z(0)
=
c_0 g^{-2}\, \Omega^{-1}
\big\langle e^{{\mathcal F}} \big\rangle_{0} 
\tag{9.29}
\]
where $\mathcal F$ is given by (9.27) and $\Omega^{-1}$ should be
defined using some regularization prescription (e.g.\
$\Omega^{-1} \rightarrow \partial/\partial \log \vepsilon$).
$\mathcal F$ can be considered as a nonlocal (depending on all powers of
derivatives of $x^\mu$ integrated over moduli) addition to the free
string action. Only the leading $O(\log \vepsilon)$ infinite part of
$\mathcal F$ should be local, since it should coincide with the tadpole
counterterm (3.21)
\[
\mathcal F_\infty
=
- \sum_{n=1} b_n \hat O_n \log \vepsilon
+ O(\log^{2}\vepsilon).
\]
Equation (9.29) represents a resummation of the usual perturbation
theory\foot{There are two possible strategies of how to try to ``resum'' the usual
perturbative expansion for the string partition function. The first is
to represent $\hat Z$ as a path integral for a simple free string action
but over a complicated ``universal moduli space'' [{69}]
which includes all moduli spaces of finite genera as subspaces. The
second is to represent $\hat Z$ as a path integral simply over a sphere
but for a complicated 2-d action. According to some previous
suggestions [{70}] this action may depend nontrivially on
some additional to $x^\mu$ variables (ghost fields). According to our
ansatz (9.29), this action may depend only on $x^\mu$ but in a
complicated nonlocal way.
  }.

\subsection*{3.}

 To illustrate how the functional $\mathcal F$ in (9.27) and (9.29)
is constructed let us consider the $O(g^{4})$ term in $\mathcal F$
(9.27). Expanding $e^{\mathcal F}$ in powers of $g$ we should reproduce
the ordinary perturbation expansion in genus $n$. Namely, we should have
\[
\hat Z_{2}(0)
=
c_{0}\Omega^{-1}
\Big\langle
\hat H_{2}
+ \tfrac{1}{2}(\hat H_{1})^{2}
\Big\rangle_{0},
\tag{9.30}
\]
where $\hat Z_{2}(0)$ is the usual genus 2 contribution to the string
partition function, given, e.g., in the Schottky parametrization, by
Eq.~(8.10). Thus we get the following equation for the modular measure
in $\hat H_{2}$ (9.28)\foot{Note once again the crucial role of the use of a parametrization in
which the M\"obius volume factorizes.  }
\[
\int d\hat\mu_{2}
=
\int d\mu_{2}
-
\frac{1}{2}\, q_{2}
\int d\mu_{1}^{(1)}
\int d\mu_{1}^{(2)}
\Big\langle
: e^{h_{1}^{(1)}} :\ 
: e^{h_{1}^{(2)}} :
\Big\rangle_{0},
\qquad
q_{2} = p_{1}^{2} c_{2}^{-1},
\tag{9.31}
\]
where we have noted that
$\langle : e^{h_{2}} : \rangle_{0} = 1$ and used the superscripts
$(1,2)$ to denote the two different sets of genus one moduli.
$d\mu_{1}$ is given by (9.24), (7.8) and $h_{1}$ by (9.21).

The expectation value of the product of the two operators
$Q = : e^{h} :$ in  (9.14) can be easily computed in general using
(9.7), (9.12), (9.4)
\begin{align}
&\langle Q^{(1)}[x] Q^{(2)}[x] \rangle_{0}
= 
N^{(1)} N^{(2)}
\Big\langle
e^{- \frac{1}{2} x (\tilde\Delta^{(1)} + \tilde\Delta^{(2)}) x}
\Big\rangle_{0}\no 
\\
&=
\exp \Big\{
- \tfrac{1}{2}D\,
\operatorname{tr}
\log \!\bigl[(1 + \Delta_{0}\tilde{\mathcal G}^{(1)})
(1 + \Delta_{0}\tilde{\mathcal G}^{(2)})\bigr]
-
\tfrac{1}{2}D\,
\operatorname{tr}
\log \!\bigl[1 + (\tilde\Delta^{(1)} + \tilde\Delta^{(2)}) \cdot \mathcal G_{0}\bigr]
\Big\}
\no\\
&=
\exp \Big[
- \tfrac{1}{2}D\,
\operatorname{tr}
\log \!\bigl(
1 - \Delta_{0}\tilde{\mathcal G}^{(1)} \cdot
\Delta_{0}\tilde{\mathcal G}^{(2)}
\bigr)
\Big]
\equiv \mathcal M_{2}.
\tag{9.32}
\end{align}
Substituting  the expression for $\tilde{\mathcal G}_1$ (7.9) or (7.11)
we determine $\mathcal M_2(\xi^{(1)},\eta^{(1)},k^{(1)};
\xi^{(2)},\eta^{(2)},k^{(2)})$.
Note that $\mathcal M_2$ depends nontrivially on both sets of moduli
(the expectation value couples them).

The expression in the exponent can be simplified by using (once) the relation
$\Delta_0\tilde{\mathcal G} = -\frac{1}{4\pi^2}\omega_a(z)t^{-1}_{ab}
\bar\omega_b(\bar z)$ (see (6.8)).
This relation can be used twice in the first term of the expansion of
$\mathcal M_2$ in powers of $\tilde{\mathcal G}$ (cf.\ (7.62))
\begin{align}
\mathcal M_2
&= 1 + \tfrac{1}{2}D\,
\mathrm{tr}(\Delta_0\tilde{\mathcal G}^{(1)}
\Delta_0\tilde{\mathcal G}^{(2)}) + \cdots
\nonumber\\
&= 1 +
\frac{D\,|(\xi_1-\eta_1)(\xi_2-\eta_2)|}
{2(2\pi)^2 \log|k_1| \log|k_2|}
\int
\frac{d^2 z_1\, d^2 z_2}
{\big|(z_1-\xi_1)(z_1-\eta_1)
 (z_2-\xi_2)(z_2-\eta_2)\big|^2}
+ \cdots ,
\tag{9.33}
\end{align}
where we have applied (7.10) and changed the superscripts (1, 2) for
the subscripts for correspondence with (8.10). Using the formal
projective invariance of $\mathcal M_{2}$ and the measures $d\mu_{n}$
we may fix it in the same way as we did in Sec.~8: $\xi_{1}=\infty$,
$\eta_{1}=1$, $\eta_{2}=0$, $\xi_{2}=\xi$. The limit $\xi \to 0$ then
corresponds to the factorization of the genus 2 surface on two tori
(see (8.11)--(8.18)). According to (7.33) the second term in (9.31)
contains the logarithmic singularity which appears to cancel the
corresponding singularity (8.18) in $\int d\mu_{2}$ if
$\tfrac{1}{2}q_{2}=(2\pi)^{4}$, cf.\ (8.19), (8.20). The same condition
guarantees the cancellation of the quadratic tachyonic infinities
originating from the leading order terms in the expansions of the
measures.\foot{Note the presence in (9.33) of the same factor $D$ which we have found
in Sec.~8 from the expansion of the determinant of the period matrix
factor in $d\mu_{2}$.}

The cancellation of the tadpole divergence in $\int d\mu_{2}$ against
the singularity in the expectation value of the square of the one-handle
operator in (9.31) does \emph{not} imply that all leading multiple
divergences in the generating functional $\hat Z$ may be represented by
multiple insertions of $\hat H_{1}$. In the case of a nontrivial
background ($I_{\text{int}} \neq 0$) one is to consider factorizations
with external ``legs'' on one part of a diagram. However, if a
generalization of (9.29) with $\mathcal F \rightarrow \mathcal F -
I_{\text{int}}$ is true at least as some approximation (cf.\ (9.23),
(9.26) and (9.27)) which should be sufficient for the analysis of the
RG, the singularities which appear in $\mathcal F$ and hence the
corresponding counterterms should be universal. 

Treating $\hat H_{n}$
as a kind of vertex operators, we expect to find singularities when the
integrands of (arbitrary number of) $\hat H_{n}$ and (or) of the
ordinary vertex operators in $I_{\text{int}}$ ``collide''. This easily
explains the exponentiation of the external leg divergences  [{18}]
and also implies that the loop-corrected $\beta$-functions should
contain terms of all powers in the massless fields.\foot{At this point we disagree with Refs.~17 and 32 where no higher order
terms in $\beta$ are found (these authors do not account for the
contributions of momentum dependent singularities of the amplitudes
(see Sec.~3)).  }
The question of renormalizability of $\hat Z$ is thus reduced to that
of renormalizability of the theory $\hat I = I - \mathcal F$ on the
sphere.

There are some subtleties in the renormalization procedure at the
$\log^{k}\vepsilon$, $k \ge 2$, level. Consider, for example, the genus 2
contributions. Splitting $\hat H_{1}$, $\hat H_{2}$ into a finite piece
and
a logarithmic divergence (proportional to $\O_n$),
$\hat H_n = \hat H_{n f} + \hat H_{n\infty}$, we find for the part of
$\hat Z_2$ which is divergent due to the tadpole singularities
\begin{equation}
\hat Z_{2\infty} \sim
\langle e^{-I_{\rm int}} \hat H_{2 f} \rangle_0
+ \langle e^{-I_{\rm int}} \hat H_{1 f}\hat H_{1\infty} \rangle_0
+ \langle e^{-I_{\rm int}} \hat H_{1\infty}^2 \rangle_0\ .
\tag{9.34}
\end{equation}
The first two terms here are $O( \log \vepsilon)$ while the third is
$O( \log^2 \vepsilon)$. From the analysis of factorization one, however,
expects to find an additional $O( \log^2 \vepsilon)$ term corresponding to
the 2-point function on the torus connected with the tadpole on the
torus and the sphere with external leg insertions. 

One possibility
could be that the combinatorics of string diagrams is such that this
extra term combines together with the term corresponding to the two
separate tadpole insertions on the sphere to be in agreement with the
last term in (9.34). In fact, there is an additional
$O(g^4  \log^2 \vepsilon)$ term present in the exponent which produces an
additional $ \log^2 \vepsilon$ term in (9.34) (see Ref.~77). This is
suggested by the observation that we are also to renormalize the string
coupling which appears in $\mathcal F$. To check this directly it is
desirable to study the OPE relations for the operators $\hat H_n$.

We finish with a remark, that the approach discussed in this section
seems to be potentially very powerful. Combined with the requirement of
renormalizability it may give a clue to a more fundamental formulation
of the theory. To see if this is actually so one is still to get a
deeper understanding of how the RG is actually realized at higher
($n=2,3,\ldots$) loop orders and to understand in general the role of
higher $ \log^k \vepsilon$, $k \ge 2$ divergences  [{77}]. 
There also remain some open questions concerning a prescription for
subtraction of the Möbius volume divergence.

\section*{10.\ Concluding remarks}

In this paper we have discussed some issues related to the problem of
extension of the ``renormalization group'' approach to string loop
level. We have emphasized the central role played by the string
generating functional $\hat Z$ in the first quantized formulation of
the string theory and have checked the renormalizability property of
$\hat Z$ on some particular examples. It was found that in order to
have the renormalizability of $\hat Z$ with respect to both ``local''
and ``modular'' infinities one should define $\hat Z$ by using a
special prescription for subtraction of Möbius infinities and also by
employing the ``extended'' (M\"obius non-fixed) Schottky parametrization
of the moduli space.

There are of course  many  open problems. We still do not have a
fully satisfactory regularization prescription which simultaneously
cuts off both the ``local'' and ``modular'' infinities. While the
``local'' divergences can be automatically regularized by inserting a
cutoff in the 2-d propagator we can only cut off the modular integrals
``by hand'' (e.g.\ by imposing a condition that the points on the
complex plane cannot come closer than at distance $ \vepsilon$, etc.).
The problem of a cutoff prescription is closely related to that of an
off-shell extension. It is necessary to go off the tree level conformal
point in order to obtain the loop corrections to $\beta$-functions and
hence to string equations of motion. Though it appears that the string
generating functional (defined in such a way that to preserve the
basic low energy space-time symmetries) provides a possible off-shell
extension, there remain ambiguities related to field redefinitions,
choice of world sheet
parametrization and also a question of correspondence between a
particular off-shell extension and a particular cutoff prescription.

The  loop expansion in the bosonic string theory is not well-defined because of the tachyonic instability of the tree
level vacua. It is unlikely to be possible to get rid of the tachyon by
making small shifts of the ``massless'' couplings. This difficulty is
absent in all tachyon-free superstring theories in which all the loop
divergences can be given ``a renormalization group'' interpretation,
i.e.\ can be cancelled out by a renormalization of the ``massless''
couplings (see [10,71]).\foot{The $\operatorname{Im}\,\tau \to \infty$ singularities which sometimes
appear in the superstring effective actions (see e.g.\ Ref.\ 72) are of
IR nature (are due to the massless loops). They appear only within the
small momentum expansion (which, in the absence of other dimensional
parameters, is equivalent to the $\alpha'$ expansion). These
singularities correspond to the UV infinities in the ``$\alpha' = 0$''
massless field theory (see also [73]). The loop corrections to the
effective action are, in fact, finite but in general nonlocal (contain,
e.g., $\log(\alpha' k^2)$ structures) and hence cannot be regularly
expanded in powers of momenta. Other references discussing loop
corrections to superstring effective actions include [10, 12, 55, 74, 75].}
 Hence most of
the discussion above  is, strictly speaking, true only in
the superstring case.\foot{From the technical point of view, the superstring case is also
distinguished by the absence of power divergences, what makes the
analysis of regularized theory less ambiguous.
}

\newpage

\section*{Appendix}

Below we shall compute the leading terms in the string generating
functional $\hat Z$ which depend on the space-time metric.
We shall not be interested in a particular prescription of subtraction
of M\"obius infinities and shall not specify the modular measure.
Hence we shall consider essentially the computation of the
$\sigma$-model partition function $Z$ on a corresponding compact
2-surface.

In general, $Z$ should be defined in such a way that preserves general
covariance and other ``low energy'' space-time symmetries.
The definition of $Z$ includes a choice of the path integral measure and
a choice of a regularization and renormalization prescription
(see Refs.~7, 21 and~22).
The computational procedure we shall use is in direct correspondence with
the usual vertex operator approach for computing string amplitudes.

Namely, we shall choose the trivial (space-time metric $G_{\mu\nu}$
independent) path integral measure and compute $Z$ by expanding near the
flat space, i.e.\ expanding in powers of
$h_{\mu\nu}=G_{\mu\nu}-\delta_{\mu\nu}$.
We shall introduce a cutoff through the 2d Green function.
To obtain the covariant expression for $Z$ it is then necessary to
specify a particular prescription of subtraction of power (quadratic)
short distance 2d infinities.

We shall start with ($2\pi\alpha' = 1$) 
\begin{equation}
Z=\int [Dx]\,
\exp\!\Big[
-\frac{1}{2}\int d^{2}z\,\sqrt{g}\,
g^{ab}\partial_{a}x^{\mu}\partial_{b}x^{\nu}
G_{\mu\nu}(x)
\Big],
\tag{A.1}
\qquad \ \ \ 
G_{\mu\nu}=\delta_{\mu\nu}+h_{\mu\nu}\ .
\end{equation}
Separating the integral over the constant zero mode of the Laplacian
($x^{\mu}=y^{\mu}+\eta^{\mu}(z)$, see Sec.~2) we get ($ \langle 1\rangle = 1$)
\begin{align}
\  &\qquad \qquad Z = c \int d^{D}y\, \mathcal L\  , \tag{A.2} \\
\mathcal L &= 1
-\frac{1}{2}\Big\langle
\int d^{2}z\,\sqrt{g}\, g^{ab}
\partial_{a}\eta^{\mu}\partial_{b}\eta^{\nu}
h_{\mu\nu}(y+\eta)
\Big\rangle
+\frac{1}{8}\Big\langle
\Big(
\int d^{2}z\,\sqrt{g}\, g^{ab}
\partial_{a}\eta^{\mu}\partial_{b}\eta^{\nu}
h_{\mu\nu}(y+\eta)
\Big)^{2}
\Big\rangle
+ \cdots 
\no 
\end{align}
Introducing a cutoff into the Green's  function
$\langle \eta^{\mu}(z)\eta^{\nu}(z')\rangle
= \delta^{\mu\nu} \G(z,z')$
(see (2.5) and (2.6)) we find, using integration by parts
(cf.\ (7.54)--(7.59))
\begin{align}
&\int d^{2}z\,\sqrt{g}\, g^{ab}
\langle \partial_{a}\eta^{\mu}\partial_{b}\eta^{\nu}\rangle 
= \delta^{\mu\nu}\lim_{1\to 2}
\int d^{2}z_{1}\,\sqrt{g}\,\nabla^{a}_{1}\nabla_{a2}\,  \G_{12}
\nonumber\\
&= \delta^{\mu\nu}\lim_{1\to 2}
\int d^{2}z_{1}\,\sqrt{g}\,
\Big[\delta^{(2)}(z_{1},z_{2})-\frac{1}{V}\Big]
= \delta^{\mu\nu}(\bar\delta-1),
\tag{A.3}\\
 &
V \equiv \int d^{2}z\,\sqrt{g},
\qquad
\bar\delta \equiv \int d^{2}z\,\sqrt{g}\,\delta^{(2)}(z,z'),
\qquad
\G_{12}\equiv \G(z_{1},z_{2}) \ , \no 
\\
&\Big\langle
(\partial\eta^{\mu}\bar\partial\eta^{\nu})_{1}
(\partial\eta^{\alpha}\bar\partial\eta^{\beta})_{2}
\Big\rangle
= A_{1}\,\delta^{\mu\nu}\delta^{\alpha\beta}
+ A_{2}\,(\delta^{\mu\alpha}\delta^{\nu\beta}
+ \delta^{\mu\beta}\delta^{\nu\alpha}), \qquad \ \  A_{1}=(\bar\delta-1)^{2},
\tag{A.4}\\
& A_{2}
= \int d^{2}z_{1}\sqrt{g(z_{1})}
\int d^{2}z_{2}\sqrt{g(z_{2})}\,
\nabla_{1}^{a}\nabla_{2}^{b}\, \G_{12}
\nabla_{1a}\nabla_{2b}\, \G_{12}
\nonumber\\
&\qquad = \int d^{2}z_{1}\sqrt{g(z_{1})}
\int d^{2}z_{2}\sqrt{g(z_{2})}
\Big[\delta^{(2)}(z_{1},z_{2})-\frac{1}{V}\Big]^{2}
= \bar\delta - 1\ .\no 
\end{align}
The regularized expression for $\bar\delta$ is
\begin{equation}
\bar\delta = \frac{a_1}{ \vepsilon^2} + a_2 + O( \vepsilon).
\tag{A.5}
\end{equation}
For example, in the heat kernel regularization $a_1 = 1$,
$a_2 = \tfrac{1}{6}\chi$ where $\chi$ is the Euler number of the 2-space.
If we set $\bar\delta = 0$ as our renormalization prescription
(cf.\ Refs.~21 and~22), i.e.\ if we drop the quadratic divergence
together with the finite part in (A.5) we get the covariant expression
for the derivative independent term in $\mathcal L$ in (A.2) 
\begin{equation}
\mathcal L
= 1 + \frac{1}{2} h^\mu{}_\mu
+ \frac{1}{8}(h^\mu{}_\mu)^2
- \frac{1}{4} h_{\mu\nu} h^{\mu\nu}
+ \cdots
= \sqrt{G}\,(1+\cdots)\ .
\tag{A.6}
\end{equation}
The prescription $\bar\delta = 0$ corresponds to setting the regularized
total number of the eigenmodes of the Laplacian equal to zero.
$\sqrt{G}$ is thus the ``contribution'' of the zero mode.

Next, consider the derivative dependent terms in $\mathcal{L}$. Dropping for simplicity the terms
with $h^{\mu}{}_{\mu}$ and $\partial_\mu h_{\mu\nu}$ we get
\begin{equation}
\mathcal{L}
= 1 + c_1\, h_{\mu\nu}\,\partial^2 h_{\mu\nu}
  + c_2\, \partial_\lambda h_{\mu\nu}\,\partial_\lambda h_{\mu\nu}
  + \ldots ,
\tag{A.7}
\end{equation}
\[
c_1
= \frac{1}{4}
  \int_{1}\!\!\int_{2}
  \,\mathcal{G}_{11}\,
  \nabla_{1a}\nabla_{2b}\,
  \mathcal{G}_{12}\,
  \nabla_{1}^{a}\nabla_{2}^{b}\,
  \mathcal{G}_{12}\,,
\qquad \ \ \ 
c_2
= \frac{1}{4}
  \int_{1}\!\!\int_{2}
  \,\mathcal{G}_{12}\,
  \nabla_{1a}\nabla_{2b}\,
  \mathcal{G}_{12}\,
  \nabla_{1}^{a}\nabla_{2}^{b}\,
  \mathcal{G}_{12}\,.
\]
The two terms in (A.7) correspond to the $O(k^2)$ terms in the off-shell 2-graviton
amplitude. It is straightforward to check that if $\bar{\delta}=0$ the coefficients
$c_i$ can be related to the coefficients which appear in the 3-graviton on-shell
amplitude, i.e.\ in the $O(\partial^2 h^3)$ term in $\mathcal{L}$. Observing that
\begin{align}
\int d^{D}y\,\sqrt{G}\,R
= \int d^{D}y \Big[ &
\frac{1}{4}\,h_{\mu\nu}\,\partial^2 h_{\mu\nu}
-\frac{1}{4}\!\left(
h_{\mu\nu}\,h_{\alpha\beta}\,\partial_\mu\partial_\nu h_{\alpha\beta}
+ 2\,h_{\mu\rho}\,\partial_\mu h_{\nu\lambda}\,\partial_\nu h_{\rho\lambda}
\right)\no \\
&+ O\!\left(h_{\mu\nu},\,\partial^2 h\, h h,\,\partial_\mu h_{\mu\nu}\right)
\Big],
\tag{A.8}
\end{align}
we can rewrite $\mathcal{L}$ as follows
\begin{align}
&\mathcal L
= \sqrt{G}\,\big[1+\gamma_1 R + O(\partial^3 h)\big] \ , 
\tag{A.9}
\\
&\gamma_1 = c_1 - c_2
= \int_{1}\!\!\int_{2}
(\mathcal{G}_{11} - \mathcal{G}_{12})
\nabla_{1a}\nabla_{2b}\mathcal{G}_{12}\,
\nabla_{1}^{a}\nabla_{2}^{b}\mathcal{G}_{12}\, .
\tag{A.10}
\end{align}
Thus if we drop all power divergences using the $\bar{\delta}=0$ prescription, all the
$O(\partial^{2}h)$ terms in $\mathcal{L}$ group together to form the Ricci scalar.\footnote{For example, 
the $hh\partial\partial h$ term in (A.8) appears in $\mathcal{L}$ with the coefficient
$\int_{1}\!\int_{2}\!\int_{3}
\big(\nabla_{1}(\mathcal{G}_{12}-\mathcal{G}_{13})\big)^{3}
\big(\nabla_{2}\nabla_{3}\mathcal{G}_{23}\big)^{2}$.
Integrating by parts and using (2.6) and $\bar{\delta}=0$ we can relate this coefficient to $\gamma_1$.}
Note that dropping terms proportional to $\bar{\delta}$ we thus drop some particular
combinations of the $1/\varepsilon^{4}$, $1/\varepsilon^{2}$, $1/\varepsilon^{2}\log\varepsilon$,
$\log\varepsilon$, etc.\ terms.

Let us compute the logarithmically divergent part of $\gamma_1$. Integrating by parts we find
\[
\qquad \qquad c_2
= -\frac{3}{2}\int_{1}\!\!\int_{2}
\Big[
\delta^{(2)}(z_1,z_2)\,
\nabla_{1}\mathcal{G}_{12}\,
\nabla_{1}\mathcal{G}_{12}
+\frac{1}{2}\left(\nabla_{1}\mathcal{G}_{12}\right)^2
\Big]
-\mathcal{G}_{11}
\]
\[
= -\frac{3}{2}\int_{1}\!\!\int_{2}
\delta^{(2)}(z_1,z_2)\,
\nabla_{1}\mathcal{G}_{12}\,
\nabla_{1}\mathcal{G}_{12}
-\frac{1}{2}\mathcal{G}_{11}\, .
\tag{A.11}
\]
Here we have set  again  $\bar{\delta}=0$ and assumed for simplicity that
$\mathcal{G}_{11}=\text{const}$ (which  is true for homogeneous spaces).
The first term in (A.11) is finite (modulo quadratic divergences).
This is easy to check explicitly, e.g.\ for the cases of the 2-sphere $S^2$  and the
2-torus $T^2$. For $S^2$ we get 
\[
\partial \mathcal{G}
\sim \frac{\bar{z}_{12}}{|z_{12}|^{2}+\varepsilon^{2}},
\qquad
\delta^{(2)}\,\partial \mathcal{G}\,\bar{\partial}\mathcal{G}
\rightarrow 0 .
\]
For $T^2$ we may start from the original expression for $c_2$ and note that
\[
\partial\bar{\partial}\mathcal{G}
= \frac{1}{4\tau_2},
\qquad
\partial^{2}\mathcal{G}
= \frac{1}{4\pi}\frac{1}{\bar{z}^{2}},
\qquad
\tau_2 = \mathrm{Im}\,\tau ,
\]
\[
\mathcal{G}\big|_{z_1\to z_2}
= -\frac{1}{4\pi}\log |z_{12}|^{2}
+ \frac{1}{4\tau_2}|z_{12}|^{2}
+ O(z^{2},\bar{z}^{2}) .
\]
Finally, using that
\[
\mathcal{G}_{11}
= -\frac{1}{4\pi}\log \varepsilon^{2}
+ \text{finite},
\]
we get (dropping again $\bar{\delta}$ in $c_1$) 
\[
\gamma_1
= -\frac{1}{2}\mathcal{G}_{11}
+ \text{finite}
= \frac{1}{4\pi}\log \varepsilon
+ \text{finite}, \]
\[\tag{A.12}
\qquad \ \ \ \ \ 
\mathcal{L}
= \sqrt{G}\Big(1 + \frac{1}{2}\alpha' \log\varepsilon\, R + \ldots \Big)\ . 
\]
Here we have restored the dependence on $\alpha'$ ($2\pi\alpha'=1$ before).

The same result is found, e.g., for the unit disc.
We would like to emphasize again that this result is true only if one
uses the covariant renormalization prescription ($\bar\delta=0$).
If one computes the graviton correlators on the disc and simply
regularizes them by some arbitrary short distance cutoff,  one in general
would be unable to group particular terms into $R$ unless one drops some
properly chosen terms. Also, the $\log \vepsilon$ coefficient of some
representative term in $R$ will look ambiguous: the presence of the
$1/ \vepsilon^2$ terms implies that the value of $\log \vepsilon$ coefficient
depends on how one subtracts the power divergences.

Let us note also that, in principle, the noncovariant expression for the
string generating functional one finds using an arbitrary regularization
scheme is not necessarily in contradiction with the covariant expression
for the string effective action
$
S \sim \int ( \sqrt{G}\, R +\lambda  \sqrt{G}\,+ \cdots) .
$
In fact, one can get different results for the field theory amplitudes
using different prescriptions of how to take the on-shell and
$\lambda\to\rm const$ limits (cf.\ Ref.~63).

For example, the result for the 2-graviton amplitude found in Ref.~76
using ``noncovariant'' regularization prescription can be reproduced
starting from the covariant effective action and accounting for the
tadpole contribution which is nonvanishing if the limit
$\lambda\to\rm const$ is taken after one cancels out the graviton pole
against the $k^{-2}$ factor in the tree 3-graviton vertex.
The resulting 2-graviton field theory amplitude is of course gauge
 dependent as is the 2-graviton string amplitude (which
depends on how one subtracts the power divergences).
Our point of view is that one should use the simplest and natural
prescription under which:
(i) in field theory $\lambda$ is taken constant from the very beginning
and hence the field theory 2-graviton amplitude is given simply by the
contact term in $\sqrt{G}$,  and
(ii) in string theory one uses the particular covariant subtraction
prescription which gives the covariant expression for the graviton
amplitudes (e.g.\ the value of 2-graviton amplitude is consistent with
the $\sqrt{G}$ term in $\hat Z$).

\section*{Acknowledgments}

The author is grateful to D.~Amati, O.~Andreev, R. Metsaev  and J.~Russo for useful
discussions.

\newpage 

\small

\section*{References}

\begin{enumerate}

\item C.~Lovelace, ``Strings in curved space,'' Phys.\ Lett.\ B135 (1984) 75; ``Stability of String Vacua. I. A New Picture of the Renormalization Group,'' Nucl.\ Phys.\ B273 (1986) 413.

\item E.~S.~Fradkin and A.~A.~Tseytlin, ``Effective Field Theory from Quantized Strings,'' Phys.\ Lett.\ B158 (1985) 316; ``Quantum String Theory Effective Action,'' Nucl.\ Phys.\ B261 (1985) 1.

\item C.~Callan, D.~Friedan, E.~Martinec and M.~Perry, ``Strings in Background Fields,'' Nucl.\ Phys.\ B262 (1985) 593.

\item A.~Sen, ``Equations of Motion for the Heterotic String Theory from the Conformal Invariance of the Sigma Model,'' Phys.\ Rev.\ D32 (1985) 2102.

\item P.~Candelas, G.~Horowitz, A.~Strominger and E.~Witten, ``Vacuum Configurations for Superstrings,'' Nucl.\ Phys.\ B256 (1985) 46.

\item T.~Banks and E.~Martinec, ``The Renormalization Group and String Field Theory,'' Nucl.\ Phys.\ B294 (1987) 733.

\item A.~A.~Tseytlin, ``Sigma model approach to string theory,'' Int.\ J.\ Mod.\ Phys.\ A4 (1989) 1987.

\item A.~A.~Tseytlin, ``On the Renormalization Group Approach to String Equations of Motion,''  Int.\ J.\ Mod.\ Phys.\ A4 (1989) 4249.

\item W.~Fischler and L.~Susskind, ``Dilaton Tadpoles, String Condensates and Scale Invariance,'' Phys.\ Lett.\ B171 (1986) 383; ``Dilaton Tadpoles, String Condensates and Scale Invariance II,'' Phys.\ Lett.\ B173 (1986) 262.

\item C.~Callan, C.~Lovelace, C.~Nappi and S.~Yost, ``String Loop Corrections to beta Functions,'' Nucl.\ Phys.\ B288 (1987) 525; ``Adding Holes and Crosscaps to the Superstring,'' Nucl.\ Phys.\  B293 (1987) 83; 
``Loop Corrections to Superstring Equations of Motion,''
 Nucl. Phys. B308 (1988) 221; 
 ``Loop Corrections to Conformal Invariance for Type 1 Superstrings,''  Phys.Lett.B206 (1988) 41;

J. Polchinski  and Y.  Cai ``Consistency of Open Superstring Theories,''
Nucl. Phys. B{296} (1988)  91.

\item S.~Das and S.~J.~Rey, ``Dilaton condensates and loop effects in open and closed bosonic strings,'' Phys.\ Lett.\ B186 (1987) 328. 

\item   K.~Inami and H.~Nishino, 
``Superstring loop corrections to sigma-model beta-functions,'' Phys.\ Lett.\ B196 (1987) 151.

\item W.~Fischler, I.~Klebanov and L.~Susskind, ``String loop divergences and effective lagrangians,'' Nucl.\ Phys.\ B306 (1988) 271; I.~Klebanov, SLAC preprint (1988). 

\item A.~A.~Tseytlin, ``String theory effective action: string loop corrections,'' Int.\ J.\ Mod.\ Phys.\ A3 (1987) 365. 

\item A.~Polchinski, ``Factorization of bosonic string amplitudes,'' Nucl.\ Phys.\ B307 (1988) 61. 

\item A. A. Tseytlin, ``Mobius infinity subtraction and effective action in $\sigma$ model approach to closed string theory,'' Phys.\ Lett.\ B208 (1988) 221. 

\item S.~Das, ``Renormalizing handles and holes in string theory,'' Phys.\ Rev.\ D38 (1988) 3105. 

\item H.~Ooguri and N.~Sakai, ``String Loop Corrections From Fusion of Handles and Vertex Operators,'' Phys.\ Lett.\ B197 (1987) 109; ``String multiloop corrections to equations of motion,'' Nucl.\ Phys.\ B312 (1989) 435.

\item A.~A.~Tseytlin, ``Graviton amplitudes, effective action and string generating functional on the disk,"  Int.\ J.\ Mod.\ Phys.\ A4 (1989) 3269.

\item J.~Liu and J.~Polchinski,  ``Renormalization of the Mobius Volume,'' 
Phys.\ Lett.\ B203 (1988) 39.

\item A.~A.~Tseytlin, ``Mobius Infinity Subtraction and Effective Action in $\sigma$ Model Approach to Closed String Theory,''  
Phys.\ Lett.\ B208 (1988) 221.

\item A.~A.~Tseytlin, ``Partition Function of String 
$\sigma $ Model on a Compact Two Space,''  Phys.\ Lett.\ B223 (1989) 165.

\item (a) C.~Lovelace, ``Simple n-reggeon vertex,'' Phys.\ Lett.\ B32 (1970) 490; ``M-loop generalized Veneziano formula,''
 Phys.\ Lett.\ B32 (1970) 703;

D.~Olive, ``Operator vertices and propagators in dual theories,'' Nuovo Cimento 3A (1971) 399;

M.~Kaku and L.~P.~Yu, ``Unitarization of the dual-resonance amplitude. I. planar n-loop amplitude,''  Phys.\ Rev.\ D3 (1971) 2992;

M.~Kaku and J.~Scherk, ``Divergence of the two-loop planar graph in the dual-resonance model,''  Phys.\ Rev.\ D3 (1971) 430;

(b) V.~Alessandrini, ``A general approach to dual multiloop diagrams,''  Nuovo Cimento 2A (1971) 321;

V.~Alessandrini and D.~Amati, ``Properties of dual multiloop amplitudes,'' 
Nuovo Cimento 4A (1971) 793.


\item E.~Gava, R.~Iengo, T.~Jayraman and R.~Ramachandran,  
``Multiloop Divergences in the Closed Bosonic String Theory,''  
Phys.\ Lett.\ B168 (1986) 207. 

\item A.~A.~Belavin and V.~G.~Knizhnik,  ``Algebraic Geometry and the Geometry of Quantum Strings,''
Phys.\ Lett.\ B168 (1986) 201;  ``Complex geometry and the theory of quantum strings,''  
Sov.\ Phys.\ JETP 91 (1986) 364. 

\item E.~Martinec,  
``Conformal Field Theory on a (Super)Riemann Surface,''  
Nucl.\ Phys.\ B281 (1987) 157.

\item S.~Weinberg,  
``Cancellation of One Loop Divergences in SO(8192) String Theory,''
Phys.\ Lett.\ B187 (1987) 278. 

\item N.~Marcus,  
``Unitarity and regularized divergences in string amplitudes,''  
Phys.\ Lett.\ B219 (1989) 265. 

\item L.~Ford, {\it ``Automorphic functions''}  
(Chelsea, New York, 1951).

\item S.~Mandelstam, ``The interacting string picture and functional integration," 
 in {\it Unified string theories},  
Proceedings of the Santa Barbara Workshop,  
eds.\ M.~Green and D.~Gross  
(World Scientific, Singapore, 1986), p.~46.

\item A.~Neveu and P.~West, ``Group theoretic approach to the perturbative string   S matrix,''
Phys.\ Lett.\ B193 (1987) 187; ``Group theoretic approach to the open bosonic string multiloop S matrix,''
Commun.\ Math.\ Phys.\ 114 (1988) 613.

 P.~Di~Vecchia, R.~Nakayama, J.~Petersen, J.~Sidenius and S.~Sciuto,
``BRST invariant N reggeon vertex,'' Phys.\ Lett.\ B182 (1986) 164;
``Covariant N string amplitude,'' Nucl.\ Phys.\ B287 (1987) 621;
``N String, g  Loop Vertex for the Bosonic String,''
Phys.\ Lett.\ B206 (1988) 643;

P.~Di~Vecchia, M.~Frau, A.~Lerda and S.~Sciuto,
``A simple expression for the multiloop amplitude in the bosonic string,'' 
Phys.\ Lett.\ B199 (1987) 49; ``N String Vertex and Loop Calculation in the Bosonic String,''
Nucl.\ Phys.\ B298 (1988) 527;

J.~Petersen and J.~Sidenius, ``Covariant loop calculus for the closed bosonic string,''
Nucl.\ Phys.\ B301 (1988) 247.

\item U.~Ellwanger and M.~G.~Schmidt,  ``Loop Corrected String Field Equations From World Sheet Weyl Invariance,'' 
 Z.Phys.C 43 (1989) 485;
U.~Ellwanger, 
``The exact renormalization group and string  loop corrections  for the NSR   superstring,''
Nucl. Phys. B{332}  (1990)  300

\item D.~H.~Friedan, 
``Nonlinear Models in Two Epsilon Dimensions,''
Phys. Rev. Lett. {45} (1980) 1057;
``Nonlinear Models in Two + Epsilon Dimensions,''
Annals Phys. {163} (1985) 318

\item S.~Randjbar-Daemi, A.~Salam and J.~A.~Strathdee,
``$\sigma$ Models and String Theories,''
Int. J. Mod. Phys. A {2}   (1987)  667 

\item G. Keller and A. Silvotti,
``Quantum Measure and Weyl Anomaly of Two-Dimensional Bosonic Nonlinear Sigma Models,''
Ann. Phys. 183 (1988) 269;

I.~Gerstein, R.~Jackiw, B.~Lee and S.~Weinberg, ``Chiral loops,''
Phys.\ Rev.\ D3 (1971) 2486.

\item A. A. Tseytlin,
``Conformal Anomaly in Two-Dimensional Sigma Model on Curved Background and Strings", 
Phys. Lett. B178 (1986) 34.

\item A.~M.~Polyakov, ``Quantum Geometry of Bosonic Strings,''   Phys.\ Lett.\ B103 (1981) 207.

\item 
H. Osborn,
``Renormalization and Composite Operators in Nonlinear Sigma Models,'' 
Nucl. Phys. B294 (1987) 595.

\item A. B. Zamolodchikov,
``Irreversibility of the Flux of the Renormalization Group in a 2D Field Theory,''
JETP Lett. 43 (1986) 730, Pisma Zh.Eksp.Teor.Fiz. 43 (1986) 565.

A.~M.~Polyakov, ``Directions in string theory,'' Phys.\ Scripta T15 (1987) 191.

\item A.~A.~Tseytlin, ``Vector Field Effective Action in the Open Superstring Theory,'' Nucl.\ Phys.\ B276 (1986) 391; 
``Conditions of Weyl Invariance of Two-dimensional $\sigma$ Model From Equations of Stationarity of 'Central Charge' Action,'' Phys.\ Lett.\ B194 (1987) 63.

\item 
H.~Osborn,
``String theory effective actions from bosonic $\sigma$ models,''
Nucl.\ Phys.\ B308 (1988) 629; 
``Renormalization group and two point functions in bosonic $\sigma$ models,''
Phys.\ Lett.\ B214 (1988) 555.

\item
B.~E.~Fridling and A.~Jevicki,
``Nonlinear $\sigma$ models as S matrix generating functionals of strings,''
Phys.\ Lett.\ B174 (1986) 75.

\item 
D.~Shevitz and A.~Strominger, ``Two-dimensional approach  to string field theory,'' 
Nucl. Phys. B327 (1989) 621

\item 
S.~R.~Das and B.~Sathiapalan, ``String propagation in a tachyon background,'' 
Phys.\ Rev.\ Lett.\ 56 (1986) 2664; 57 (1987) 1511; 

R.~Akhoury and Y.~Okada, ``Unitarity constraints for string propagation in the presence of background fields,'' 
Phys.\ Lett.\ B183 (1987) 65. 

C.~Ito and Y.~Watanabe, ``Nonperturbative effect in the beta functions and the equations of motion for string,'' 
Phys.\ Lett.\ B198 (1987) 486; 

R.~Brustein, D.~Nemeschansky and S.~Yankielowicz, ``Beta functions and S matrix in string theory,'' 
Nucl.\ Phys.\ B301 (1988) 224; 

I.~Klebanov and L.~Susskind, ``Renormalization group and string amplitudes,'' 
Phys.\ Lett.\ B200 (1988) 446; 

V.~Periwal, ``Renormalization and string amplitudes,'' Mod.\ Phys.\ Lett.\ A4 (1989) 33;

\item 
D.~G.~Boulware and L.~S.~Brown,
``Tree graphs and classical fields,''
Phys.\ Rev.\ 172 (1968) 1628; 

I.~Ya.~Arefeva, L.~D.~Faddeev and A.~A.~Slavnov, 
``Generating functional for the S matrix in gauge theories,''
Theor.\ Math.\ Phys.\ 21 (1975) 1165, Teor.\ Mat.\ Fiz.\ 21 (1974) 311.  

A.~Jevicki and C.~Lee,
``S-matrix generating functional and effective action,''
Phys.\ Rev.\ D37 (1988) 1485.

\item T.~Kubota and G.~Veneziano,
``Off-shell Effective Actions in String Theory'',
Phys.\ Lett.\ B207 (1988) 419.

\item V.~G.~Knizhnik, ``Analytic fields on riemann surfaces. II,''
Commun.\ Math.\ Phys.\ 115 (1987) 567;

L.~Dixon, D.~Friedan, E.~Martinec and S.~Shenker,
``The conformal field theory of orbifolds,''
Nucl.\ Phys.\ B282 (1987) 13; 

M.~Bershadsky and A.~Radul,
``Conformal field theories with additional Z(n) symmetry,''
Int.\ J.\ Mod.\ Phys.\ A2 (1987) 165;
``g-Loop Amplitudes in Bosonic String Theory in Terms of Branch Points,''
Phys.Lett.B193 (1987) 213.

\item K.~Amano  and A.~Tsuchiya, ``Mass Splittings and the Finiteness Problem of Mass Shifts in the Type {II} Superstring at One Loop,''
Phys.\ Rev.\ D39 (1989) 565.

\item S.~J.~Rey,  ``Unified view of BRST anomaly and its cancellation in string amplitudes,''  
Nucl.\ Phys.\ B316 (1989) 197.

\item E.~Cohen, H.~Kluberg-Stern and R.~B.~Peschanski,
``One Loop Regularization of the Polyakov String Functional,''
Nucl. Phys. B{328}  (1989)  499.

\item S.~Weinberg, ``Radiative corrections  in string theory,''
{\it The Oregon Meeting, Proceedings of the Annual Meeting of the DPF of APS},
Eugene, Oregon, 1985, ed.\ R.~Hua
(World Scientific, Singapore, 1986), p.~850;

N.~Seiberg,
``Anomalous dimensions and mass renormalization in string theory,''
Phys.\ Lett.\ B187 (1987) 56; 

A.~Sen,
``Mass renormalization and cancellation of  BRST  anomaly in string theories,''
Nucl.\ Phys.\ B304 (1988) 403. 

\item S.~Weinberg, ``Coupling Constants and Vertex Functions in String Theories,'' 
Phys.\ Lett.\ B156 (1985) 309.

\item E.~Witten, ``Some properties of O(32) superstrings,'' Phys.\ Lett.\ B149 (1984) 351; 
``Dimensional reduction of superstring models,''
Phys.\ Lett.\ B155 (1984) 151.

\item M.~Dine and N.~Seiberg, ``Couplings and Scales in Superstring Models,'' 
Phys.\ Rev.\ Lett.\ 55 (1985) 366. 

\item N.~Sakai and Y.~Tanii,
``One loop amplitudes and effective action in superstring theories,''
Nucl.\ Phys.\ B287 (1987) 457. 

\item A.~Polchinski,
``Evaluation of the one loop string path integral,''
Commun.\ Math.\ Phys.\ 104 (1986) 37.

\item R.~Rohm, 
``Spontaneous supersymmetry breaking in supersymmetric string theories,''
Nucl.\ Phys.\ B237 (1984) 553. 

\item J.~Shapiro, 
``Loop graph in the dual tube model,''
Phys.\ Rev.\ D5 (1972) 1945.

\item J.~H.~Schwarz, ``Superstring theory,''
Phys.\ Reports 89C (1982) 223;

E.~D'Hoker and D.~Phong,  ``The Geometry of String Perturbation Theory,''
Rev.\ Mod.\ Phys.\ 60 (1988) 917.

\item L.~G.~Koh and H.~J.~Shin, ``World-sheet topology and target manifold in string theory,''
Phys.\ Rev.\ D36 (1987) 1773.

\item A.~A.~Tseytlin,
``Renormalization of Mobius Infinities and Partition Function Representation for String Theory Effective Action,''
Phys.\ Lett.\ B202 (1988) 81;

O.~D.~Andreev and A.~A.~Tseytlin,
``Partition Function Representation for the Open Superstring Effective Action: Cancellation of Mobius Infinities and Derivative Corrections to Born-Infeld Lagrangian,''
Nucl.\ Phys.\ B311 (1988) 205.

\item J.~Dai and J.~Polchinski, private communication.

\item J.~Cline, ``Disk and RP2  corrections to the  string effective Lagrangian'',  
Int. J. Mod. Phys. A{4}, 5293-5319 (1989)

\item D.~Fay,
{\it Theta Functions on Riemann Surfaces},
Springer Notes in Mathematics 352 (Springer, 1973);

S.~Hamidi and C.~Vafa, ``Interactions on orbifolds,''
Nucl.\ Phys.\ B279 (1987) 465.

\item S.~de~Alwis, ``The dilaton vertex in the path integral formulation of strings,''
Phys.\ Lett.\ B168 (1986) 59.

\item K.~Roland, ``Two and three loop amplitudes in covariant loop calculus,''
Nucl.\ Phys.\ B313 (1989) 432.

\item A.~A.~Belavin, V.~G.~Knizhnik, A.~Morozov and A.~Perelomov,
``Two and three-loop amplitudes in the bosonic string theory,''
Phys.\ Lett.\ B177 (1986) 324; 

A.~Morozov,
``Explicit Formulae for One, Two, Three and Four Loop String Amplitudes,''
Phys. Lett. B184 (1987) 171;

G.~Moore, ``Modular forms and two-loop string physics,''
Phys.\ Lett.\ B176 (1986) 369;

A.~Kato, Y.~Matsuo and S.~Odake, ``Modular invariance and two loop bosonic string vacuum amplitude,''
Phys.\ Lett.\ B179 (1986) 241.

\item J.~Hughes, J.~Liu and J.~Polchinski, ``Virasoro-Shapiro From Wilson,''
Nucl.\ Phys.\ B316 (1989) 15.

\item D.~Friedan and S.~Shenker, ``The Analytic Geometry of Two-Dimensional Conformal Field Theory,''
Nucl.\ Phys.\ B281 (1987) 509.

\item V.~G.~Knizhnik,   ``Algebraic geometry and string theory,''
Kiev preprint ITP-87-62P (1987).

\item H.~S.~La and P.~Nelson,  ``Unambiguous fermionic string amplitudes,''
Phys.\ Rev.\ Lett.\ 63 (1989) 24;
``Effective Field Equations for Fermionic Strings,''
Nucl. Phys. B{332}  (1990) 83;

P.~Nelson, ``Covariant Insertion of General Vertex Operators,''
Phys.\ Rev.\ Lett.\ 62 (1989) 99.

\item J.~Minahan,  ``One Loop Amplitudes on Orbifolds and the Renormalization of Coupling Constants,''
Nucl.\ Phys.\ B298 (1988) 36;

V.~Kaplunovsky, ``One Loop Threshold Effects in String Unification,''
Nucl.\ Phys.\ B307 (1988) 145;

W.~Lerche, ``Elliptic Index and Superstring Effective Actions,''
Nucl.\ Phys.\ B308 (1988) 102;

Z.~Bern and D.~Kosower, ``A New Approach to One Loop Calculations in Gauge Theories,''
Phys.\ Rev.\ D38 (1988) 1888;
``Absence of Wave Function Renormalization in Polyakov Amplitudes,''
Nucl. Phys. B{321} (1989)  605-628;

O.~Yasuda,  ``Nonrenormalization Theorem for the Green-schwarz Counterterm and the Low-energy Effective Action,''
Phys.\ Lett.\ B218 (1989) 455.

\item R.~Metsaev and A.~Tseytlin, ``On loop corrections to string theory effective actions,'' 
Nucl.\ Phys.\ B298 (1988) 109.
 
\item J.~Ellis, P.~Jetzer and L.~Mizrachi,
 ``No Renormalization of the 
N=1 Supergravity Theory Derived From the Heterotic String,''
Phys.\ Lett.\ B196 (1987) 492;
``One Loop String Corrections to the Effective Field Theory,''
Nucl. Phys. B{303}  (1988)  1;

J.~Ellis and L.~Mizrachi, ``Multiloop No Renormalization Theorems for the Ten-dimensional Heterotic String,''
Nucl. Phys. B{327}  (1989)  595;

W.~Lerche, B.~Nilsson and A.~N.~Schellekens,  ``Heterotic String Loop Calculation of the Anomaly Cancelling Term,''
Nucl.\ Phys.\ B289 (1987) 609;

W.~Lerche, B.~Nilsson, A.~Schellekens and N.~Warner,
``Anomaly Cancelling Terms From the Elliptic Genus,''
Nucl. Phys. B{299}  (1988) 91;

D.~L\"ust, S. Theisen and G.~Zoupanos,
``Four-dimensional Heterotic Strings and Conformal Field Theory,''
Nucl.\ Phys.\ B296 (1988) 800;

K.~Meissner, J.~Pawelczyk and S.~Pokorski,
``One Loop Corrections to Four Graviton Interaction in SST II and Heterotic String Theory,''
Phys.\ Rev.\ D38 (1988) 1144;

M.~Abe, H.~Kubota and N.~Sakai,  ``Loop Corrections to the Heterotic String Effective Lagrangian,''
Phys.\ Lett.\ B200 (1988) 461;  ``Loop Corrections to the E8 x E8 
Heterotic String Effective Lagrangian,''
Nucl.\ Phys.\ B306 (1988) 405;

R.~Iengo and C.-J.~Zhu, ``Notes on Nonrenormalization Theorem in Superstring Theories,''
Phys.\ Lett.\ B212 (1988) 309;
``Two Loop Computation of the Four Particle Amplitude in Heterotic String Theory,''    Phys.\ Lett.\ B212 (1988) 313.

\item M.~Dine, W.~Fischler and N.~Seiberg,  ``Small Handles and Auxiliary Fields,''
in: 2nd Meeting on Quantum Mechanics of Fundamental Systems (CECS), 1988, p.59
Texas preprint UTIG-08-88.

\item J.~Minahan,    ``Calculation of the One Loop Graviton Mass Shift in Bosonic String Theory,''
Nucl. Phys. B{333}   (1990) 525-535.

\item J.~Russo and A.~Tseytlin,
``Renormalization of multiple infinities and renormalization group in
string loops'', Nucl.Phys.B340 (1990) 113.

\end{enumerate}

\end{document}